\documentclass[hyperref,publist,addack,firstparindent,usenames,svgnames,nohebrew]{thesis_style_and_class/technionThesis}

\usepackage{commands}

\usepackage{thesis_style_and_class/technionThesisSetup} 
\usepackage{thesis_style_and_class/technionThesisMacros} 

\setboolean{addacknowledgments}{true}

\begin{document}


    \makeTitleEnglish
    
    \chapter{Abstract}

Simulating many-body quantum systems on a classical computer is a
difficult problem due to the large number of degrees of freedom,
which causes the computational complexity to grow exponentially with
the size of the system. Tensor Networks (TN) is a framework that
breaks down large tensors into a network of smaller tensors
and allows an efficient simulation of certain many-body quantum systems. 
To calculate expectation values of local observables, or to simulate the effect of nearest-neighbors interactions, a
contraction of the entire network is needed. This is a known hard problem \cite{PEPS:schuch2007computational}, which cannot be done exactly for general systems with spatial dimension 
$D>1$ 
and is the major bottle-neck in all tensor-network based algorithms.
In the past decade, various approximate-contraction algorithms have been suggested, 
all of which have their strengths and weaknesses.
Nevertheless, contracting a 2D TN is still a major numerical challenge, which limits the use of TN techniques for many interesting systems.

Recently \cite{Duality:TN-PGM}, a close connection between TN and
Probabilistic Graphical Models (PGM) which is a graphical way of
representing classical multivariate probability distributions, has
been shown.  In the PGM framework, marginals of complicated
probability distributions can be approximated using iterative
message passing algorithms such as the Belief Propagation (BP). The
BP algorithm can be adapted to the framework of TN to be used as an
efficient contraction algorithm \cite{ItaiArad:alkabetz2021tensor}. 
While the BP algorithm is extremely efficient and easy to parallelize, it often yields inaccurate results for highly correlated quantum states, or for systems with a large amount of frustration. To overcome this, we
suggest the 
\blockBP{}
algorithm, which coarse-grains the system into blocks and performs BP between the blocks.

This thesis focuses on: (i) development and implementation of the \blockBP{} algorithm for infinite lattices; (ii) using
this algorithm to study the anti-ferromagnetic Heisenberg model on the Kagome lattice in the thermodynamic limit - a frustrated 2D model that is difficult to simulate using existing numerical method.

\section*{Abbreviations}
\begin{longtable}[l]{ll}
    TN  & Tensor Network \\ 
    PGM  & Probabilistic Graphical Model \\ 
    BP  &  Belief Propagation \\ 
    BlockBP  & Block Belief Propagation \\ 
    ITE & Imaginary Time Evolution \\
    AFM & Anti-FerroMagnetic \\
    MPS & Matrix Product State (type of TN) \\
    PEPS & Projected Entangled Pairs State (type of TN) \\
\end{longtable}

    \chapter{Introduction}

The study of many-body quantum systems represents one of the most challenging areas in modern physics. 
The complexity of simulating these large systems classically increases exponentially
with their size, which significantly impedes their study.
To exemplify this fact, consider a system of $N$ particles with local Hilbert-space dimension $d$. The state
of this system in some tensor-product basis is given by
\begin{equation}
\label{eq:intro:tensor-networks1}
    \Ket{\psi} 
    =     
    \sum_{i_1 i_2 \ldots  i_N = 0}^{d-1}
    C_{i_1 i_2 \ldots      i_N} 
    \Ket{i_1 i_2 \ldots i_N} ,
\end{equation}
where $i_k \in \{ 0, 1, \ldots, d-1 \}$.
Representing such a state on a classical computer,
would require us to store $d^N$ complex amplitudes for all
possible states of the system, a number which grows exponentially
with the system size. It would therefore be impossible to represent
even a moderate sized system of $100$ spins this way, let alone
perform any computations on it such as time-evolving the state, computing an observable or
finding the ground state --- This representation is inefficient. 

A rapidly growing approach to address these problems is the framework of Tensor Networks (TN), which we will use throughout this work.

\section{Tensor Networks}
\label{sec:intro:TN}

\subsection{Foundations of Tensor-Networks}
\label{subsec:intro:TN:foundations}

In the realm of quantum many-body physics, understanding the complex interactions and entanglements among a vast number of particles poses a formidable challenge. Traditional computational methods often fall short due to the exponential growth of the Hilbert space with increasing particle numbers. This is where Tensor Networks (TNs) emerge as a transformative tool, offering a novel approach to efficiently represent and analyze quantum states.

We can represent the wave-function $\ket{\psi}$ of a discrete quantum sate as a tensor. E.g., a quantum state of $N$ "$d$-qudits" (i.e., $N$ sites each of a Hilbert space of dimension $d$) can be represented by Eq.~\ref{eq:intro:tensor-networks1} using a large tensor %
$
C_{i_1 i_2 \ldots      i_N}
$, %
and thus requires storing $d^N$ complex numbers.
The TN approach breaks this single tensor $C$ into multiple tensors. This can be done in multiple ways, but one of the simplest to explain is via the singular-value decomposition (SVD), which breaks (or decomposes) a matrix $M$ into $2$ unitary matrices $U, V^{\dagger}$ and $1$ diagonal matrix $\Sigma$. 
Splitting the quantum state into multiple smaller tensors sometimes allows an efficient simulation of previously intractable systems~\cite{TN:Foundations:orus2014practical, PEPS:orus2009simulation,PEPS:cirac2021matrix}.
To understand this claim, imagine that the big rank-$N$ tensor %
$C_{i_1 i_2 \ldots      i_N}$ is broken into $N$ tensors, each of limited rank and dimensions, thus requiring only $\mathcal{O}(N)$ complex numbers in memory. Without getting now into the details of how this can be done (this should become clear while reading this introduction), the smaller space-complexity alone makes previously impossible computations on large quantum systems, seem at least plausible.

\subsubsection{SVD}
\label{subsec:intro:TN:SVD}

SVD can not only decompose matrices, but also tensors
\footnote{A process commonly known throughout the field of Quantum Information as the Schmidt Decomposition.}%
. This is done by fusing together two groups of legs thus reshaping the tensor into a matrix, then performing SVD on the matrix and finally reshaping the tensor back to a matrix by splitting the fused legs. This method transforms %
Eq.~\ref{eq:intro:tensor-networks1} into Eq.~\ref{eq:intro:tensor-networks2}:%
\begin{equation}
\label{eq:intro:tensor-networks2}
    \ket{\psi}
    =
    \sum_{
        i_1, i_2, \cdots i_N
        = 0
    }^{d-1}
    \sum_{j=0}^{D-1}
    U_{
        i_1, i_2, \cdots i_{k-1}, j
    }
    \Sigma_{
        j, j
    }
    V^{\dagger}_{
        j, i_{k}, i_{k+1} \cdots i_N
    }
    \ket{
        i_1, i_2, \cdots i_N
    }
\end{equation}
\begin{figure}[htbp]
    \centering
    \begin{tikzpicture}[scale=1, every node/.style={scale=1}]
    
    \node at (-3.2,0) {\huge $\ket{\psi}$};
    
    \node at (-2.3,0) {$=$};
    
    \node[draw, circle, minimum size=1cm, fill=blue!20] (C) at (0,0) {$C$};
    \draw[thick] (C.135) -- ++(-0.5,0.5) node[left] {$i_1$};
    \draw[thick] (C.180) -- ++(-0.7,0) node[left] {$i_2$};
    \draw[thick] (C.225) -- ++(-0.5,-0.5) node[left] {$i_3$};
    \draw[thick] (C.315) -- ++(0.5,-0.5) node[right] {$i_N$};
    \draw (C.270) ++(0,-0.15) coordinate (dots) node[below] {$\cdots$};
    
    \node at (1.8,0) {$=$};
    \node at (1.8,0.5) {SVD};
    
    \node[draw, circle, minimum size=1cm, fill=red!20] (U) at (4,0) {$U$};
    \node[draw, rectangle, minimum size=0.6cm, fill=green!20] (S) at (5.5,0) {$\Sigma$};
    \node[draw, circle, minimum size=1cm, fill=yellow!20] (V) at (7,0) {$V^\dagger$};
    
    \draw[thick] (U.135) -- ++(-0.5,0.5) node[left] {$i_1$};
    \draw[thick] (U.180) -- ++(-0.7,0) node[left] {$i_2$};
    \draw[thick] (U.270) -- ++(0,-0.5) node[left] {$i_{k-1}$};
    \draw (U.225) ++(-0.2,-0.05) coordinate (dotsU) node[below, rotate=-30] {$\cdots$};
    
    \draw[thick] (U.east) -- (S.west) node[midway, above] {$j$};
    \draw[thick] (S.east) -- (V.west) node[midway, above] {$j$};
    
    \draw[thick] (V.45) -- ++(0.5,0.5) node[right] {$i_{k}$};
    \draw[thick] (V.0) -- ++(0.7,0) node[right] {$i_{k+1}$};
    \draw[thick] (V.270) -- ++(0,-0.5) node[right] {$i_N$};
    \draw (V.315) ++(0.2,-0.05) coordinate (dotsV) node[below, rotate=30] {$\cdots$};
    
    \end{tikzpicture}
\caption{SVD on tensors.}
\label{fig:intro:tn:svd}
\end{figure}
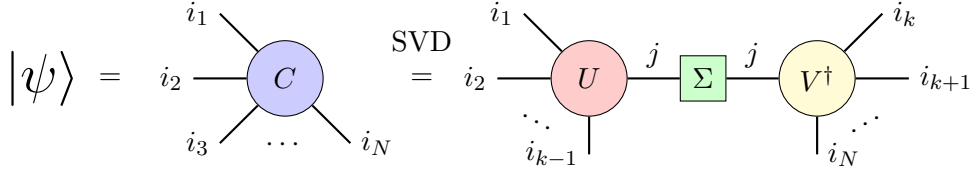
These two equations are summarized in Fig. \ref{fig:intro:tn:svd}.
Note that here we had the freedom to distribute the indices between tensors $U$ and $V^{\dagger}$. The amount of indices each tensor has, affects the dimension of the common index $j$, an index shared between the diagonal matrix $\Sigma$ and the two tensors $U$ and $V^{\dagger}$. This index does not appear inside the $\ket{ket}$s representing the quantum state. The dimension $D$ of the index $j$ is commonly referred to as the "bond dimension" and sometime as the "virtual dimension". We can choose to neglect some of the singular values in $\Sigma$, thus decreasing the bond dimension, a process called "truncation". If the bond dimension is truncated bellow the Schmidt rank, it means that we definitely neglected some positive values in the Schmidt decomposition, thus losing some of the crucial components that describe the full entangled state.

\subsubsection{TN Graph representation and Contraction}

\begin{figure}[htbp]
    \centering
    \begin{tikzpicture}[scale=1, every node/.style={scale=1}]
    
    \node at (-3.2,0) {\huge $\ket{\psi}$};
    
    \node at (-1.8,0) {$=$};
    
    \node[draw, circle, minimum size=1cm, fill=blue!20] (T) at (8,0) {$T$};
    \draw[thick] (T.135) -- ++(-0.5,0.5) node[left] {$i_1$};
    \draw[thick] (T.180) -- ++(-0.7,0) node[left] {$i_2$};
    \draw[thick] (T.225) -- ++(-0.5,-0.5) node[left] {$i_3$};
    \draw[thick] (T.315) -- ++(0.5,-0.5) node[right] {$i_N$};
    \draw (T.270) ++(0,-0.15) coordinate (dots) node[below] {$\cdots$};
    
    \node at (5,0) {$=$};
    \node at (5,0.5) {contraction};
    
    \node[draw, circle, minimum size=1cm, fill=red!20] (U) at (0,0) {$A$};
    \node[draw, circle, minimum size=1cm, fill=yellow!20] (V) at (2,0) {$B$};
    
    \draw[thick] (U.135) -- ++(-0.5,0.5) node[left] {$i_1$};
    \draw[thick] (U.180) -- ++(-0.7,0) node[left] {$i_2$};
    \draw[thick] (U.270) -- ++(0,-0.5) node[left] {$i_{k-1}$};
    \draw (U.225) ++(-0.2,-0.05) coordinate (dotsU) node[below, rotate=-30] {$\cdots$};
    
    \draw[thick] (V.45) -- ++(0.5,0.5) node[right] {$i_{k}$};
    \draw[thick] (V.0) -- ++(0.7,0) node[right] {$i_{k+1}$};
    \draw[thick] (V.270) -- ++(0,-0.5) node[right] {$i_N$};
    \draw (V.315) ++(0.30,-0.25) coordinate (dotsV) node[below, rotate=30] {$\cdots$};

    \draw[thick] (U) -- (V) node[midway, above, text width=2em, align=center] {$j$};
    
    \end{tikzpicture}

\caption[Graphical representation of TN Contraction]{Graphical representation of TN Contraction. Two tensors sharing an index $j$ can be contracted according to the equation %
$
    \sum_{j}
    A_{
        i_1, i_2, \cdots i_{k-1}, j
    }
    B_{
        j, i_{k}, i_{k+1} \cdots i_N
    }
    =
    T_{
        i_1, i_2, \cdots i_{N}
    }
$
.}
\label{fig:intro:tn:contractionBasics}
\end{figure}
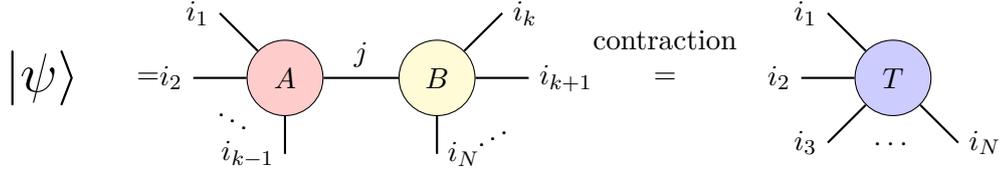

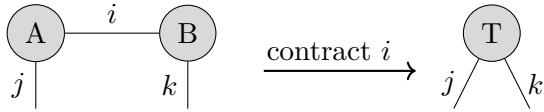
\begin{figure}[htbp]
    \centering
    \begin{tikzpicture}
        \node[shape=circle, draw=black, fill=gray!30] (A) at (0,0) {A};
        \node[shape=circle, draw=black, fill=gray!30] (B) at (2,0) {B};
        \draw (A) -- node[above] {$i$} (B);
        \draw (A) -- node[left]  {$j$} (0, -1);
        \draw (B) -- node[left]  {$k$} (2, -1);

        \draw[->, thick] (3,-0.5) -- (5,-0.5) node[midway, above, text width=5em] {contract $i$};

        \node[shape=circle, draw=black, fill=gray!30] (AB) at (6,0) {T};
        \draw (AB) -- node[left]   {$j$} (5.5, -1);
        \draw (AB) -- node[right]  {$k$} (6.5, -1);
        
    \end{tikzpicture}
    \caption[TN contraction example-1]{A simple Tensor Network with tensors A and B connected by index $i$. Contracting the network along edge $i$ transforms the network into a single tensor $T$. Outgoing legs cannot be contracted as they do not connect nodes. 
    }
    \label{fig:TN:indexing}
\end{figure}

\begin{figure}[htbp]
    \begin{subfigure}{0.45\textwidth}
        \begin{tikzpicture}
            \node[shape=circle, draw=black, fill=gray!30] (A) at (0,0) {A};
            \node[shape=circle, draw=black, fill=gray!30] (B) at (2,0) {B};
            \node[shape=circle, draw=black, fill=gray!30] (C) at (1,1.73) {C};
    
            \draw (A) -- (B);
            \draw (B) -- (C);
            \draw (C) -- (A);
    
            \node[shape=circle, draw=black, fill=red!30] (d) at (5,0.86) {d};
    
            \draw[->, thick] (2.5,0.86) -- (4.5,0.86) node[midway, above, text width=6em] {contract all edges};
    
        \end{tikzpicture}
        \caption{Contracting an entire triangle Tensor Network. The resulting tensor is shown as node d which is a scalar.}
        \label{fig:TN:contractingEverything:TriangleToScalar}
    \end{subfigure}
    \hfill
    \begin{subfigure}{0.45\textwidth}
        \begin{tikzpicture}
            \node[shape=circle, draw=black, fill=yellow!30] (A) at (0,0) {A};
            \node[shape=circle, draw=black, fill=blue!30] (B) at (1.5,0) {B};
            \node[shape=circle, draw=black, fill=red!30] (C) at (0,1.5) {C};
            \node[shape=circle, draw=black, fill=purple!30] (D) at (1.5,1.5) {D};
    
            \draw (A) --node[left] {$i$} (C);
            \draw (C) --node[above] {$j$} (D);
            \draw (B) --node[left] {$k$} (D);
            \draw (A) --node[above] {$l$} (B);
    
            \draw[->, thick] (2.5,0.86) -- (4,0.86) node[midway, above] {contract $l$};
            
            \node[shape=circle, draw=black, fill=green!30] (AB) at (5.25,0) {AB};
            \node[shape=circle, draw=black, fill=red!30] (C2) at (4.5,1.5) {C};
            \node[shape=circle, draw=black, fill=purple!30] (D2) at (6,1.5) {D};

            \draw (AB) --node[left]  {$i$} (C2);
            \draw (C2) --node[above] {$j$} (D2);
            \draw (AB) --node[right] {$k$} (D2);
    
        \end{tikzpicture}
        \caption{A square TN with 4 nodes. Contracting only leg $l$ resulting in tensor-network with 3 nodes. A and B tensors has been contracted into Tensor AB.}
        \label{fig:TN:contractionExamples:SquareToTriangle}
    \end{subfigure}
    \caption[TN contraction example-2]{Contraction of Tensor-Networks. When two nodes are connected, the tensors they represent can be contracted along the indices represented by the legs that are connected.}
    \label{fig:TN:contractionExamples}
\end{figure}
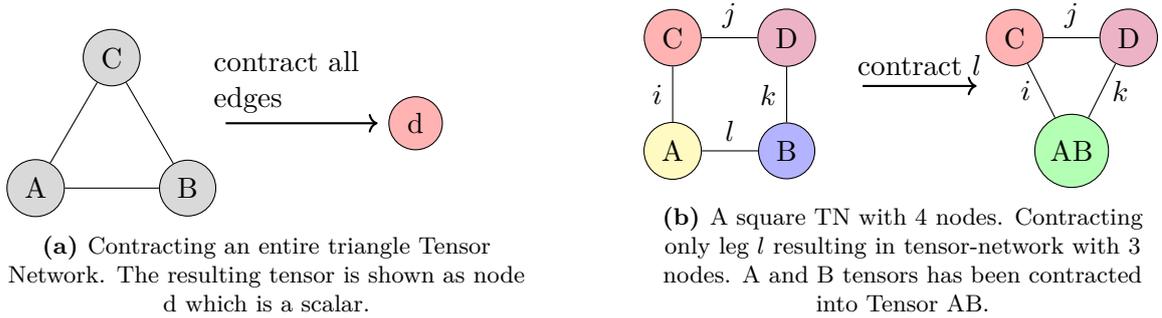

Tensor-Networks are commonly represented by a graph. Each node in a Tensor Network represents a tensor, and the edges represent indices of the tensor. When two nodes are connected via an edge, it means that these tensors share a common index. When tensors share an index an equation of the form of Eq. \ref{eq:intro:tensor-networks:contraction1} represents the quantum state.

\begin{equation}
\label{eq:intro:tensor-networks:contraction1}
    \ket{\psi}
    =
    \sum_{
        i_1, i_2, \cdots i_N
        = 0
    }^{d-1}
    \sum_{j=0}^{D-1}
    A_{
        i_1, i_2, \cdots i_{k-1}, j
    }
    B_{
        j, i_{k}, i_{k+1} \cdots i_N
    }
    \ket{
        i_1, i_2, \cdots i_N
    }
\end{equation}
Here the index $j$ of dimension $D$ is the common index shared between the tensors. Note that by summing over index $j$, we can get an equivalent representation of the same quantum state (Eq. \ref{eq:intro:tensor-networks:contraction2}):
\begin{equation}
\label{eq:intro:tensor-networks:contraction2}
    \ket{\psi}
    =
    \sum_{
        i_1, i_2, \cdots i_N
        = 0
    }^{d-1}
    T_{
        i_1, i_2, \cdots i_{N}
    }
    \ket{
        i_1, i_2, \cdots i_N
    }
\end{equation}
These equivalent representations define an equality between the connected tensors $A,B$ and the tensor $T$ (Eq. \ref{eq:intro:tensor-networks:contraction3}):
\begin{equation}
\label{eq:intro:tensor-networks:contraction3}
    \sum_{j}
    A_{
        i_1, i_2, \cdots i_{k-1}, j
    }
    B_{
        j, i_{k}, i_{k+1} \cdots i_N
    }
    =
    T_{
        i_1, i_2, \cdots i_{N}
    }
\end{equation}
This leads us to define a common operation used throughout the framework of Tensor Networks called "contraction" (see Fig. \ref{fig:intro:tn:contractionBasics}).
The mechanism of contraction can be used to represent all types of linear-maps, and hence it is very natural to use it for the study of many-body quantum systems.
Contraction transfers us from one TN to an equivalent smaller TN. 
Though the smaller network appears simpler, the operation often leads to bigger tensors containing multiple legs, causing an exponential increase in memory-space. 
More examples of TN contractions can be seen in Fig. \ref{fig:TN:indexing} and Fig. \ref{fig:TN:contractionExamples}.

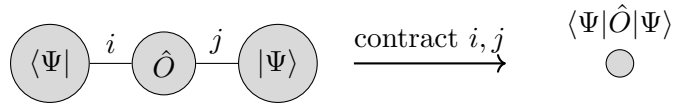
\begin{figure}[htbp]
    \centering
    \begin{tikzpicture}
        \node[shape=circle, draw=black, fill=gray!30] (Psi) at (0,0) {$\langle\Psi|$};
        \node[shape=circle, draw=black, fill=gray!30] (O) at (1.5,0) {$\hat{O}$};
        \node[shape=circle, draw=black, fill=gray!30] (PsiD) at (3,0) {$|\Psi\rangle$};
        \draw (Psi) -- node[above] {$i$} (O);
        \draw (O) -- node[above] {$j$} (PsiD);
        
        \draw[->, thick] (4,0.0) -- (6,0.0) node[midway, above] {contract $i,j$};

        \node[shape=circle, draw=black, fill=gray!30,label={$\langle\Psi|\hat{O}|\Psi\rangle$}] (observable) at (7.5,0) {};
        
    \end{tikzpicture}
    \caption{A Tensor Network representing the expectation value $\langle \Psi | \hat{O} | \Psi \rangle$.}
    \label{fig:TN:expectation_value}
\end{figure}

In quantum mechanics, calculation of expectation values can also be represented as a Tensor Network contraction. This involves contracting a network that includes tensors representing quantum states and operators. The contraction of the network shown in Figure~\ref{fig:TN:expectation_value} corresponds to the calculation of the expectation value of the operator $\hat{O}$ with respect to the quantum state $|\Psi\rangle$.

The goal of TNs, specifically in the scope of many-body quantum physics, is to represent with multiple connected tensors the same information held within a single larger tensor. Our ability to achieve this goal is
dependant on the dimension of the indices connecting the tensors. These bond dimensions are an important hyper-parameter one needs to choose when deciding to work with TNs.
One way to control this parameter is through the process of truncation discussed above, where the least significant singular values are dismissed from the SVD.
If all bond-dimensions are the same, the number is commonly denoted by $D$. Increasing $D$ increases the computational complexity of performing contractions, but also increases the representability of real physical systems 
\cite{PEPS:verstraete2006matrix}.
Quite often though, SVD, being a local process, fails to produce the best truncation in relation to the larger system. Namely, the question of which are the dominant Degrees Of Freedom (DOF) in each tensor depends largely on its relation to the environment. 
It might even be that a rather small singular value captures most of the entanglement between more distant parts of the system.

The graph representation of TN reveals another important aspect: by mimicking the geometry of physical systems, we strive to conserve some physical behavior of the system that is dependant on said geometry. For example, nearest-neighbors operations or local operations at large, are commonly used since the graph itself dictates the notion of locality. In addition, quantum states that adhere to an area-law %
(i.e., the amount of entanglement between two parts of a bi-partition, depends on the area of the boundary defining the bi-partition) %
are easily simulated by TNs
\cite{QI:eisert2008area}.

\subsection{MPS amd PEPS}
\label{subsec:intro:TN:mps_and_peps}

MPS (Matrix Product States) (Fig. \ref{fig:TN:mps})
and
PEPS (Projected Entangled Pair States) (Fig. \ref{fig:TN:peps}) 
\cite{PEPS:cirac2021matrix,PEPS:verstraete2004renormalization, MPS:schollwock2011density} 
are two commonly used types of tensor networks that have gained significant attention in the field of quantum physics, particularly in the study of quantum many-body systems. These mathematical frameworks are powerful tools for representing quantum states in a computationally efficient manner, especially in systems with large numbers of particles.

\newlength{\pepsSep} \setlength{\pepsSep}{35pt}  
\newlength{\pepsLineWidth} \setlength{\pepsLineWidth}{0.6mm}  
\def\pepsW{5} 
\def\pepsH{3} 

\pgfmathtruncatemacro{\pepsWminusOne}{\pepsW - 1}  
\pgfmathtruncatemacro{\pepsHminusOne}{\pepsH - 1}

\begin{figure}[htbp]
\begin{subfigure}{0.45\textwidth}    
    \centering
    \begin{tikzpicture}
        \tikzset{tensor/.style={circle, draw=black, fill=blue!10, minimum size=10pt, inner sep=0pt}}
        
        \foreach \x in {1,...,\pepsW}
        {
            \node[tensor]  (\x) at (\pepsSep*\x,0) {$A_{\x{}}$};
        }
    
        \foreach \x in {1,...,\pepsWminusOne}
        {
            \pgfmathtruncatemacro{\nextx}{\x + 1} 
            \draw[line width=\pepsLineWidth] (\x) -- (\nextx) node[midway, above, text width=2em, align=center] {\small $\alpha_{\x{}}$};
        }
        
        \foreach \x in {1,...,\pepsW}
        {
            \pgfmathsetmacro{\tempDiagX}{\pepsSep*(\x)}
            
            \pgfmathsetmacro{\tempDiagY}{-\pepsSep*(0.4)}

            \edef\diagX{\tempDiagX pt}
            \edef\diagY{\tempDiagY pt}

            \draw[line width=\pepsLineWidth] (\x) -- (\diagX, \diagY) node[anchor=west, text width=2em] {\small $i_{\x{}}$};
        }

    \end{tikzpicture}
    \caption{MPS tensor network of length $\pepsW$. The equivalent tensor $T$ is acquired by contracting the TN over all common indices ($\alpha_{i}$), i.e., 
    $
        c^{i_1, ..., i_{\pepsW{}}}
        =
        \sum_{\{\alpha\}}
        {(A_1)}_{\alpha_{1}}^{i_1}
        {(A_2)}_{\alpha_{1}\alpha_{2}}^{i_2}
        \cdots
        {(A_{\pepsW{}})}_{\alpha_{\pepsWminusOne{}}}^{i_{\pepsW{}}}
    $
    }
    \label{fig:TN:mps}
\end{subfigure}
\hfill
\begin{subfigure}{0.45\textwidth}
    \centering
    \begin{tikzpicture}
        \tikzset{tensor/.style={circle, draw=black, fill=blue!10, minimum size=10pt, inner sep=0pt}}

        \foreach \x in {1,...,\pepsW}
            \foreach \y in {1,...,\pepsH}
            {
                \node[tensor]  (\x-\y) at (\pepsSep*\x,-\pepsSep*\y) {};
            }
        
        \foreach \x in {1,...,\pepsWminusOne}
            \foreach \y in {1,...,\pepsH}
            {
                \pgfmathtruncatemacro{\nextx}{\x + 1} 
                \draw[line width=\pepsLineWidth] (\x-\y) -- (\nextx-\y);
            }
        
        \foreach \x in {1,...,\pepsW}
            \foreach \y in {1,...,\pepsHminusOne}
            {
                \pgfmathtruncatemacro{\nexty}{\y + 1} 
                \draw[line width=\pepsLineWidth] (\x-\y) -- (\x-\nexty);
            }
            
        \foreach \x in {1,...,\pepsW}
            \foreach \y in {1,...,\pepsH}
            {
                \pgfmathsetmacro{\tempDiagX}{\pepsSep*(\x-0.2)}
                
                \pgfmathsetmacro{\tempDiagY}{-\pepsSep*(\y+0.4)}
    
                \edef\diagX{\tempDiagX pt}
                \edef\diagY{\tempDiagY pt}
    
                \draw[line width=\pepsLineWidth] (\x-\y) -- (\diagX, \diagY);
            }
        
    \end{tikzpicture}
    \caption{A $\pepsH{}\times{}\pepsW{}$ PEPS tensor network.}
    \label{fig:TN:peps}
\end{subfigure}
\caption[MPS and PEPS]{Examples of MPS and PEPS Tensor-Networks that are commonly used.}
\label{fig:TN:mps_and_peps_examples}
\end{figure}
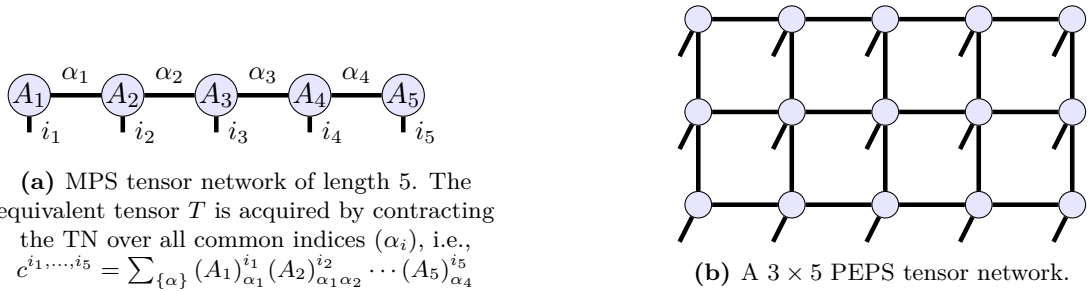

When using either MPSs or PEPSs it is common to define a single bond dimension $D$. As discussed in the previous paragraph, choosing this parameter affects both the computational time, the allocated memory and the representability of the TN.

\subsubsection{MPS}
\label{subsec:intro:TN:mps_and_peps:mps}

MPSs (Fig. \ref{fig:TN:mps}) are particularly well-suited for one-dimensional systems or systems with short-range interactions. In MPS, the quantum state of a system is represented as a sequence of rank-$3$ tensors, each corresponding to a single site in the system. The three indices of each tensor $k$ are divided into a single physical leg $i_{k}$ (corresponding to the physical dimensions of the Hilbert-space per site) and two virtual legs $\alpha_{k-1}, \alpha_{k}$ of some chosen dimension $D$ connecting the tensor to its neighbors. Note that the first and last tensors contains only a single virtual leg.

A system with $N$ sites is related to the big rank-$N$ tensor describing the full quantum state, by contraction the link of tensors along the virtual dimensions (Eq. \ref{eq:intro:tensor-networks:mps}):
\begin{equation}
\label{eq:intro:tensor-networks:mps}
    c^{i_1, ..., i_{\pepsW{}}}
    =
    \sum_{\{\alpha\}}
    {(A_1)}_{\alpha_{1}}^{i_1}
    {(A_2)}_{\alpha_{1}\alpha_{2}}^{i_2}
    \cdots
    {(A_{N})}_{\alpha_{N-1}}^{i_{N}}  
\end{equation}

The key advantage of MPS is its ability to efficiently represent states that exhibit area-law entanglement\cite{MPS:verstraete2004density}, which is typical in ground states of gapped one-dimensional systems. The MPS format is central to the success of algorithms like the Density Matrix Renormalization Group (DMRG)\cite{MPS:schollwock2011density}, which is a powerful method for finding ground states of one-dimensional quantum systems.

\subsubsection{PEPS}
\label{subsec:intro:TN:mps_and_peps:peps}

PEPSs (Fig. \ref{fig:TN:peps}), on the other hand, generalize the concept of MPS to higher dimensions. While MPS are akin to a chain of matrices, PEPS are structured as a network of tensors with more complex connectivity, which naturally maps onto the geometry of two-dimensional or higher-dimensional lattices. PEPS are particularly useful for studying quantum systems where particles are arranged in a 2D lattice and interact predominantly with their nearest neighbors \cite{PEPS:localHamiltonians}. The structure of PEPS allows them to capture the entanglement properties in these higher-dimensional systems effectively \cite{PEPS:hasikSimulatingChiral}, although the computational cost to manipulate PEPS is generally higher than that for MPS due to the increased connectivity of the tensors.

Compared to MPS, each tensor in a PEPS has multiple virtual legs, depending on the structure of the graph and its geometrical dimension, while still having a single physical leg.

\subsubsection{pMPS}
\label{subsec:intro:TN:mps_and_peps:pmps}

\begin{figure}[htbp]
    \centering
    \begin{tikzpicture}
        \tikzset{tensor/.style={circle, draw=black, fill=blue!10, minimum size=10pt, inner sep=0pt}}
        
        \foreach \x in {1,...,\pepsW}
        {
            \node[tensor]  (\x) at (\pepsSep*\x,0) {$A_{\x{}}$};
        }
    
        \foreach \x in {1,...,\pepsWminusOne}
        {
            \pgfmathtruncatemacro{\nextx}{\x + 1} 
            \draw[line width=\pepsLineWidth] (\x) -- (\nextx) node[midway, above, text width=2em, align=center] {\small $\alpha_{\x{}}$};
        }
        
        \foreach \x in {1,...,\pepsW}
        {
            \pgfmathsetmacro{\tempDiagX}{\pepsSep*(\x)}
            
            \pgfmathsetmacro{\tempDiagY}{-\pepsSep*(0.4)}

            \edef\diagX{\tempDiagX pt}
            \edef\diagY{\tempDiagY pt}

            \draw[line width=\pepsLineWidth] (\x) -- (\diagX, \diagY) node[anchor=west, text width=2em] {\small $i_{\x{}}$};
        }

        \draw[line width=\pepsLineWidth] (1) to[out=170, in=10, looseness=1.8] node[midway, above, text width=2em, align=center] {\small $\alpha_{\pepsW{}}$} (\pepsW);
        
    \end{tikzpicture}
    \caption[pMPS]{Examples of a periodic MPS (pMPS) with \pepsW{} tensors.}
    \label{fig:TN:pMPS}
\end{figure}
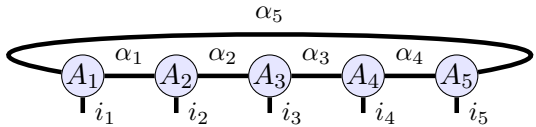

Periodic Matrix Product States (pMPS) extend the capabilities of simple Matrix Product States (MPS) by incorporating periodic boundary conditions. In a pMPS, the structure is periodic, meaning the last site is connected back to the first site, forming a closed loop
(See Fig. \ref{fig:TN:pMPS})
.
Each site in the chain is represented by a tensor with three indices, similar to a simple MPS, but without any special boundary tensors. 


Mathematically, a pMPS for a quantum state $|\psi\rangle$ of a system with $N$ sites can be written as:
\begin{equation}
    |\psi\rangle = \sum_{i_1, i_2, \ldots, i_N} \text{Tr}(A^{i_1} A^{i_2} \cdots A^{i_N}) |i_1, i_2, \ldots, i_N\rangle,
    \label{eq:TN:pMPS}
\end{equation}
where $\text{Tr}$ denotes the trace, ensuring the periodicity.

As discussed in \cite{PEPS:verstraete2008matrix}, 
this closed-loop structure allows pMPS to sometimes capture more complex entanglement patterns and longer-range correlations albeit at a computational cost: the contraction increases by a polynomial factor of the virtual bond dimension $D$.

\subsection{Contracting PEPS}
\label{subsec:intro:TN:contraction_of_PEPS}

As demonstrated in Fig. \ref{fig:TN:expectation_value}, the contraction of tensor-networks is a core concept in this framework since when one needs to compute local expectations, a contraction of the entire tensor-network must be performed.
Another common reason to contract the network, is to get a compact representation of the environment around a local patch of the system, thus allowing for simulation of local Hamiltonians.

The computational cost of contracting an arbitrary TN is exponential in the dimensions of the tensors comprising the network, and is in fact a $\#P$-hard problem
\cite{PEPS:schuch2007computational}.
Moreover, finding the best contraction order is a $NP$-hard problem, analogous to the "traveling sales man" problem.

Over the years, various TN contraction algorithms have been
suggested 
\cite{PEPS:orus2009simulation,PEPS:verstraete2008matrix,
nishino1996corner, PEPS:sandvik2007variational, PEPS:wang2011monte,
levin2007tensor, TN:jiang2008accurate, evenbly2015tensor}, which use
various approximations and strategies to facilitate the TN
contraction. However, these algorithms still face difficulties when
dealing with TNs representing two or three-dimensional systems, or
more generally, networks that contain loops. Contracting such
networks often requires very high computational resources to obtain
decent approximations.
Currently, there is no single "best" algorithm for the contraction of Tensor-Networks; all algorithms play a balancing act between speed and accuracy --- reducing run-time increases the errors. 

\subsubsection{The boundary-MPS method}
\label{sec:intro:bubblecon}

\begin{figure}[htbp]
    \centering
    \includegraphics[width=0.65\textwidth]{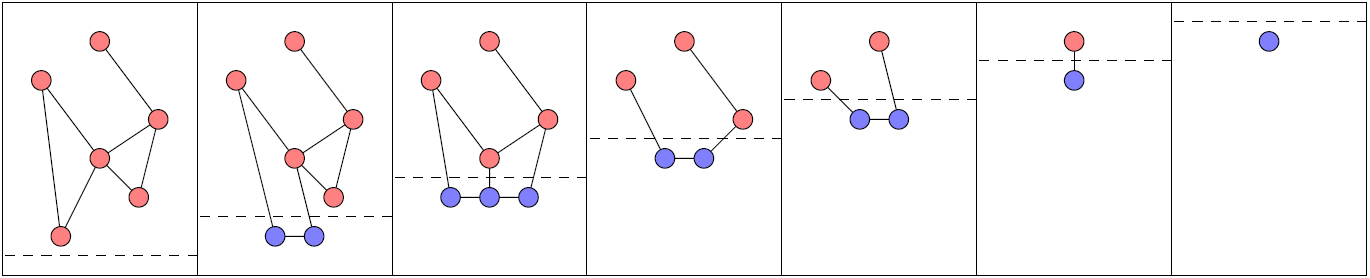}    
    \caption[The boundary MPS method]{Illustration of the boundary MPS method on a PEPS tensor network. Image from \cite{bubblecon}.}
    \label{fig:TN:bubbleConIllustartion}
\end{figure}

A commonly used contraction technique which we utilize throughout this study, is the boundary-MPS method 
\footnote{%
This technique is sometimes called "bubble-contraction", while its author called named it "the sweep line contraction algorithm". 
}
\cite{bubblecon}. 
An illustartion can be seen in Fig. \ref{fig:TN:bubbleConIllustartion}.

The method can be thought of as a bubble that expands and swallows the TN tensor-by-tensor. Every time a tensor is swallowed, it is contracted into an MPS that summarizes the information of the TN over the entire area covered by the bubble.
The method is done in a few steps that are repeated until the entire region of interest is covered.
Starting from a small region (a bubble), the contracted area is expanded sequentially by incorporating neighboring tensors into the contracted region. At each step, an approximation or truncation is typically performed to maintain the manageability of the tensor dimensions. Namely, if after contracting a new tensor, the number of legs passing the bubble-boundary exceeds the number of tensors in the MPS, an SVD is performed to split one tensor into two so to keep the bubble a valid MPS 
\footnote{In this context, it means having a single out-going leg per tensor in the MPS}
.
As described in Subsec.
\fullsubsubref{subsec:intro:TN:SVD}%
, performing SVD is a great opportunity to reduce the virtual dimensions between tensors%
\footnote{
It is worth mentioning that as discussed above, truncating the dimension by considering the singular values alone, is sub-optimal. Taking the environment into account is crucial for maintaining entanglement within the system, a feature that will be discussed in further details in Subsec \fullsubsubref{subsec:ite:4PEPS:iteStep}.
}
. 
If virtual dimensions are always truncated to be kept beneath some predetermined threshold, we call this threshold the \textbf{truncation bond-dimension} usually denoted by the letter $\chi$.

\begin{figure}
    \centering
    \begin{subfigure}{0.40\textwidth}    
        \centering
        \includegraphics[width=1.0\linewidth]{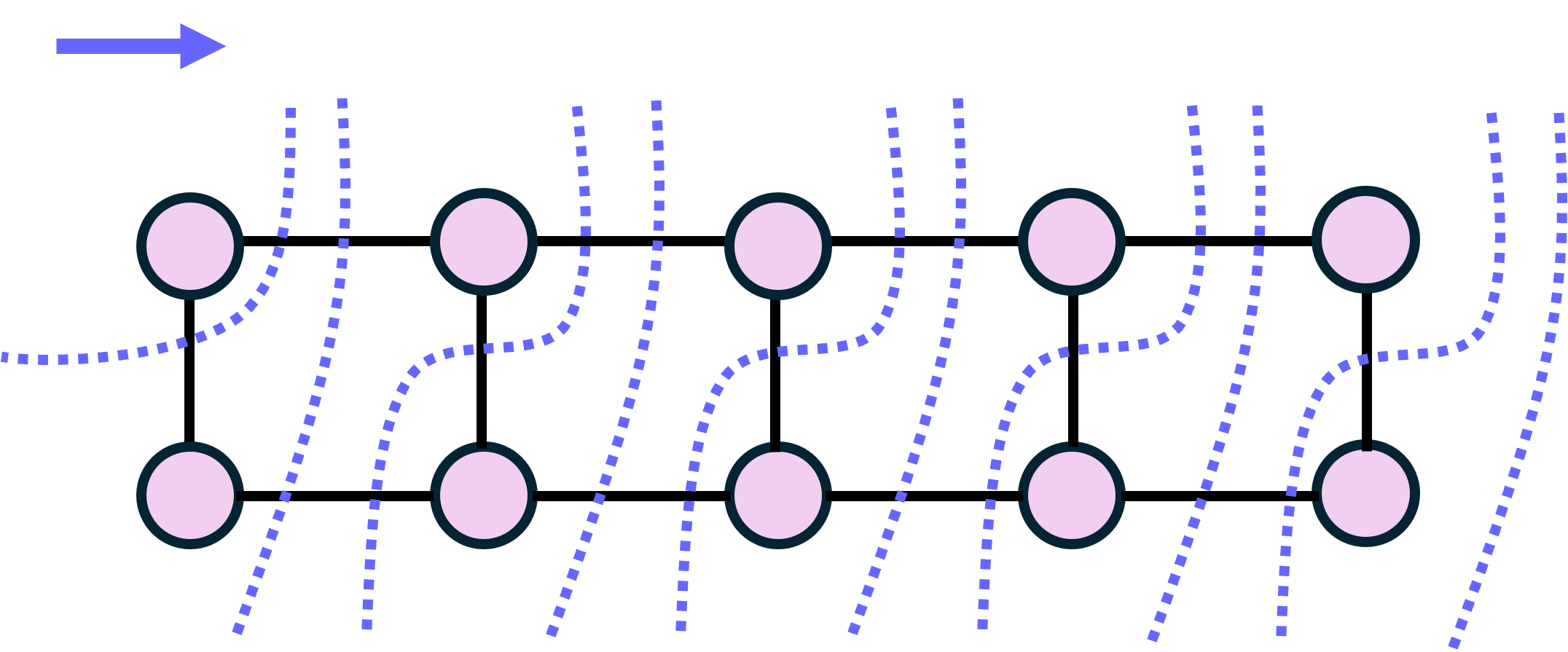}
        \caption{The efficient contraction order.}
        \label{fig:intro:tn:contraction:bubbleconQuasi1D:good}
    \end{subfigure}
    \hfill
    \begin{subfigure}{0.50\textwidth}    
        \centering
        \includegraphics[width=1.0\linewidth]{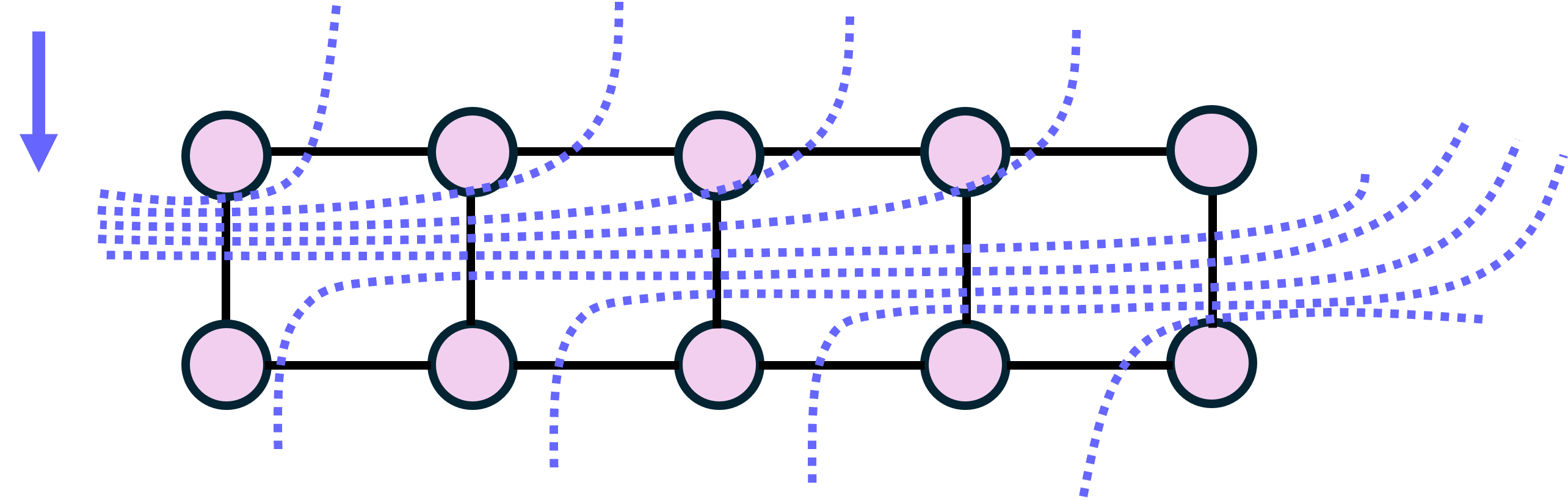}
        \caption{The worst contraction order.}
        \label{fig:intro:tn:contraction:bubbleconQuasi1D:bad}
    \end{subfigure}
    \caption[Best and worst boundary-MPS orders on a 1D TN]{An efficient \bubblecon{} order for a quasi-1D TN. (\subref{fig:intro:tn:contraction:bubbleconQuasi1D:good}) Best to expand the bubble along the main dimension, and at each depth, contract all shallower dimensions. (\subref{fig:intro:tn:contraction:bubbleconQuasi1D:bad}) The worst order is when the bubble expands along the deepest dimension - first.}
    \label{fig:intro:tn:contraction:bubbleconQuasi1D}
\end{figure}

For some Tensor Network structures, we can find a smart direction for the expansion of the bubble-contraction that will help up us maintaining a small number of tensors, and hence be more efficient in terms of computational cost.
For example, a quasi-1D structure (see Fig. 
\ref{fig:intro:tn:contraction:bubbleconQuasi1D}) is most efficiently contracted along its longest dimension last, meaning that the bubble expands primarily along the deepest dimension, filling the shallower dimensions at each step. This keeps the number of tensors involved at each step limited by the shallower dimensions.

\begin{figure}[htbp]
    \centering
    \includegraphics[width=0.65\textwidth]{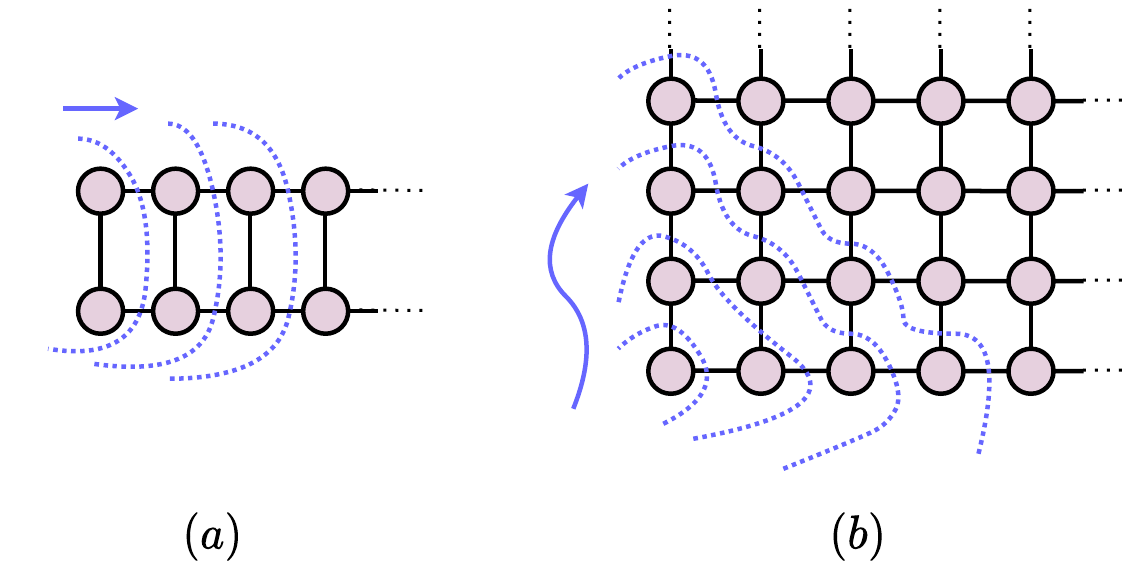}    
    \caption[boundary-MPS on 1D vs 2D PEPS]{While for some TNs, a smart direction for the expansion of the bubble can be found (a), contraction of a general PEPS (b) using this method is challenging.}
    \label{fig:TN:bubbleConIllustartionComplexity}
\end{figure}

For other TNs, like general PEPSs, such "easy" direction does not exist (see Fig. \ref{fig:TN:bubbleConIllustartionComplexity}). This causes whoever uses this method to choose between two options: 1) paying in longer computational time while preserving the quantum properties of the system through its representation as an MPS; or 2) faster computation achieved by performing SVD with tighter $\chi$, paying with a result that only approximates the true state of the system.

It should be noted that the time complexity (analyzed in \cite{bubblecon}) of performing \bubblecon{} on a double-layered ($\braket{bra|ket}$) TN with $n$ tensors, bond-dimension $D$ and a truncation bond-dimension $\chi \geq D^2$ --- is (Eq. \ref{eq:bubblecon:timeConplex}):
\begin{equation}
    \label{eq:bubblecon:timeConplex}
    \texttt{time: }
    \mathcal{O}\left(
        n \chi^3
    \right)
\end{equation}
If throughout the process the MPS contains at most $k$ tensors (depends on the TN's structure and direction of contraction), the space complexity is known to be:
\begin{equation}
    \label{eq:bubblecon:spaceConplex}
    \texttt{space: }
    \mathcal{O}\left(
        k \chi^2
    \right)
\end{equation}

The scientific community that uses tensor-networks is still looking for efficient algorithms that can quickly contract PEPS of dimensions higher than $1$, while reliably preserving the information that was encoded in the large TN system.

\section{PGM-TN Duality and the Belief-Propagation Algorithm}
\label{sec:intro:PGM-TN}

\subsection{Probabilistic Graphical Models}

Probabilistic Graphical Models (PGMs) are statistical frameworks designed to represent complex joint multivariate probability distributions through the use of graphs \cite{koller2009probabilistic, mezard2009information}. There are multiple closely related graphical representations for such distributions. The most versatile representation, which is pertinent to our discussion, is the factor graph (FG). A factor graph is a bipartite graph $G(V,F,E)$, comprising two distinct types of nodes: variable nodes $v \in V$ and factor nodes $f \in F$.

Variable nodes correspond to the random variables within the distribution, while factor nodes illustrate the relationships or dependencies among these variables. Each factor node is linked to a factor function, which encodes the dependencies between the variables connected to that node. The global multivariate distribution is represented as the \emph{product} of all these factor functions.

\begin{figure}[htbp] 
    \centering 
    \includegraphics[scale=0.6]{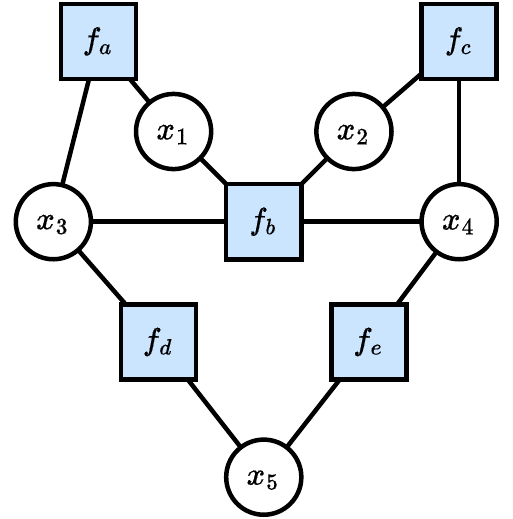} \caption{Example of a factor graph} \label{fig:intro:pgn_tn:pgn_example} 
\end{figure}

For instance, the FG shown in \Fig{fig:intro:pgn_tn:pgn_example} represents the distribution \begin{align*} P(x_1, x_2, x_3, x_4, x_5)=\frac{1}{Z} f_a (x_1,x_2) f_b(x_1,x_2,x_3,x_4) f_c(x_2,x_4) f_d(x_3,x_5) f_e(x_4,x_5), \end{align*} where $Z = \sum_{x_1,x_2,x_3,x_4,x_5} P(x_1,x_2,x_3,x_4,x_5)$ is a normalization constant. It is important to note that, since the product of the factor functions must constitute a probability distribution, these functions must be non-negative.

\subsection{PGM-TN Duality}

Using a recently \cite{Duality:TN-PGM} found connection between TN and PGM, the famous Belief Propagation (BP) algorithm can be adapted from probabilistic models into the framework of TN, for an efficient contraction scheme \cite{ItaiArad:alkabetz2021tensor}. 

\begin{figure}[htbp]
    \centering
    \includegraphics[width=0.8\linewidth]{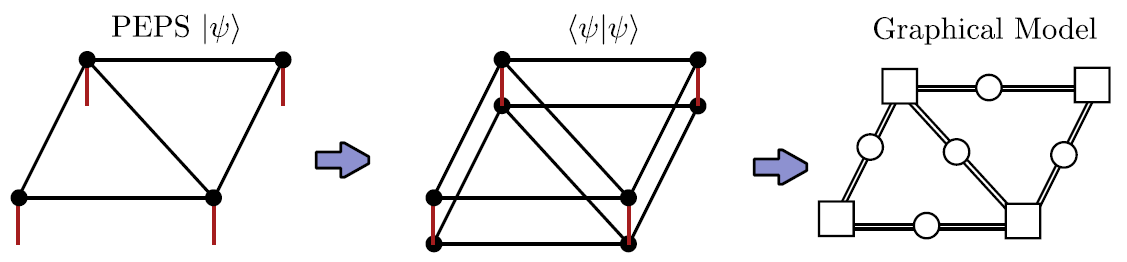}
    \caption[TN to PGM]{Mapping a tensor network to a graphical model of type double-edge factor graph. Image from \cite{ItaiArad:alkabetz2021tensor}
    .
    }
    \label{fig:intro:pgn_tn:tn_to_pgn}
\end{figure}

Given a PEPS (Subfig. \ref{fig:TN:peps}) representing the state $\ket{\psi}$,
a necessary step needed in order to adapt PGM methods, is to transform the PEPS into a TN with similar properties to a PGN. Namely, the $\bra{\psi}$ PEPS must be computed by conjugating all tensors, and then connected to the original $\ket{\psi}$ PEPS along all physical legs%
\footnote{Legs going outside the TN instead of those connected to other tensors in the network.}%
, %
to produce the $\braket{\psi|\psi}$ TN %
(See Fig \ref{fig:intro:pgn_tn:tn_to_pgn})
. 
A contraction of the TN over these physical indices is then performed, which turns each virtual index of dimension $D$ into a "double-leg" of dimension $D^2$.

Where in the classical probabilistic model, each two nodes are mitigated by a factor, here, between each two tensors $A$ and $B$ we place two matrices, one matrix from Tensor $A$ to tensor $B$ and one the other way around. For each tensor, the matrices pointed toward itself encompasses the effect of its surrounding. The matrices pointing away from it represent the effect this tensors contributes on its neighbors.

\subsection{BP, simple-update and full-update}
\label{subsec:into:png_tn:bp_su_fu}

While the BP algorithm will be explained in more depth in Sec. \ref{section:blockBP:introToBeliefPropagation}, here I want to highlight a particular aspect of BP: it surrounds each tensor with local matrices, providing a compact representation of its environment. This approach allows for efficient simulation of the system or computation of local expectation values. However, it is limited in precision when dealing with states that exhibit long-range correlations, due to its reliance on local environments. This method has been shown \cite{ItaiArad:alkabetz2021tensor} to be equivalent to the Simple-Update (SU) algorithm \cite{TN:jiang2008accurate,structureMatrix:Orus:jahromi2019universal}.

Where the BP (SU) algorithm struggles with highly correlated quantum states, the full-update algorithm is often used. This method is designed to provide a more accurate representation of quantum states by considering the full environment of each tensor during the optimization process. This involves contracting the entire tensor network, which can be computationally intensive, especially for large systems or those with high bond dimensions.

In this thesis, I utilize a newly emerged method called the \blockBP{} \cite{Itai:BlockBP} algorithm (more on that in section \ref{subsec:blockBP:BlockBPAlgo}). This method sits somewhere in the middle between the simple-update and full-update approaches, by coarse-graining the system into blocks and performing BP between these blocks. The simplicity of this algorithm, as well as its promise to capture long-distance correlations, makes it attractive for use in highly-correlated systems where previous algorithms struggle. In this work, we apply it to study the antiferromagnetic Heisenberg model on the infinite Kagome lattice—a difficult problem studied extensively over the past few decades using various numerical techniques. Even after extensive research, the nature of its ground state remains strongly debated.

\section{The Kagome Lattice}
\label{sec:intro:kagome}

The \kagome{} lattice is a repetitive graph with
global node degree of 4 \footnote{4 neighbors for each node}.
\begin{figure}[htbp]
    \includegraphics[width=0.5\textwidth]{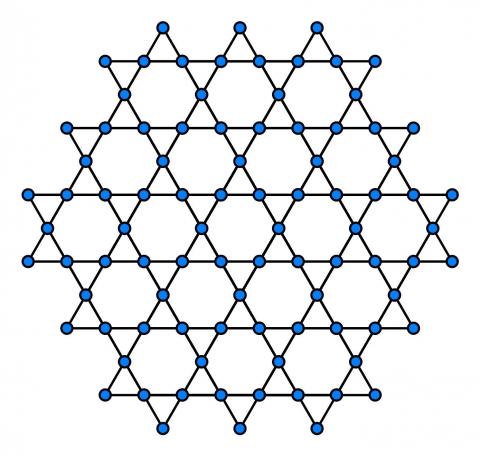}
    \centering
    \caption{The Kagome lattice.} 
    \label{fig:KagomaLattice}
\end{figure}
Deriving its name from a traditional Japanese basket-weaving pattern, the \kagome{} lattice features a two-dimensional network of corner-sharing triangles (see Fig.~\ref{fig:KagomaLattice}) . This configuration leads to significant geometric frustration in antiferromagnetic systems, where spins prefer to align antiparallel to each neighbor. However, the triangular arrangement prevents all spins from simultaneously satisfying antiparallel alignment, resulting in a highly degenerate ground state with no static magnetism.
One of the most captivating aspects of the \kagome{} lattice under antiferromagnetic interaction is the numerical observation that the ground state of the system is a quantum spin liquid (QSL) state, i.e., a state devoid of magnetic long-range order and characterized by novel types of excitations. 
Experimentally, realizing a perfect \kagome{} lattice poses challenges due to structural imperfections. Nevertheless, materials like Herbertsmithite (ZnCu$_3$(OH)$_6$Cl$_2$) are approximated by a \kagome{} lattice of copper ions and are believed to manifest a quantum spin liquid \cite{kagome:shores2005structurally}. 
Despite extensive theoretical and experimental studies, many properties of spin systems on the \kagome{} lattice with antiferromagnetic Heisenberg interactions remain not fully understood. Key open questions include the definitive experimental confirmation of the QSL state and the nature of its excitations, which continue to be significant research frontiers in condensed matter physics.

In this thesis, I will describe our
adaptation of the BP algorithm to the TN world and show how it can be
used to efficiently approximate the contraction of the infinite \kagome{} lattice, represented as a TN. We then use this algorithm to find the ground state of the antiferromagnetic Heisenberg model on the \kagome{} lattice in the thermodynamic limit 
via imaginary time evolution (\ITE{}).

    \chapter{Block Belief Propagation}
\label{chapter:blockBP}

The study of two-dimensional (2D) quantum systems plays a vital role in various fields of physics and material science. For example, the field of solid-state physics heavily relies on the study of 2D lattices and it is where the technology and theory of semi-conductors has risen from \cite{historyOfSemiConductors, theoryOfSemiConductors}. 
Accurately investigating these systems numerically, particularly those exhibiting strong correlations, presents a significant challenge due to computational limitations. While Quantum Monte Carlo (QMC) \cite{QMC:SimulationSolids} methods have proven successful, they are often plagued by the sign problem \cite{QMC:signProblem}, hindering precise calculations. Projected entangled pair states (PEPS) represent a powerful class of TNs, capable of handling both finite (fPEPS) and infinite (iPEPS) 2D systems \cite{PEPS:verstraete2004renormalization, PEPS:orus2019tensor, PEPS:cirac2021matrix}. However, a key bottleneck for PEPS algorithms lies in the efficient and stable computation of local environment of tensors, used for simulations %
(see discussion in Subsec. \ref{subsec:into:png_tn:bp_su_fu})
.
This task can become computationally intractable (i.e., \#P-hard) in worst-case scenarios for 2D and higher-dimensional systems \cite{PEPS:schuch2007computational}. Although existing approaches like boundary matrix product state (\bubblecon{}) \cite{bubblecon,PEPS:verstraete2008matrix}, corner transfer matrix (CTM) \cite{PEPS:orus2009simulation}, and Monte Carlo sampling methods \cite{PEPS:sandvik2007variational, PEPS:wang2011monte} exist, they can be computationally expensive or require advanced numerical schemes.

\section{Introduction to Belief-Propagation}
\label{section:blockBP:introToBeliefPropagation}

Belief Propagation (BP) is a well-established algorithm in statistical inference \cite{bp:pearl1988probabilistic} for computing local marginal probabilities of complex networks in multivariate probability models. Notably, BP achieves exact results for tree-like networks. However, for networks with a significant number of loops or long-range correlations, BP's performance deteriorates. This limitation arises from its consideration of only direct neighbors for each node and its treatment of a node's environment in a separable manner. To overcome these shortcomings, a common approach is to group neighboring nodes and handle small loops exactly, albeit at the cost of increased computational resources and more intricate bookkeeping.

Belief propagation for Tensor-Networks \cite{ItaiArad:alkabetz2021tensor} is an iterative algorithm that finds the effective environment of tensors. %
To utilize BP for a given PEPS (Fig. \ref{fig:bp:bpExample:sub1}), we first derive the "braket" TN by contracting each "ket" tensor $t_i$ from the PEPS, with its complex-conjugate "bra" tensor $t^*_i$, along their physical indices, creating a tensor with only virtual indices%
\footnote{Each such virtual index that previously had a dimension $D$ is now of dimension $D^2$}%
, $T_i$ (Fig. \ref{fig:bp:bpExample:sub2}). %
From this double-layered tensor network representing $\braket{\psi|\psi}$, the algorithm initiates random positive semi-definite matrices around each tensor, for each of its edges (Fig. \ref{fig:bp:bpExample:sub3}).
Then, at each iteration at time $l$, the message
$
    m_{a\rightarrow b}^{(l)} 
$
from tensor $T_a$ to $T_b$, is calculated from the messages at iteration $l-1$ by equation \ref{eq:BP:DefiningEquation}, in which $T_a$ is contracted with all incoming messages towards it, except the one arriving from $T_b$ %
(Fig. \ref{fig:bp:bpExample:sub4})%
.
\begin{equation}
    m_{a\rightarrow b}^{(l)} 
    = Tr \left(
        T_a 
        \prod_{
            c\in N_a \setminus  \{ b \} 
        }
        m_{c\rightarrow a}^{(l-1)} 
    \right)
\label{eq:BP:DefiningEquation}
\end{equation} 
Here, $m_{c\rightarrow a}^{(l-1)}$ are messages at iteration $l-1$ from the tensor at some site $c\in N_a\setminus \{b\}$, meaning a neighbor of $a$ that is not $b$ itself, to the tensor at site $a$. 
The messages are matrices that are calculated iteratively by following Eq. \ref{eq:BP:DefiningEquation} until a convergence criterion is achieved.
A graphical example of the $\braket{bra|ket}$ TN and Eq. \ref{eq:BP:DefiningEquation} is given in Fig. \ref{fig:bp:bpExample}.

\newlength{\subwidth}
\setlength{\subwidth}{0.49\linewidth}  

\begin{figure}[htbp]
    \centering
    \begin{minipage}{0.95\textwidth}  
        \centering
        \setlength{\subwidth}{0.49\linewidth}  
        \begin{subfigure}[b]{0.85\subwidth}
            \centering
            \includegraphics[width=0.7\linewidth]{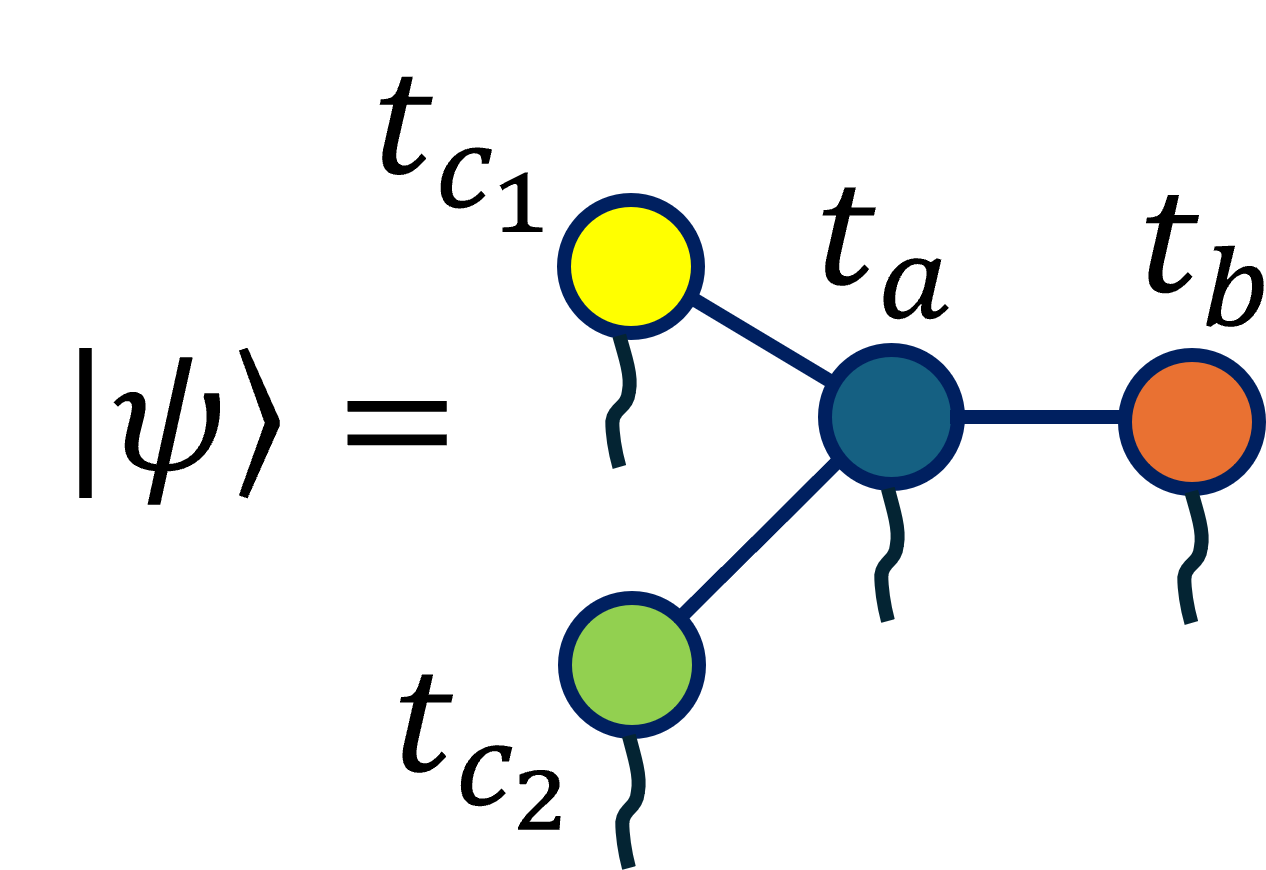} 
            \caption{A PEPS tensor network representing the state $\ket{\psi}$. Here the tensor at site $a$ is neighboring $3$ tensors.}
            \label{fig:bp:bpExample:sub1}
        \end{subfigure}
        \hfill
        \begin{subfigure}[b]{1.15\subwidth}
            \centering
            \includegraphics[width=0.98\linewidth]{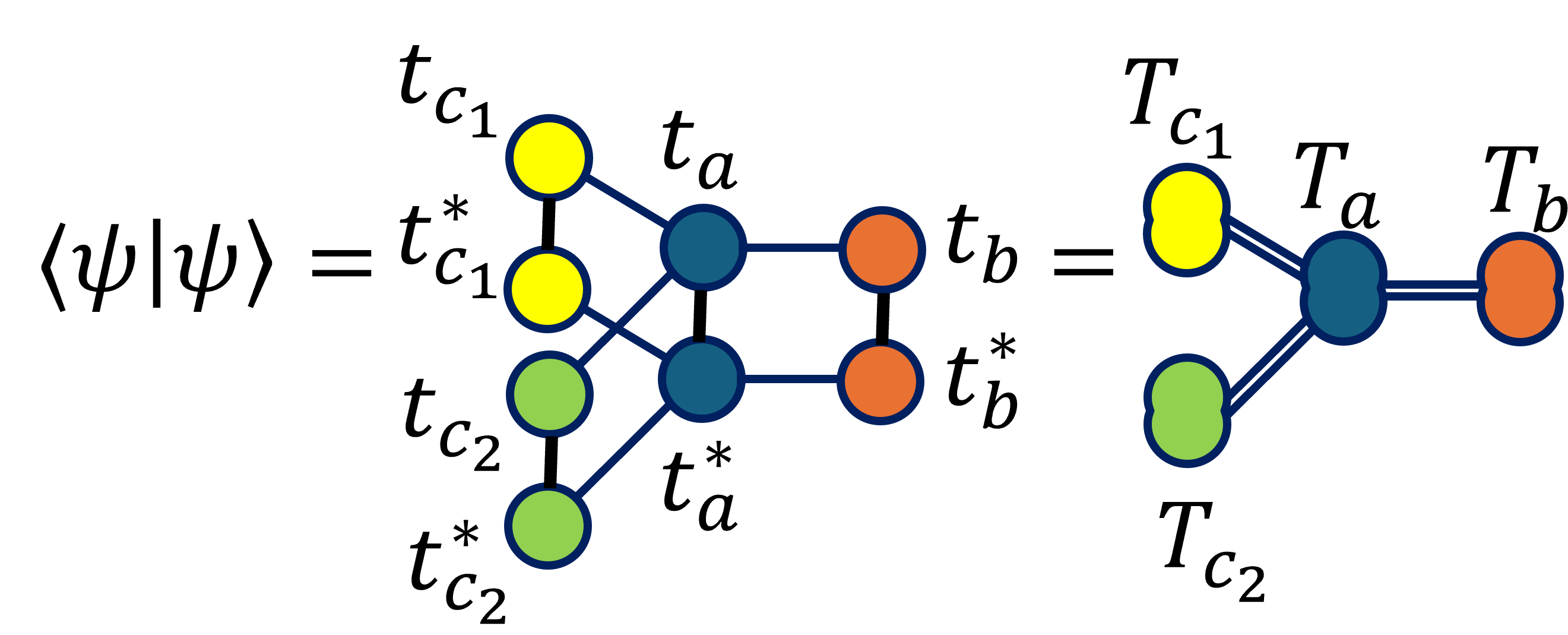} 
            \caption{%
            The contraction of the original TN with its complex conjugate along the physical dimension results in the TN representing $\braket{\psi|\psi}$. Each virtual leg is doubled in this process.
            }
            \label{fig:bp:bpExample:sub2}
        \end{subfigure}
        
        \medskip 
        
        \begin{subfigure}[b]{1.1\subwidth}
            \centering
            \includegraphics[width=0.7\linewidth]{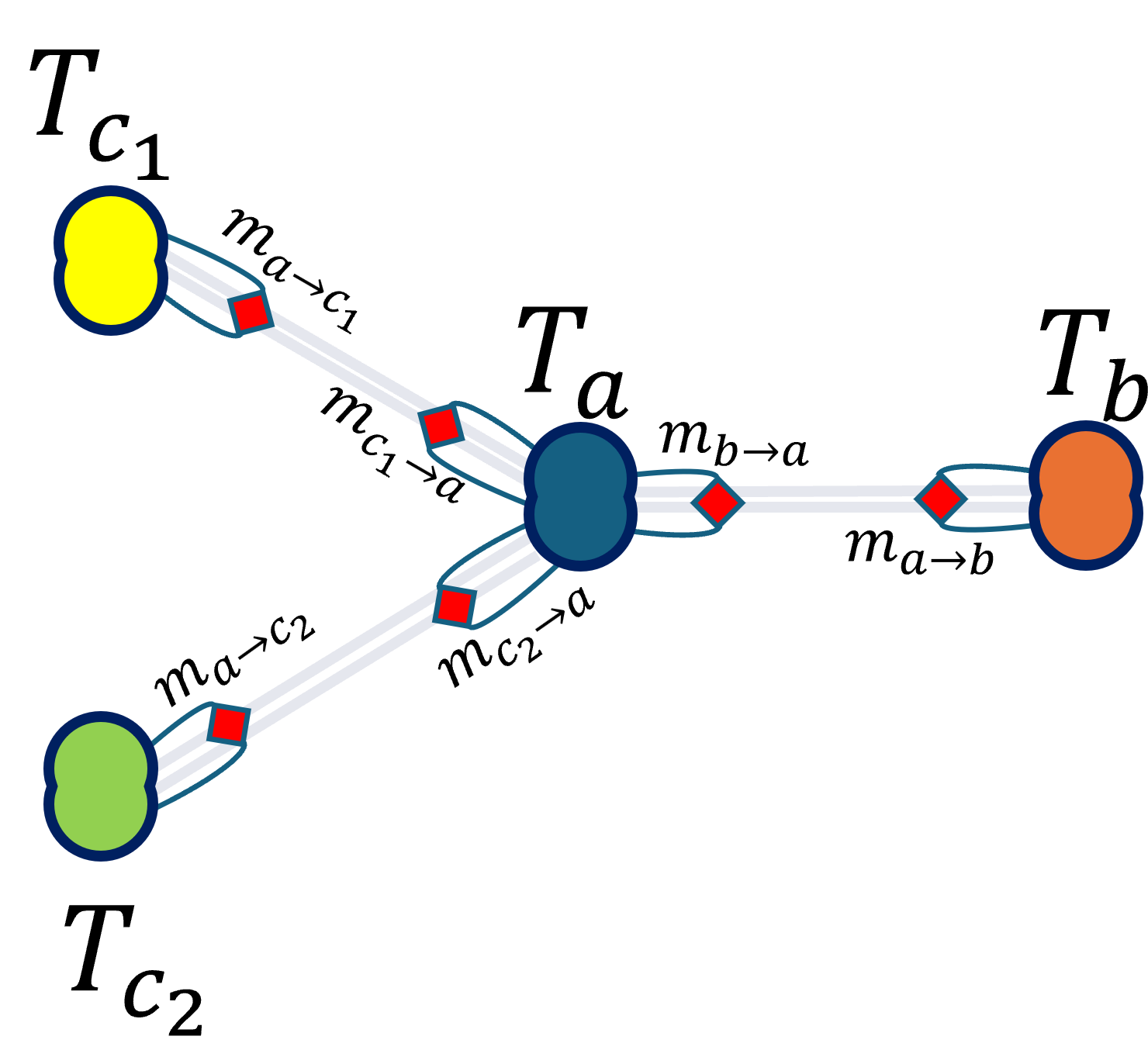}  
            \caption{Each virtual leg in $\braket{\psi|\psi}$ connecting two tensors, is replaced with messages coming from and going towards every other tensor. These messages are matrices.}
            \label{fig:bp:bpExample:sub3}
        \end{subfigure}
        \hfill
        \begin{subfigure}[b]{0.9\subwidth}
            \centering
            \includegraphics[width=0.70\linewidth]{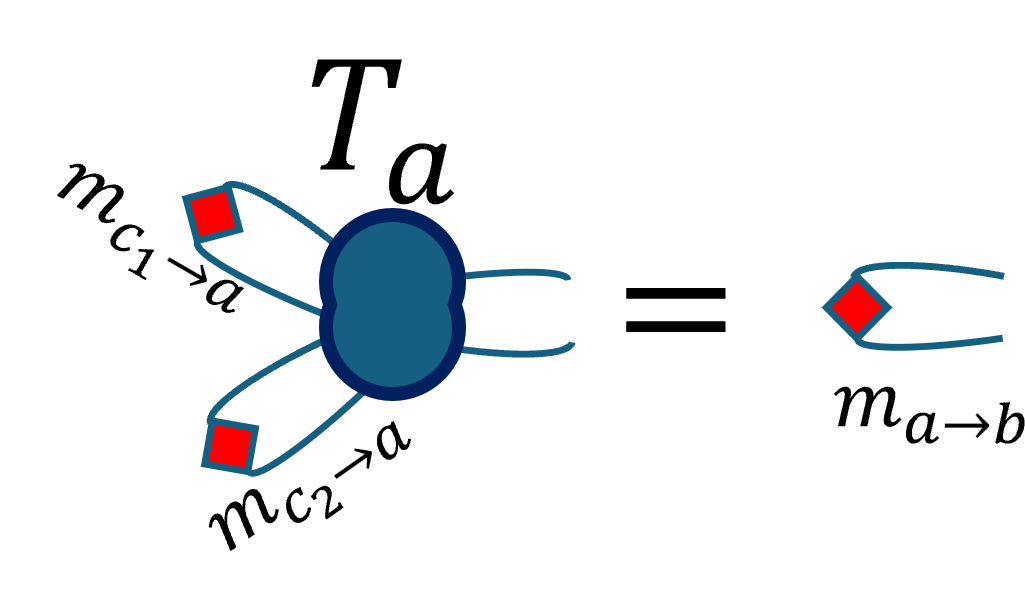}  
            \caption{
                Graphical representation of Eq. \ref{eq:BP:DefiningEquation}
                At each iteration $l$, the message
                $
                    m_{a\rightarrow b}^{(l)} 
                $
                from tensor $T_a$ to $T_b$, is calculated from the messages at iteration $l-1$ by contracting $T_a$ with all incoming messages towards it, except the one arriving from $T_b$.
            }
            \label{fig:bp:bpExample:sub4}
        \end{subfigure}
        \caption[Belief-propagation for Tensor-Networks]{%
            Belief propagation for Tensor-Networks \cite{ItaiArad:alkabetz2021tensor}. Given a PEPS (\subref{fig:bp:bpExample:sub1}) we generate double-layered tensor network representing $\braket{\psi|\psi}$ (\subref{fig:bp:bpExample:sub2}). Then the algorithm initiates random semi-definite matrices around each tensor (\subref{fig:bp:bpExample:sub3}).
            Then, at each iteration at time $l$, the message
            $
                m_{a\rightarrow b}^{(l)} 
            $
            from tensor $T_a$ to $T_b$, is calculated from the messages at iteration $l-1$ (\subref{fig:bp:bpExample:sub4}) by following equation \ref{eq:BP:DefiningEquation}.
        }
        \label{fig:bp:bpExample}
    \end{minipage}
\end{figure}

When dealing with networks without-loops (tree-graphs), BP is not an iterative process but rather an exact process \cite{ItaiArad:alkabetz2021tensor}.

Most PEPS contain loops, thus requiring an iterative protocol that updates all local environments according to Eq. \ref{eq:BP:DefiningEquation} until a convergence is achieved.
The convergence criterion for this iterative process is derived from the mean square error between two messages at consecutive steps. I.e., 
\begin{equation}
    e^{(l)}_{a\rightarrow b}
    =
    \sqrt{
        \sum_{i,j} 
        \left(
            \left(
                m_{a\rightarrow b}^{(l)} 
                -
                m_{a\rightarrow b}^{(l-1)} 
            \right)
            ^2
        \right)
        _{i,j}
    }
\label{eq:bp:convergenceCriterium}
\end{equation}
When all errors $e^{(l)}_{a\rightarrow b}$ between all neighbors $a,b$ have reached a desired threshold, the iterative process stops.

\section{The Block Belief-Propagation Algorithm}
\label{subsec:blockBP:BlockBPAlgo}

This thesis expands on the novel method named 
block belief propagation (\blockBP{}) presented in 
\cite{Itai:BlockBP} and inspired by 
classical probabilistic models. It is designed 
specifically for the approximate contraction of 
2D TNs. This method offers high parallelizability 
and can be flexibly applied to both finite and 
infinite 2D systems. The underlying principle in 
belief propagation lies in the close resemblance 
between quantum many-body systems and classical 
multivariate probabilistic models. Both share the 
characteristic of states residing in 
exponentially high-dimensional spaces, 
necessitating the (approximate) computation of 
local properties. In the context of quantum 
systems, these properties translate to local 
expectation values, while in classical models, 
they represent local marginal probabilities 
\cite{wainwright2008graphical}. Consequently, 
advancements in one field can often benefit the 
other. 

Previous attempts to leverage BP for TN 
contraction have been reported 
\cite{TN:sahu2022efficient, 
ItaiArad:alkabetz2021tensor}. When applied to 
PEPS with imaginary time evolution (\ITE{}), it was 
demonstrated \cite{ItaiArad:alkabetz2021tensor} 
to be equivalent to the simple-update algorithm, 
in which the environment
is approximated in a separable, mean-field way 
\cite{TN:jiang2008accurate}.

\begin{figure}[htbp]
    \centering
    \includegraphics[width=.55\linewidth]{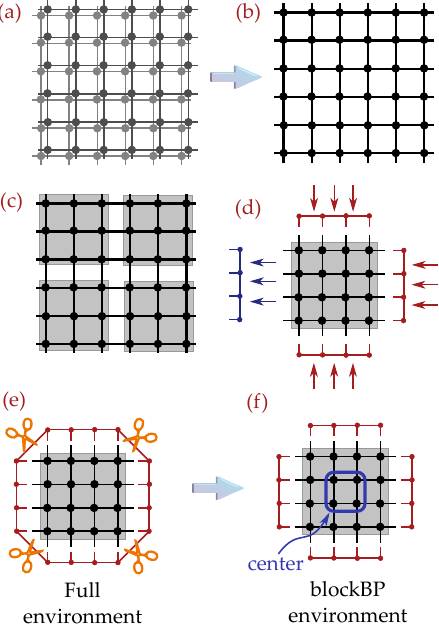}
    \caption[BlockBP]{Iterative process of the \blockBP{} algorithm and the environment around the block. A TN with physical legs \captionStyle{a} \ can be fused \captionStyle{b} \ and sectioned into blocks \captionStyle{c}. The effect (under contraction of the TN) of one block on another can be approximated using "message" MPSs \captionStyle{d}. This approximation breaks some of the entanglement in the environment \captionStyle{e,f}. Image taken from \cite{Itai:BlockBP}.}
    \label{fig:BlockBP}
\end{figure}

\blockBP{} replaces the tensors with blocks of tensors and the messages matrices with Matrix-Product-States (MPSs). Figure \ref{fig:BlockBP} shows how the \blockBP{} algorithm arranges tensors into blocks, and after fusing the physical legs into "braket" $\braket{\psi|\psi}$ form, approximates the environment of that block encoded in the surrounding MPSs. Those messages cannot encode fully the environment of a large TN, since the division into several messages breaks any entanglement that could have existed between different parts of the environment (Fig. \ref{fig:BlockBP}(e,f)). 

\blockBP{} was benchmarked in \cite{Itai:BlockBP} against the Heisenberg model and the transverse-field Ising (TI) model on simple 2D lattices and was shown to achieve results that are comparable in precision to current state-of-the-art methods. %
Notably, \blockBP{} exhibits exceptional parallelizability%
\footnote{A feature that I will detail in the next section.}%
, leading to a substantial improvement in computational efficiency compared to other leading TN-based approaches. Furthermore, the method can be easily applied to both finite and infinite systems with translational invariance. Additionally, its applicability extends to systems with different unit-cells and geometries. This versatility suggests the potential application of \blockBP{} to problems in quantum chemistry involving less regular structures.

Fig. \ref{fig:BlockBP}(d) is a visual representation of the core mechanism of the \blockBP{} algorithm, i.e. each block sends messages, represented by MPSs, to its neighboring blocks. The same way Eq. \ref{eq:BP:DefiningEquation} is the defining equation for the \emph{belief-propagation} algorithm for TNs, Eq. \ref{eq:blockBP:DefiningEquation}
is the defining equation for the \blockBP{} algorithm:

\begin{equation}
    M^{(l)}_{a\rightarrow b}
    =
    \texttt{contract} \left(
        B_{a}
        \prod_{
            c\in N_{a} \setminus \{b\}
        }
        M^{(l-1)}_{c\rightarrow a}
    \right)
    \label{eq:blockBP:DefiningEquation}
\end{equation}

Compared to the BP equation (Eq. \ref{eq:BP:DefiningEquation}), here the simple matrices at step $(l-1)$
$
    m^{(l-1)}_{c\rightarrow a}
$
are replaced with a product of matrices, i.e. the MPS
$
    M^{(l-1)}_{c\rightarrow a}
$
; Instead of the simple contraction of a tensor and some matrices around it, here the contraction is over a more complex TN comprised of the block $B_{a}$ and the MPSs sent from its neighbors 
$
    M^{(l-1)}_{c\rightarrow a}
$%
.
Since the environment is represented by multiple MPSs, instead of a single pMPS (Fig. \ref{fig:TN:pMPS}), entanglement that may have existed between different regions in the lattice --- breaks.

\subsection{BlockBP for infinite lattices }

In this work we intend on studying how \blockBP{} can be used to investigate properties of the antiferromagnetic (AFM) Heisenberg model on the infinite \kagome{} lattice. 
While this algorithm is straightforward for large finite systems, where multiple blocks interact with one-another, the study of infinite lattices requires careful adjustment.
We can conjecture that one block is sufficient 
to represent an infinite lattice where a unit-cell is repeated both throughout the lattice, and the single block that we have chosen. The goal, as I will describe more thoroughly later, is
to calculate an environment of the unit-cell that encompasses the effect of an infinite lattice thereupon. 
The method to achieve this goal will be to have a single block that sends and
receives messages to itself, thus mimicking the case where it is
surrounded by infinite copies of itself. 
Fig. \ref{fig:blockBP:infinite} is a schematic representation of how an infinite-BlockBP algorithm would work for a square lattice with a unit-cell of 4 sites. A contraction is needed in order to generate an MPS message --- this could be replaced by the more efficient \bubblecon{} algorithm (see section \ref{sec:intro:bubblecon}).

\newlength{\figheight}
\setlength{\figheight}{12.5em}

\begin{figure}[htbp]
    \centering
    \begin{minipage}{0.98\textwidth}  
        \centering
        
        \setlength{\subwidth}{0.49\linewidth}  
        
        \begin{subfigure}[b]{\subwidth}
            \centering
            \includegraphics[height=\figheight]{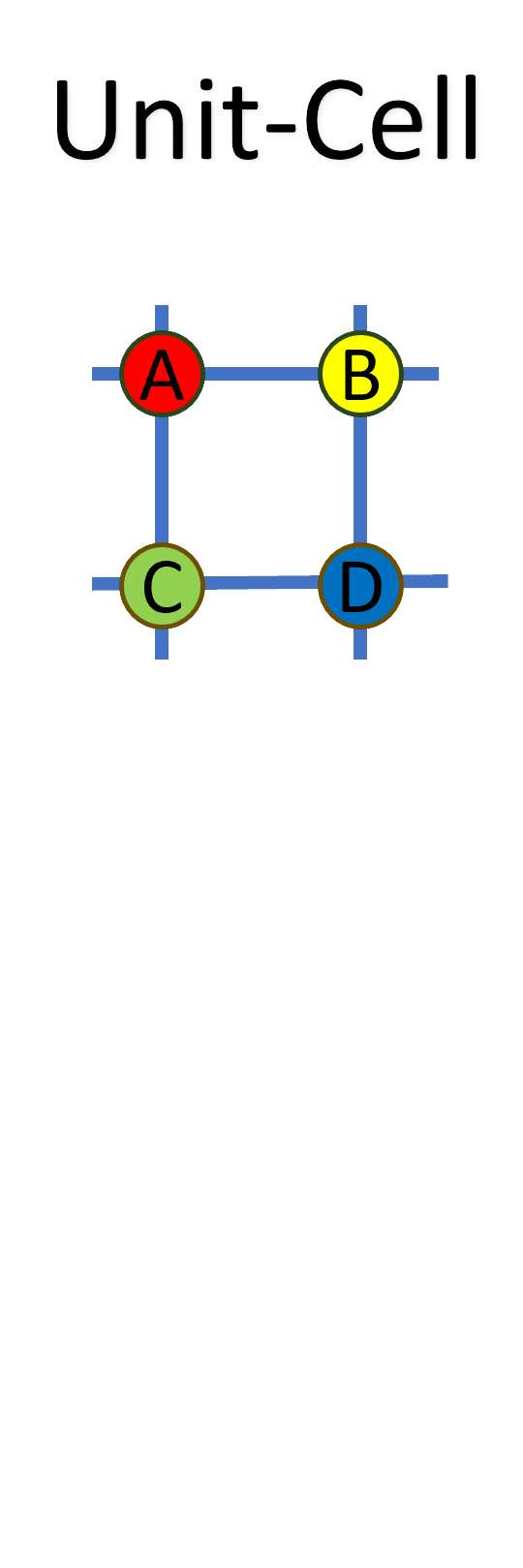} 
            \caption{A unit-cell comprised of 4 different tensors in square lattice structure.}
            \label{fig:blockBP:infinite:sub1}
        \end{subfigure}
        \hfill
        \begin{subfigure}[b]{\subwidth}
            \centering
            \includegraphics[height=\figheight]{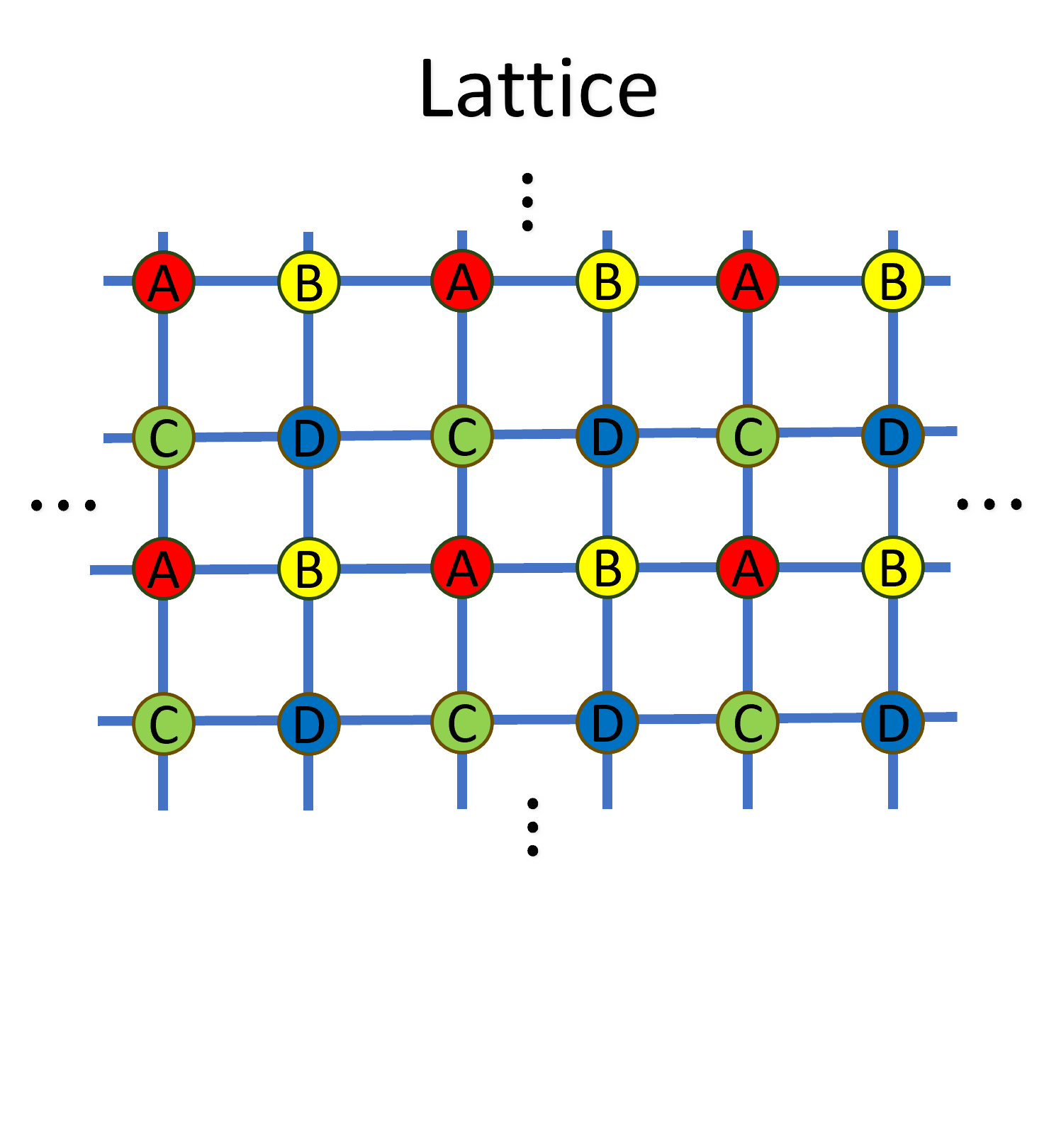} 
            \caption{The infinite square lattice can be spanned using repetitions of the unit-cell.}
            \label{fig:blockBP:infinite:sub2}
        \end{subfigure}
    
        \medskip 
        
        \begin{subfigure}[b]{\subwidth}
            \centering
            \includegraphics[height=\figheight]{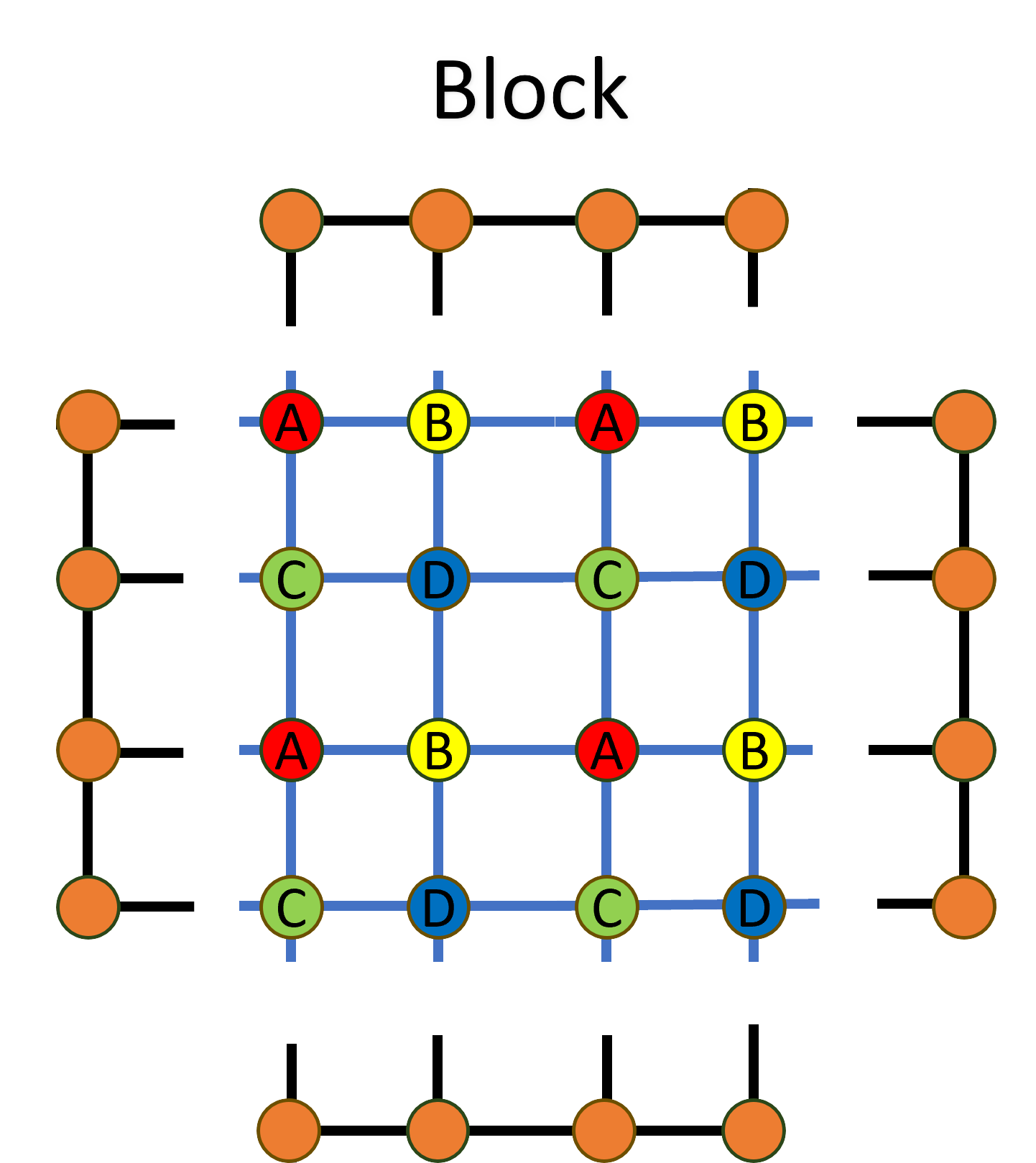} 
            \caption{A Block is chosen where the unit-cell appears with 1 or multiple repetitions. MPS messages are randomized such that they fit the geometrical boundaries of the block.}
            \label{fig:blockBP:infinite:sub3}
        \end{subfigure}
        \hfill
        \begin{subfigure}[b]{\subwidth}
            \centering
            \includegraphics[height=\figheight]{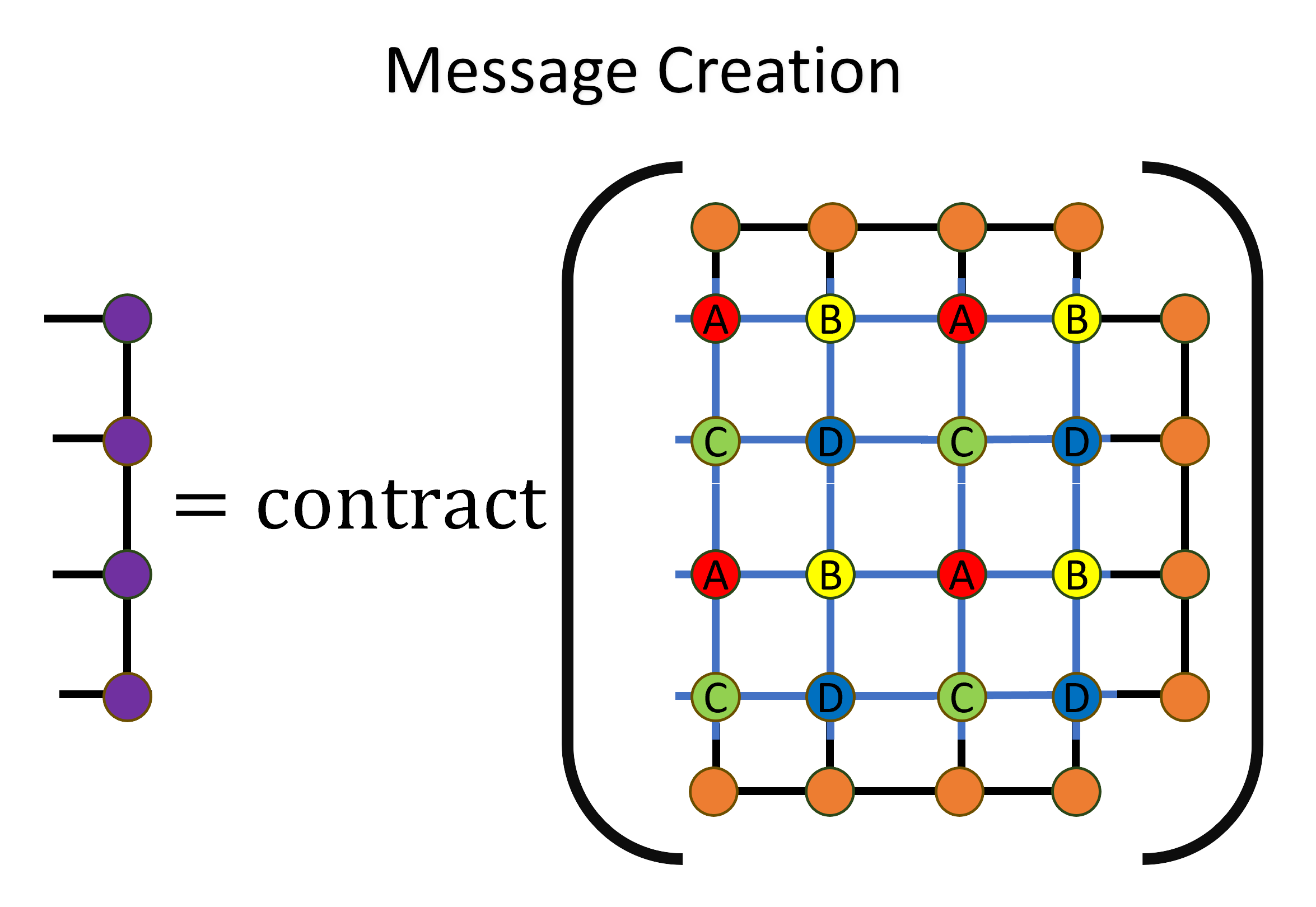} 
            \caption{At each iteration, an outgoing MPS message is the result of the contraction of the block connected to all but a single message.}
            \label{fig:blockBP:infinite:sub4}
        \end{subfigure}
        \begin{subfigure}[b]{\subwidth}
            \centering
            \includegraphics[height=\figheight]{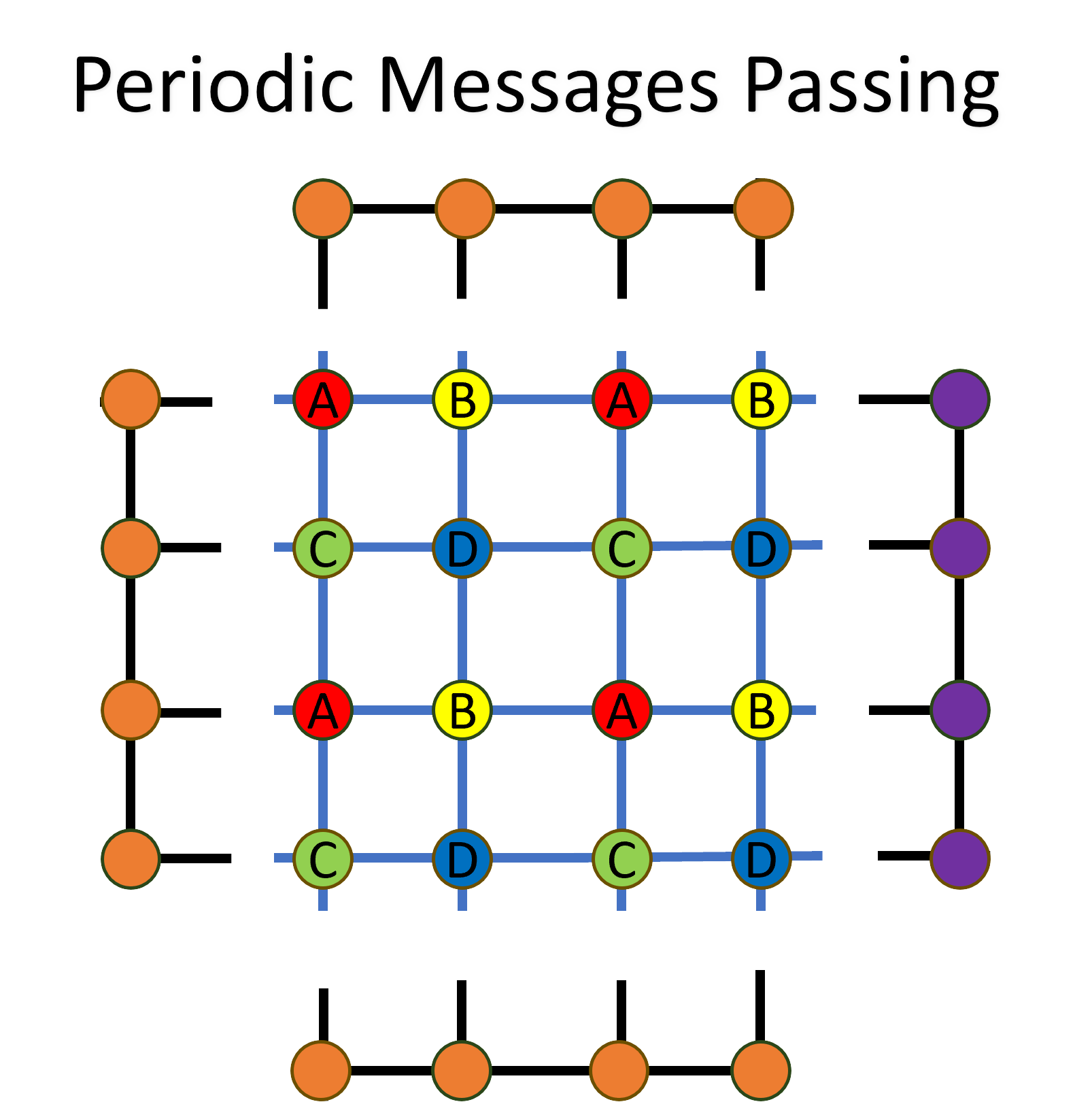} 
            \caption{%
                The updated message replaces the old message at the opposite face of the block.
                \text{   }
                \text{   }
            }
            \label{fig:blockBP:infinite:sub5}
        \end{subfigure}
        
        \caption[BlockBP for infinite lattices]{%
            Using a single block to simulate the effect of an infinite lattice. (d) shows how the messages are updated at each iteration. Whenever a contraction of the TN is needed to generate an MPS, using \bubblecon{} can be more efficient.%
        }
        \label{fig:blockBP:infinite}
    \end{minipage}
\end{figure}

The \blockBP{} algorithm with a single block sending messages to itself is outlined here in Algorithm \ref{alg:BlockBP_singleBlock}.

Note that the iteration over all block sides (Line \ref{alg:BlockBP_singleBlock:line:for_side} in Algorithm \ref{alg:BlockBP_singleBlock}) performs contraction over the same set of tensors, omitting only a single MPS each time. Thus, this step is a great candidate for multi-threading --- performing multiple calculation on the same data in parallel. Indeed, in this work I have implemented this part of the algorithm using multi-threading.

\subsubsection{Time and Space Complexity}
\label{subsec:blockBP:complexity}

For a block containing $n$ tensors, each with bond dimension $D$, $m$ messages each of length $k$ ($k$ tensors per MPS) and a truncation bond-dimension for the \bubblecon{} process of $\chi>D$ --- the overall time complexity for each iteration (each repetition of line \ref{alg:BlockBP_singleBlock:line:repeat})
is %
$
    \mathcal{O}\left(
        m n \chi^3    
    \right)
$%
.
This is due to the time cost of the \bubblecon{} process (Eq. \ref{eq:bubblecon:timeConplex}), multiplied by the amount of messages $m$.
Using multi-threading, we can compute all messages at the same time, removing $m$ from the equation. Typical number of iteration until convergence is $5$-$15$, adding a constant multiplicative factor of mere $\mathcal{O}(10)$, resulting in an overall time complexity of (Eq. \ref{eq:blockBP:timeComplexity}): %
\begin{equation}
    \label{eq:blockBP:timeComplexity}
    \texttt{time: }
    \mathcal{O}\left(
        n \chi^3    
    \right)
\end{equation} %

The space complexity is derived simply by the cost of storing the TN and the messages in memory. 
If our TN is a repetition of a unit-cell with $r$ sites, storing the TN can be efficiently done with only 
$
    \mathcal{O}\left(
        r d D^p  
    \right)
$, %
where $p$ is the number of neighbors per site, and $d$ is the physical-dimension%
\footnote{Instead of storing the tensors in their physical form, i.e., with physical legs; it's possible to keep the $\braket{bra|ket}$ form, at a space-cost of 
$
    \mathcal{O}\left(
        r D^{2p}
    \right)
$ for storing the unit-cell.
}%
.
Storing $m$ messages cost %
$
    \mathcal{O}\left(
        m k \chi^2    
    \right)
$, %
due to the space complexity of storing each message-MPS (Eq. \ref{eq:bubblecon:spaceConplex}).
Thus, %
the overall space complexity is (Eq. \ref{eq:blockBP:spaceComplexity}): %
\begin{equation}
    \label{eq:blockBP:spaceComplexity}
    \texttt{space: }
    \mathcal{O}\left(
        r d D^p  + m k \chi^2   
    \right)
\end{equation} %

\SetKw{iIn}{in}
\SetKw{iFor}{for}
\newcommand{\cmnt}[1]{\text{ } \tcp{#1}}
\newcommand{\Cmnt}[1]{\tcp{** #1 ** //}}

\begin{algorithm}
\caption{Block-BP algorithm for single block sending messages to itself}
\label{alg:BlockBP_singleBlock}
\DontPrintSemicolon
\LinesNumbered 
    \KwIn{$Block$} 
    \KwOut{$Messages$}
    \Cmnt{Initialize:}
    $i\gets 0$\;
    $Messages^{(0)} \gets $ random $MPS$ \iFor side \iIn $Block$.sides\;
    \Cmnt{Main loop:}
    \Repeat{$error \leq threshold$}{
        \label{alg:BlockBP_singleBlock:line:repeat}
        $Messages^{(i+1)} \gets \{\}$\;
        \Cmnt{Implement Eq. \ref{eq:blockBP:DefiningEquation}:}
        \For{$side$ \iIn all-block-sides}{ 
            \label{alg:BlockBP_singleBlock:line:for_side}
            $TN \gets Block + \left( \text{All } Messages^{(i)} \text{ excluding message in } side \right)$\; 
            $msg_{new} \gets $ \texttt{contract}($TN$)
            \cmnt{contraction results in an MPS (Fig. \ref{fig:blockBP:infinite:sub4})}  
            $Messages^{(i+1)}\left[\text{opposite}(side)\right] \gets msg_{new}$\ \cmnt{outgoing message replaces the message from the opposite side (Fig. \ref{fig:blockBP:infinite:sub5})}
            \label{algo:line:replaceMessagesPeriodically}
        }
        $error \gets $ distance between $ Messages^{(i+1)} $ and $ Messages^{(i)} $\;
        $i\gets i+1$\;
    }
    \KwRet{$messages^{(i)}$}
\end{algorithm}

\subsection{Conditions for studying infinite lattices using BlockBP}
\label{sec:blockBP:kagomeBlock:conditions}
\label{subsec:blockBP:conditions}

One who wishes to implement this method for the investigation of infinite lattices must be careful as to how one structures the sole block. While it may not be apparent in simple lattices (e.g. square lattices as demonstrated in Fig. \ref{fig:blockBP:infinite}) as was done in \cite{Itai:BlockBP}, more complex lattices need to be carved into blocks more carefully, 
a concept that will be discussed in this subsection.

Though it was shown that the \blockBP{} algorithm can be used to simulate infinite lattices \cite{Itai:BlockBP}, while trying to implement this algorithm on the infinite \kagome{} lattice, I discovered some rather subtle conditions that need to be met, which may not be immediately apparent.

First, 
in order to achieve the feat of using a single block to represent an infinite lattice, a block must be carved out of the lattice, with the conditions that this block must tile the entire lattice. This means that the block comprised of repetitions of the unit-cell, must conserve the linear translational symmetry when a face of the block is periodically connected to the opposite face of the block. For example, if in Fig. \ref{fig:blockBP:infinite:sub3} the block is folded into a torus, site $A$ remains to the right of site $B$, even when looking at the boundary of the block.

\begin{figure}[htbp]
    \centering
    \begin{subfigure}[b]{0.45\textwidth}
        \centering
        \includegraphics[width=\textwidth]{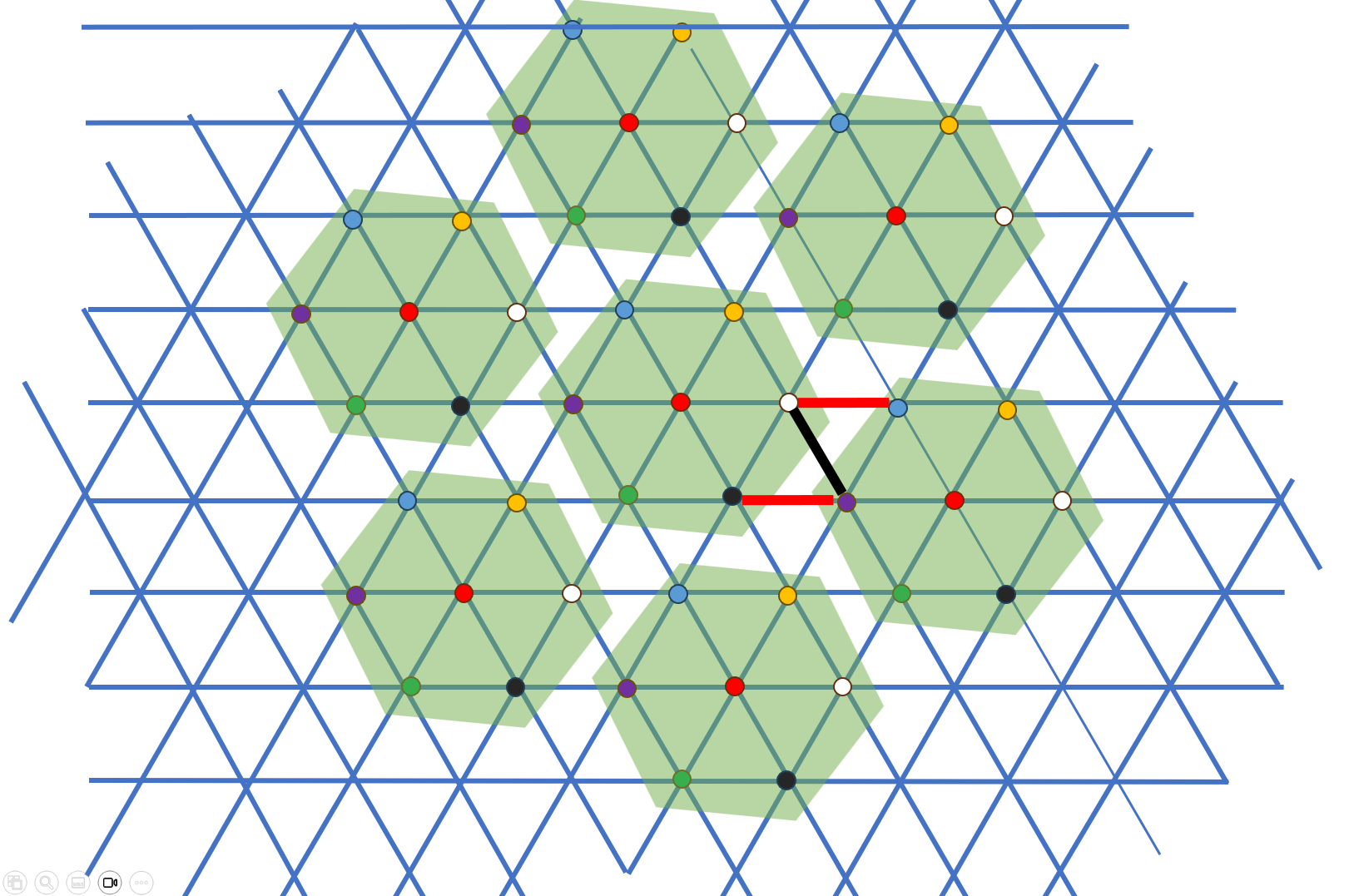}
        \caption{Wrong tiling. Edges from the same block-face pass through multiple different faces on the other side of the block.} 
        \label{fig:block:triangular:bad}
    \end{subfigure}
    \hfill
    \begin{subfigure}[b]{0.45\textwidth}
        \centering
        \includegraphics[width=\textwidth]{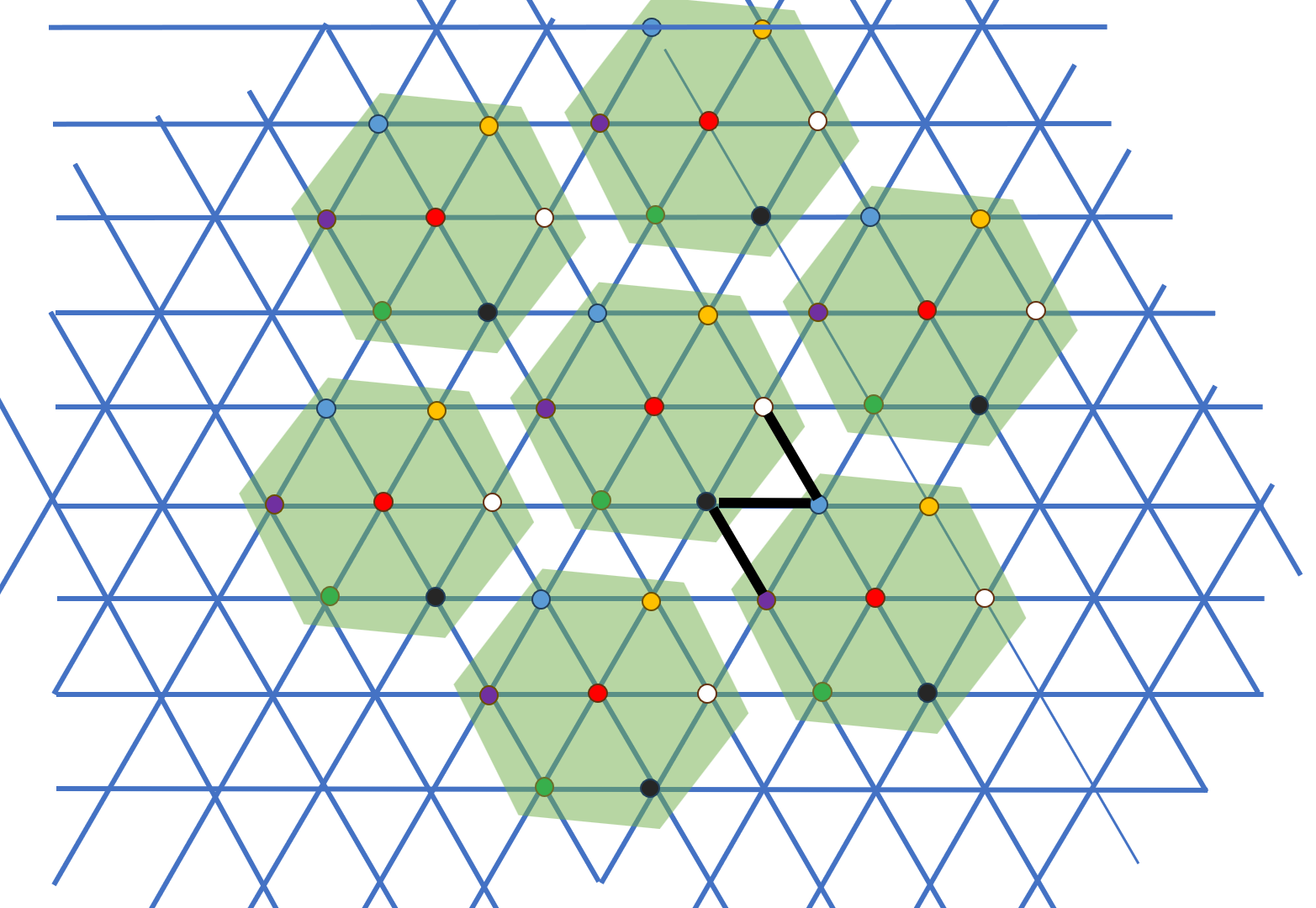}
        \caption{Correct tiling. Edges pass block faces altogether.} 
        \label{fig:block:triangular:good}
    \end{subfigure}
    \caption[Tiling a triangular lattice with a hexagonal block]{
        Tiling a triangular lattice with copies of the hexagonal block. (a) Bad example where the nodes in the lower-right face are connected to a copy of the block involving 4 different faces of the hexagonal block. (b) Good example since only two faces are crossed enabling the use of periodic boundary condition of the block to represent the infinite lattice.
    }
    \label{fig:block:triangular-lattice-hexagonal-block-good-vs-bad}
\end{figure}

Second, 
edges going out of the block, thus passing a face of the block, must pass all together and enter the face at the opposite end of the block.
This last condition, though difficult to explain in words, can be easily understood through an example on the triangular lattice. When \blockBP{} was used for the infinite triangular lattice in \cite{Itai:BlockBP}, there could have been multiple 
ways to tile the lattice. Two such ways using hexagonal blocks are demonstrated in figure \ref{fig:block:triangular-lattice-hexagonal-block-good-vs-bad}, where we draw the block multiple times on the lattice, to demonstrate how it is connected to itself in a periodic fashion. In one tiling option (subfig. \ref{fig:block:triangular:bad}) the two edges exiting the bottom-right
face must pass through two different faces (the upper and bottom left faces). 
The second option (subfig. \ref{fig:block:triangular:good}) achieves tiling of the entire lattice, while all edges also adhere to the rule of passing through opposite faces altogether.
Though it might seem as if both tilings are equivalent since both successfully span the entire lattice by repetitions of the block, choosing which edges pass which face of the block is part of what defines the block. In this sense, those two tilings are not equal. 
In the next section, where a block is carved out of the infinite \kagome{} lattice, this tasks becomes evidently more difficult, as multiple tiling suggestions fails to adhere to this rule.

This second rule is important for the core mechanism of Belief-Propagation to work: Message passing. The messages in the 'Block' variation of the BP algorithm are MPSs that align themselves along edges connecting one block to another. In the same way that nodes are connected by edges, blocks are connected by super-edges. These super-edges cannot split into multiple faces, as each MPS cannot break into multiple MPSs. %
Only if a block tiles the entire lattice with faces completely overlapping each-other, can one hope to work with \blockBP{} to simulate an infinite lattice.

\section{BlockBP for the infinite Kagome lattice}
\label{sec:blockBP:infiniteKagomeBlock}

\subsection{Defining the Kagome block}
\label{sec:blockBP:kagomeBlock:defineKagomeBlock}

\begin{figure}[htbp]
    \centering
    \begin{minipage}{0.6\textwidth}  
        \centering
        \begin{subfigure}{0.45\textwidth}
            \centering
            \includegraphics[width=\textwidth]{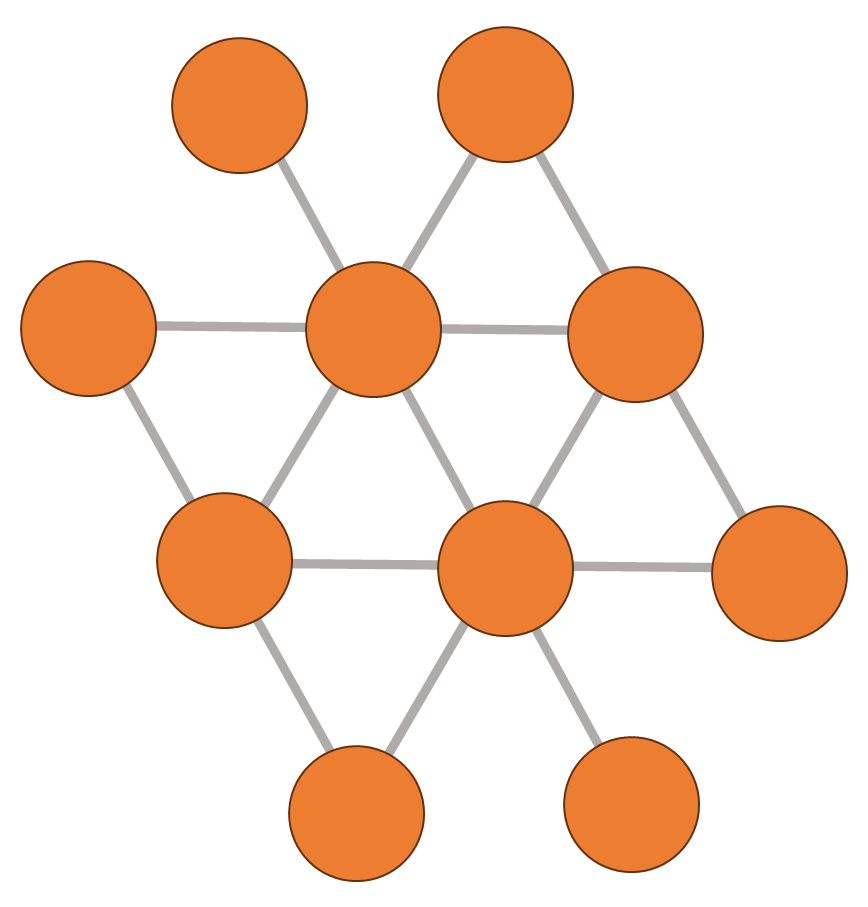}
            \caption{Start with a triangular lattice.}
        \end{subfigure}
        \hfill
        \begin{subfigure}{0.45\textwidth}
             \centering 
             \includegraphics[width=\textwidth]{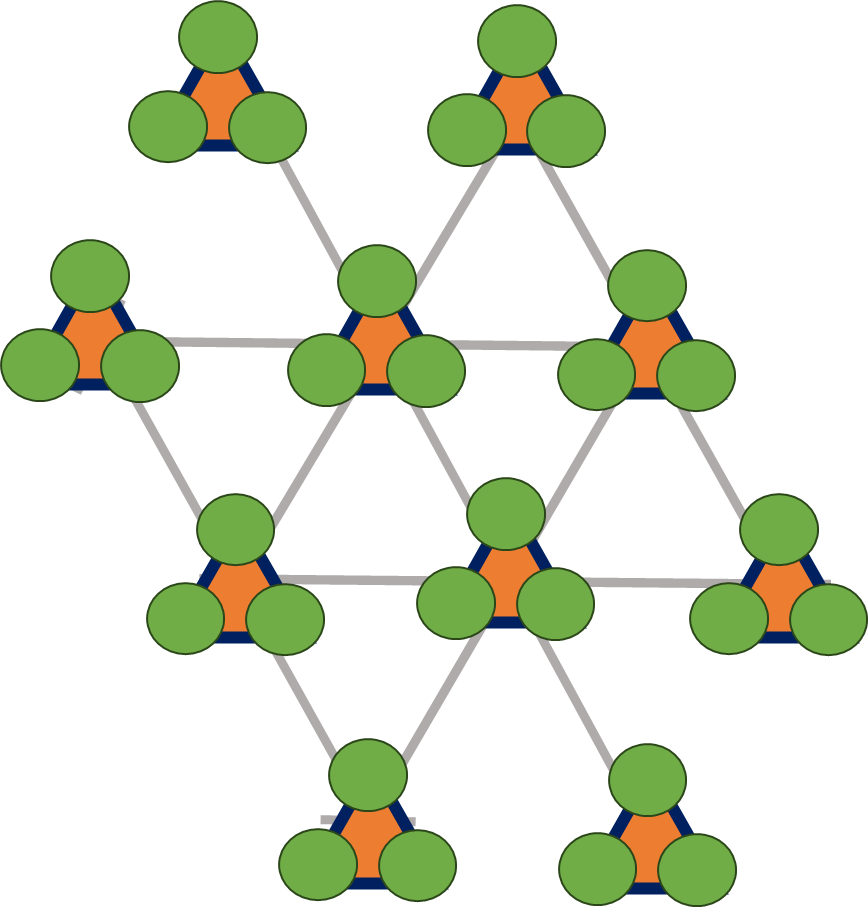}
             \caption{Replace each node with an upper-triangle. Each corner of the triangle is itself a node in the new lattice.}
        \end{subfigure}
        \begin{subfigure}{0.45\textwidth}
            \centering
            \includegraphics[width=\textwidth]{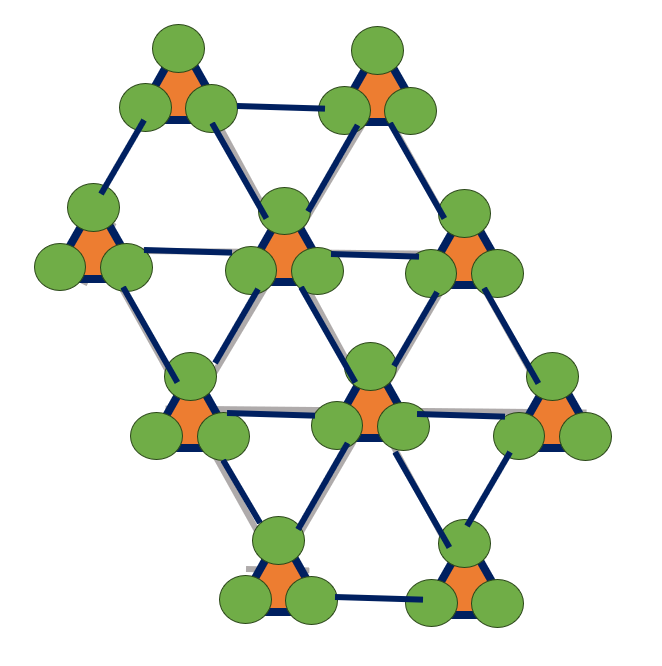}
            \caption{The corners\textbackslash{}nodes at each triangle are connected between themselves but also to corners\textbackslash{}nodes on the nearest triangle in the original lattice.}
        \end{subfigure}
        \hfill
        \begin{subfigure}{0.45\textwidth}
             \includegraphics[width=\textwidth]{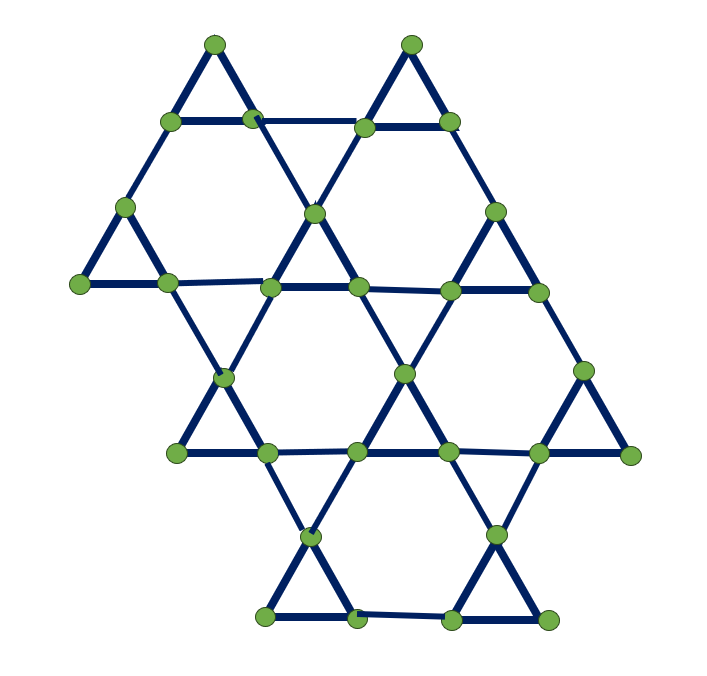}
             \caption{A \kagome{} lattice emerges.}
        \end{subfigure}
        \caption{
            Triangular lattice with embedded triangles
        }
        \label{fig:representation:embedding}
    \end{minipage}
\end{figure}

The \kagome{} lattice is a Bravais lattice with a basis of three sites forming a triangle, as demonstrated in Fig. \ref{fig:representation:embedding}. This way of looking at the \kagome{} lattice helped us choosing a block structure to adhere to the rules described in subsection \ref{subsec:blockBP:conditions}. 

\subsubsection{Hexagonal block}

\begin{figure}[htbp]
    \centering
    \includegraphics[width=0.45\textwidth]{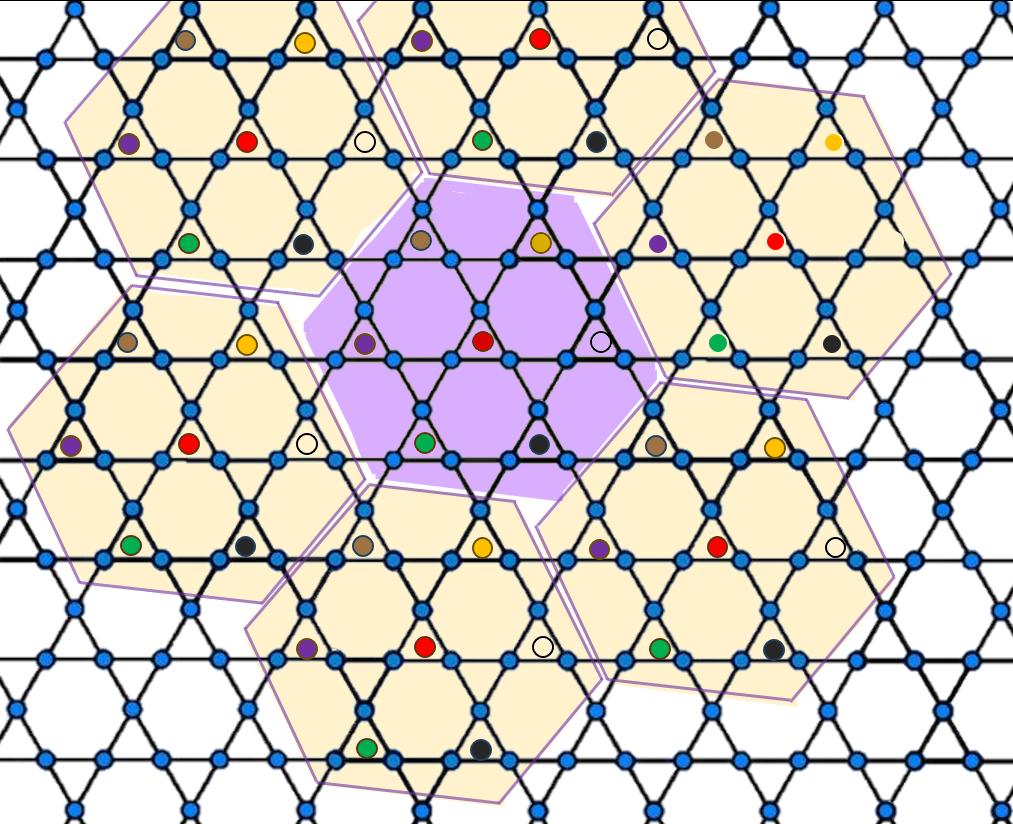}
    \caption[Hexagonal block for the Kagome lattice]{
        Hexagonal block for tiling the \kagome{} lattice. The colored dots represent unique repeated upper-triangles in the lattice. 
    }
    \label{fig:block:kagome:hexagonal}
\end{figure}

A hexagonal block is natural for the triangular lattice \cite{Itai:BlockBP} (see Fig. \ref{fig:block:triangular:good}), and hence also for the \kagome{}.
%
The resulting block with 7 upper-triangles is shown in Fig. \ref{fig:block:kagome:hexagonal}.

The block in Fig. \ref{fig:block:kagome:hexagonal}
is the smallest possible block with this shape. We can say that it has a face-length of $N=2$, defined by the number of upper-triangles on the face of the block. 
How an upper-triangle is orientated in relation to the boundary of the block matters when counting the number of tensors on that boundary, hence half
of the faces contain $4$ tensors and the other half contain $2$. Since this block has 7 upper-triangles, it has $N_{T}=7\cdot3=21$ tensors\textbackslash{}sites in total.

\begin{figure}[htbp]
    \centering        
    \includegraphics[width=0.45\textwidth]{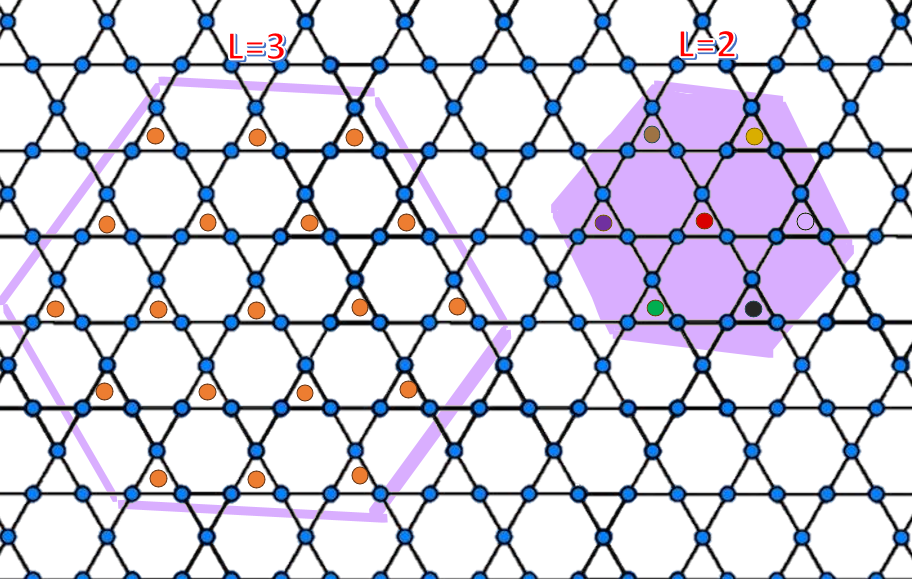}
    \caption[Hexagonal block for the Kagome lattice with $N=2,3$]{
        Hexagonal block in the \kagome{} lattice with boundary-sizes 2 and 3.
        The block with face-length $N=3$ contains 
        $
            N_{T}
            = 3(3N^2-3N+1)
            = 57
        $ 
        sites.
    }
    \label{fig:block:kagome:hexagonal-by-edge-sizes}
\end{figure}

Given a general block of face-length $N$, the number of tensors it contains is
$N_{T}=3(3N^2-3N+1)$ sites. For an example of larger blocks see figure \ref{fig:block:kagome:hexagonal-by-edge-sizes}.

\subsubsection{Parallelogram block}

\begin{figure}[htbp]
    \centering
    \includegraphics[width=0.45\textwidth]{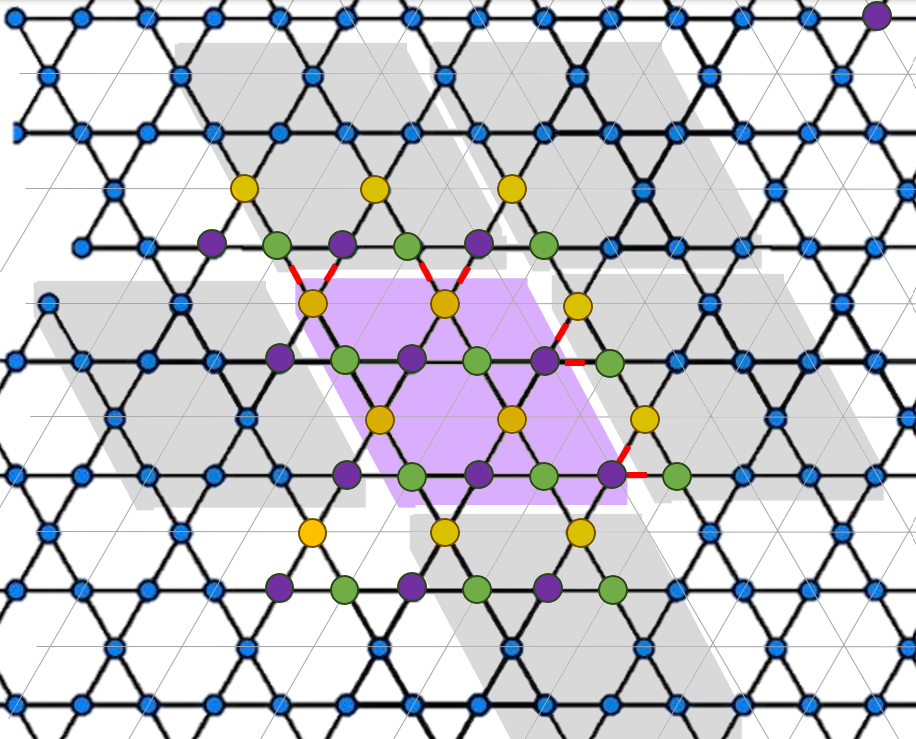}
    \caption[Hexagonal block for the Kagome lattice]{
        parallelogram block in the \kagome{} lattice 
    }
    \label{fig:block:kagome:parallelogram}
\end{figure}

Another option is to use a parallelogram\footnote{This is the same block that was used as a unit-cell in a work by Román Orús \cite{structureMatrix:Orus:jahromi2019universal}} as shown in figure \ref{fig:block:kagome:parallelogram}.
This is the smallest block of this shape, with $N_T = 12$ sites.
The same as a the number of sites in a single \kagome{} "star".  
The top and right edges have $L=2$ sites, while the bottom and left edges have $2L=4$ sites. 

\begin{figure}[htbp]
    \centering
    \includegraphics[width=0.85\textwidth]{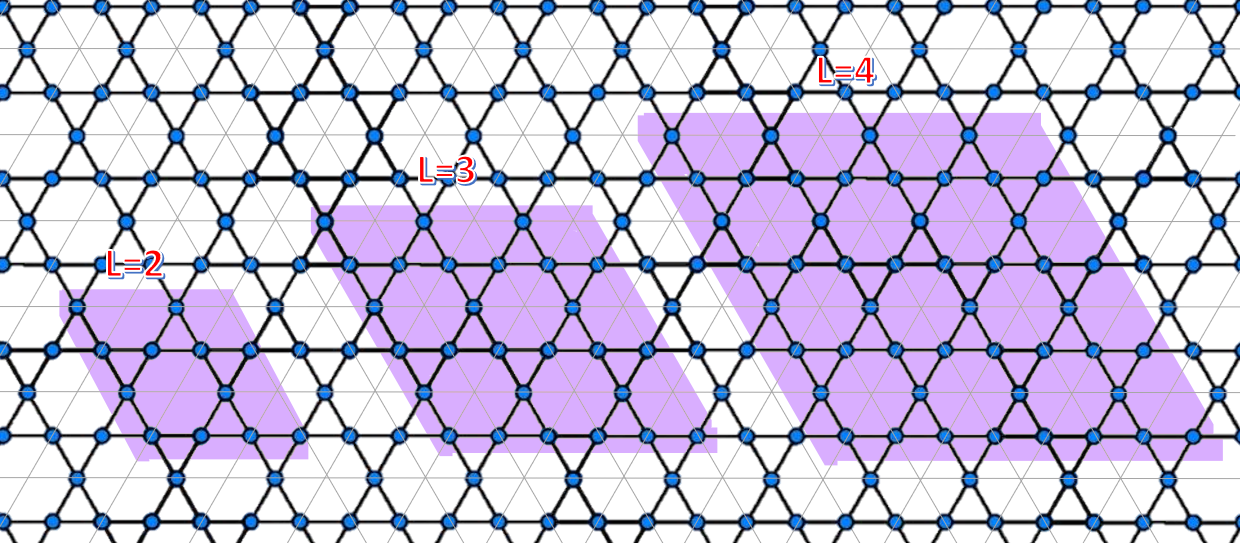}
    \caption[Hexagonal block for the Kagome lattice with $L=2,3,4$]{
        Parallelogram block in the \kagome{} lattice with edge-sizes 2, 3 and 4.
        The block with edge size $L=3$ contains 
        $
            N_{T}
            = 3N^2
            = 3\cdot9
            = 27
        $ 
        sites.
    }
    \label{fig:block:kagome:parallelogram-by-edge-sizes}
\end{figure}

Given a general block of face-length $L$, the number of tensors it contains is
$N_{T}=3L^2$ sites. For an example of larger blocks see figure \ref{fig:block:kagome:parallelogram-by-edge-sizes}.

\subsubsection{Comparison}
\label{sec:blockBP:kagomeBlock:defineKagomeBlock:comparison}

Both suggestions can tile the entire infinite lattice and adhere to the rules outlined in \ref{subsec:blockBP:conditions}. Both designs are not fully-symmetric: The amount of sites on one edge does not equal the amount of sites on the opposite edge. What stays the same is the amount of edges going out from each face, allowing the use of MPSs as message carriers in the \blockBP{} algorithm.

The hexagonal design is unique in the sense that it preserves the 6-fold rotation symmetry of the \kagome{} lattice. We will leverage this property throughout the work when rotating the unit-cell, a feature that will be discussed in Subsec. \fullsubsubref{subsec:ite:4kagome:transformations}.
When using the hexagonal block, one must maintain 6 different directions where MPSs are connected and along which to contract the entire TN (Using \bubblecon{} whenever applicable).

The first more apparent benefit of  parallelogram block, is having only 4 simpler directions in which MPSs are connected. The second is that this structure allows a further coarse-graining which transforms the \kagome{} lattice into an equivalent square lattice, as is done in \cite{Orus:spinSKagome}. This second feature allows researchers to use iPEPS algorithms built predominantly for the square lattice. 

Although the parallelogram simplicity might appear appealing, We decided to use the unique hexagonal block in this work, for its ability to preserve rotational symmetries inherent to the \kagome{} lattice.

Further details about the structure of the block and how to represent it in code, can be found in Appendix \ref{appendix:kagomeBlockIndexing}.

\subsection{Contracting the Hexagonal Kagome block}
\label{sec:blockBP:kagomeBlock:contractingKagomeBlock}

\begin{figure}[htbp]
    \centering
    \includegraphics[width=0.5\linewidth]{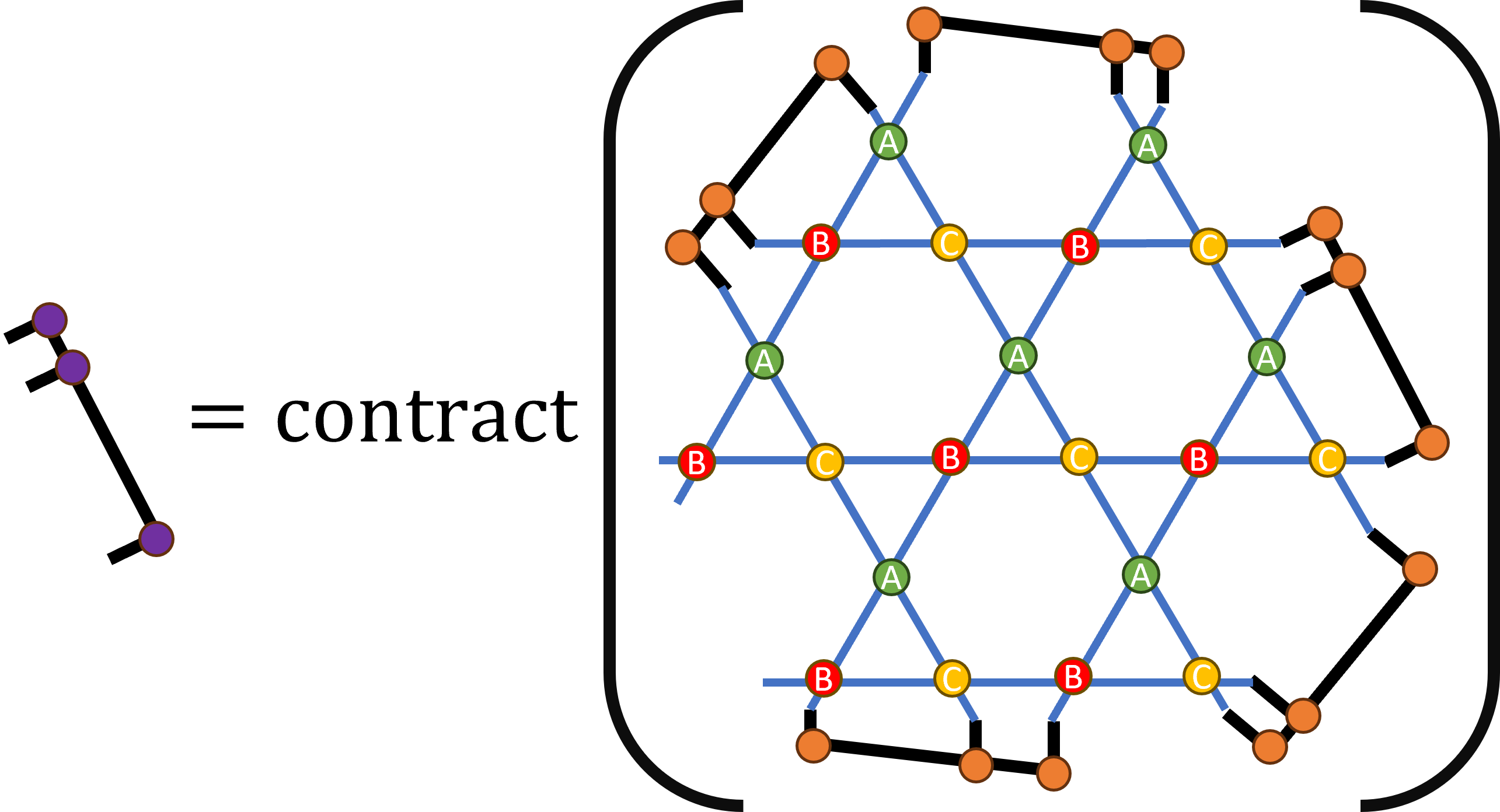}
    \caption[contraction the block results in a MPS]{%
    The contraction using \bubblecon{} of the entire block with all of its surrounding MPS messages except one, results in a MPS of the same size
    }
    \label{fig:blockBP:contraction:toMPS}
\end{figure}

As discussed in Subsec. \ref{subsec:blockBP:BlockBPAlgo},
the \blockBP{} algorithm requires the contraction of the entire block together with the messages connected to it (see Fig. \ref{fig:blockBP:infinite:sub4}).
This can be done efficiently using the \bubblecon{} algorithm (see Sec. \ref{sec:intro:bubblecon}).
If the block is contracted together will \textit{ALL} its surrounding MPS messages, a scalar is achieved. Most frequently, what is needed is a new MPS message for the update process (Eq. \ref{eq:blockBP:DefiningEquation}), a fundamental part of the algorithm. In these instances, the contraction is done up until all tensors are contracted except a single message MPS. Now, since the number of out-going legs is equal to the number of MPS legs%
\footnote{An MPS was connected in this part, so the number of legs from that MPS into the block equals the number of now out-going legs out of the block}%
,
the MPS that results from this contraction is a proper out-going message MPS (See Fig. \ref{fig:blockBP:contraction:toMPS} for a more visual example).

\subsubsection{Contraction Order}

\begin{figure}[htbp]
    \centering
    \begin{minipage}{1\textwidth}
    
        \setlength{\subwidth}{0.49\linewidth}  
        \centering
        \begin{subfigure}[b]{\subwidth}
            \centering
            \includegraphics[width=\linewidth]{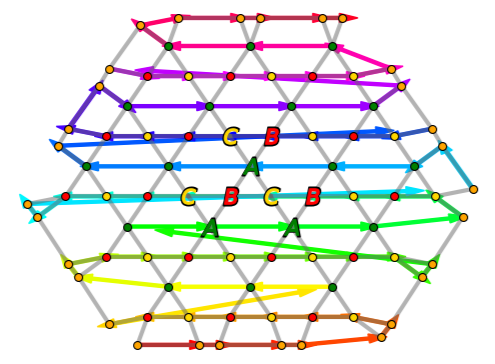}
            \caption{Contracting from the bottom up}
        \end{subfigure}
        \hfill   
        \begin{subfigure}[b]{\subwidth}
            \centering
            \includegraphics[width=\linewidth]{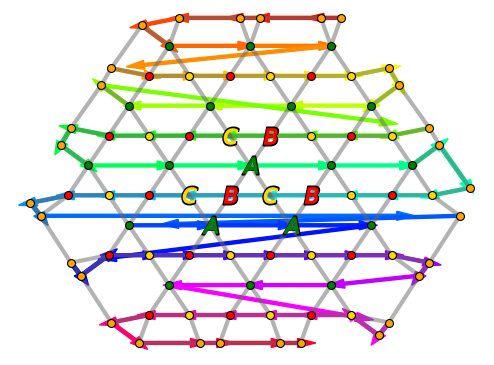}
            \caption{Contracting from the top down}
        \end{subfigure}
    
        \caption[contraction orders]{%
            $2$ out of $6$ contraction orders given to the \bubblecon{} algorithm, for $N=3$.
        }      \label{fig:blockBP:contraction:orderSchemes}
    \end{minipage}
\end{figure}

Because of the uneven number of tensors near opposite faces of the \kagome{} block and the asymmetry to reflections, contracting from below to generate the up-going message is different from (as an example) contracting from above to generate the down-going message. Thus, contraction schemes need to be thoroughly planned for all directions and all block sizes $N$. In Fig. \ref{fig:blockBP:contraction:orderSchemes}  I exemplify by providing visualizations of the contraction orders given to the \bubblecon{} algorithm, for $2$ out of $6$ possible directions, for $N=3$.

\subsubsection{Core}
\label{subsec:BlockBP:kagome:core}

Though contracting the entire block is common, frequently we would need to leave at least one unit-cell unswallowed by the \bubblecon{}.
This is due to the fact that the goal of this entire approach is to perform numerical simulations with a unit-cell that "feels" as if it is in a lattice with infinite copies of itself.
Whenever we need to perform simulations (like \ite{}, as will be discussed in Chapter \ref{chapter:ITE}) or compute expectation values, we will need to hold a single unit-cell surrounded by what must encompass the effect of an infinite environment.
The unit-cell is located preferably in the center of the block, as to mitigate any maleffects caused by the separation of the block's surrounding into multiple MPSs instead of a single periodic MPS (see Fig. \ref{fig:BlockBP}(e,f) and the discussion about pMPS in Sec. \ref{subsec:intro:TN:mps_and_peps:pmps}). Thus, a method to efficiently contract the block using \bubblecon{}, while also ending-up with a unit-cell surrounded by a periodic MPS, is required.

\begin{figure}[htbp]
    \centering
    \begin{minipage}{1\textwidth}
        \centering
        \begin{subfigure}[b]{1\linewidth}
            \centering
            \includegraphics[width=\linewidth]{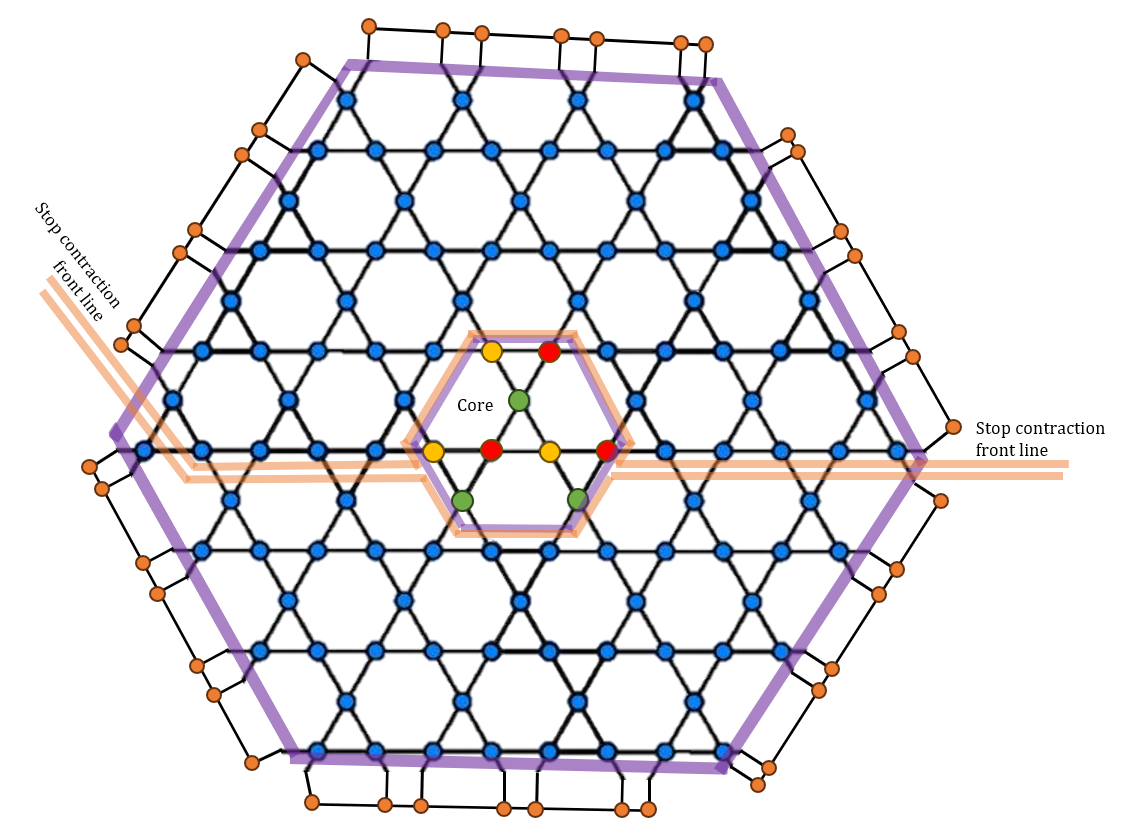}
            \caption{%
            A block of size $N=4$ is contracted from opposite sides using the \bubblecon{} algorithm. The bubbles meet in two front-lines that only do not meet in the region called the "Core"
            }
            \label{fig:blockBP:contraction:toCore:a}
        \end{subfigure}
        \begin{subfigure}[b]{0.55\linewidth}
            \centering
            \includegraphics[width=\linewidth]{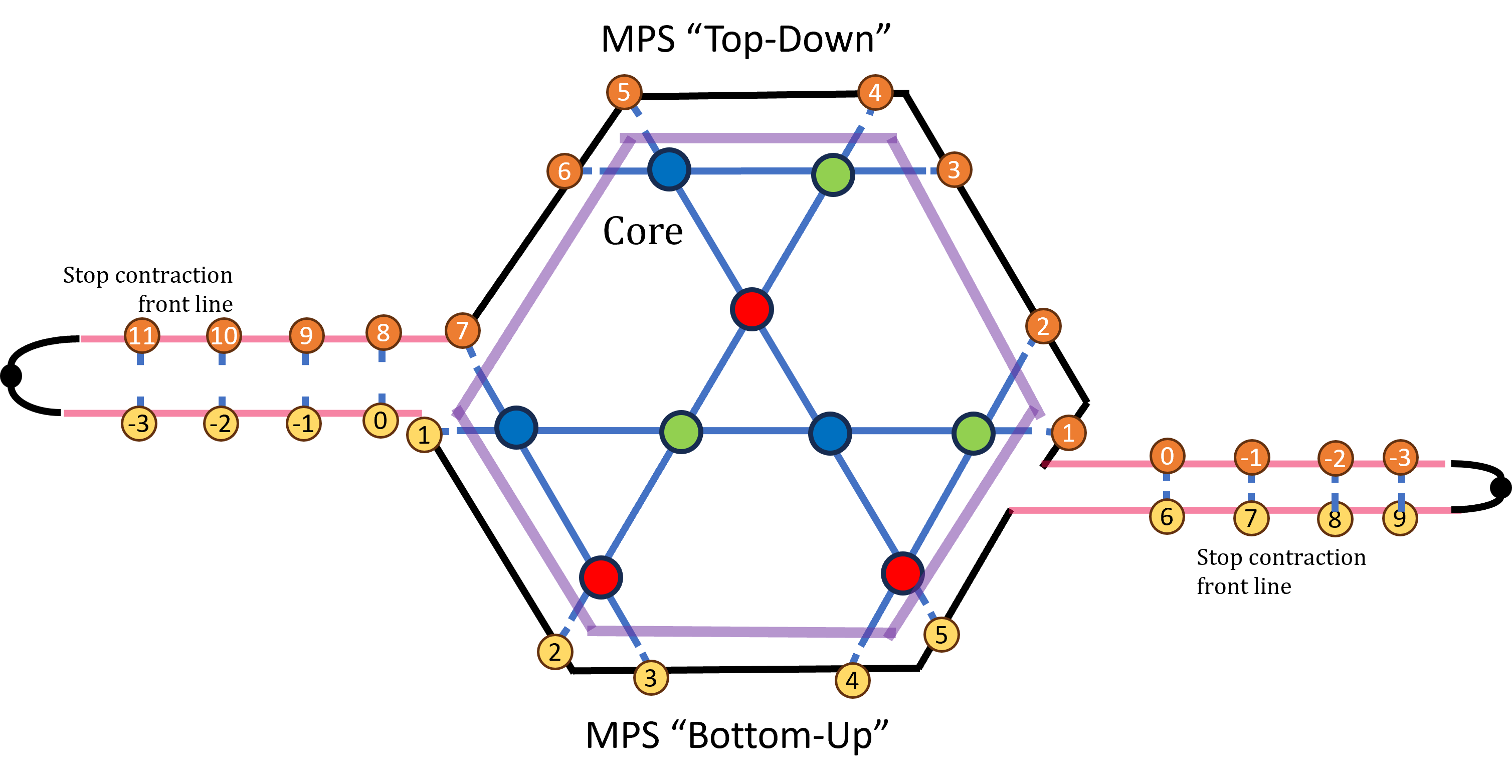}
            \caption{%
            Once everything except the core was contracted into two MPSs, we remain with two MPSs that can be contracted to generate a periodic MPS surrounding the core
            }
            \label{fig:blockBP:contraction:toCore:b}
        \end{subfigure}
        \hfill   
        \begin{subfigure}[b]{0.40\linewidth}
            \centering
            \includegraphics[width=\linewidth]{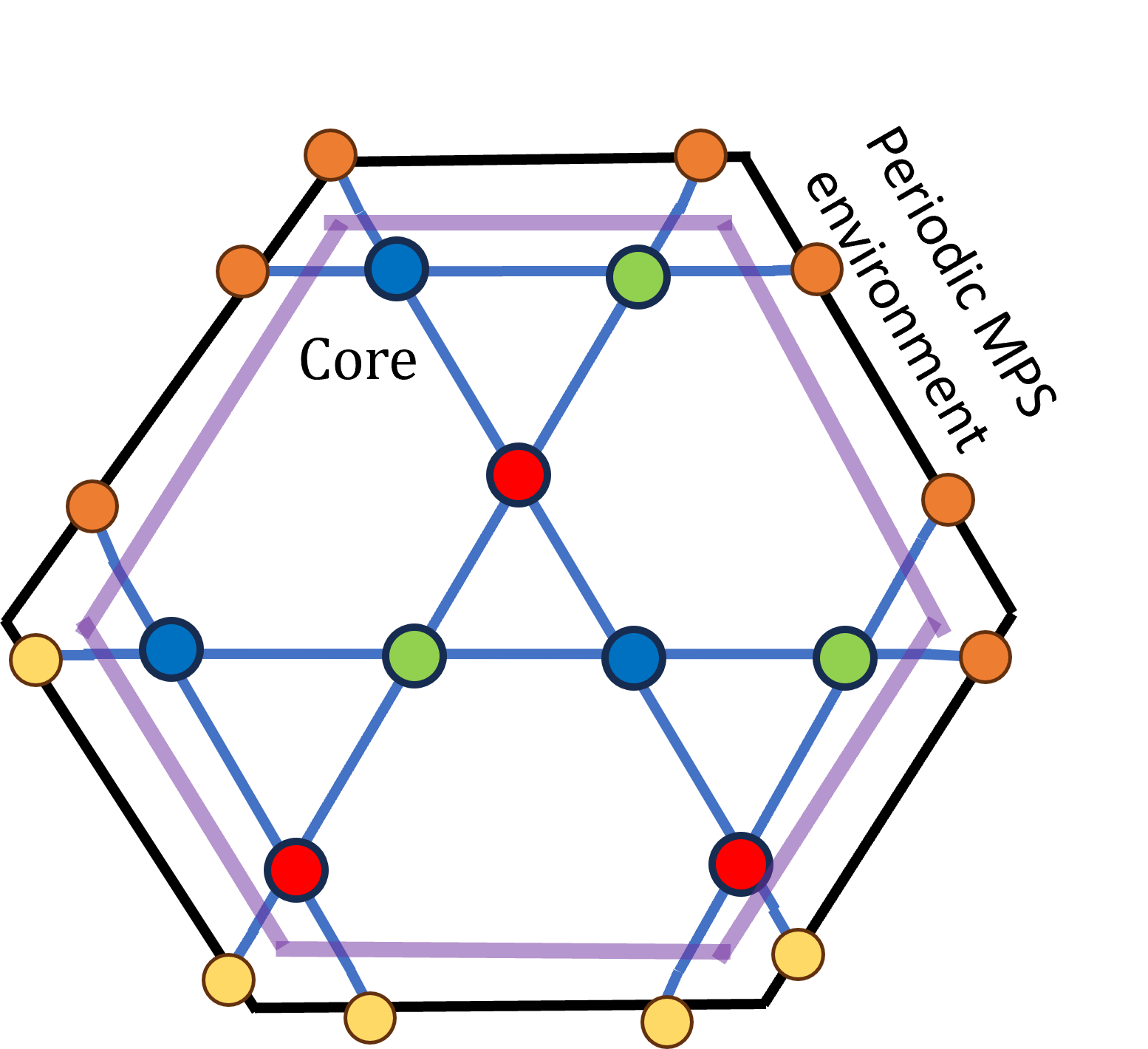}
            \caption{%
            The core surrounded by a periodic MPS. This pMPS approximates an infinite lattice.
            }
            \label{fig:blockBP:contraction:toCore:c}
        \end{subfigure}
        \caption[From BlockTN to coreTN process]{%
            A method to get a core containing the unit-cell, surrounded by a pMPS approximating the infinite lattice environment:
             Two bubbles approach each other from opposite sides of the block, meeting in the middle at $2$ front-lines. The front-lines are connected everywhere except at the core. By contracting the places the MPS are touching, the goal is achieved.
        }      \label{fig:blockBP:contraction:toCore}
    \end{minipage}
\end{figure}

The \bubblecon{} algorithm alone cannot provide the mechanism with which we can end up with a pMPS surrounding a central piece of our network. To overcome this, we have devised a method that is visualized in Fig. \ref{fig:blockBP:contraction:toCore}
.
Once we have a block with messages that converged via the \blockBP{} algorithm, we start two instances of the \bubblecon{} process from opposite sides of the block.
The bubbles follow the same contraction order
that was discussed above and
approach each other, meeting in the middle at $2$ "front-lines". The front-lines are connected everywhere except for in one disconnected region, which I from now on call the \emph{Core} (Fig. \ref{fig:blockBP:contraction:toCore:a}).
The two resulted MPSs generated from applying the \bubblecon{} algorithm twice, are effectively connected to each other in two contiguous ends. 
By contracting both MPSs to each other where applicable (an operation that looks like the motion of closing a zipper) (Fig. \ref{fig:blockBP:contraction:toCore:b}) a pMPS is formed, surrounding the core (Subfig. \ref{fig:blockBP:contraction:toCore:c}).

\subsubsection{Modes}
\label{subsec:BlockBP:kagome:modes}

\begin{figure}[htbp]
    \centering
    \begin{minipage}{1\textwidth}

        \setlength{\subwidth}{0.30\linewidth}  

        \centering
        \begin{subfigure}[b]{\subwidth}
            \centering
            \includegraphics[width=\linewidth]{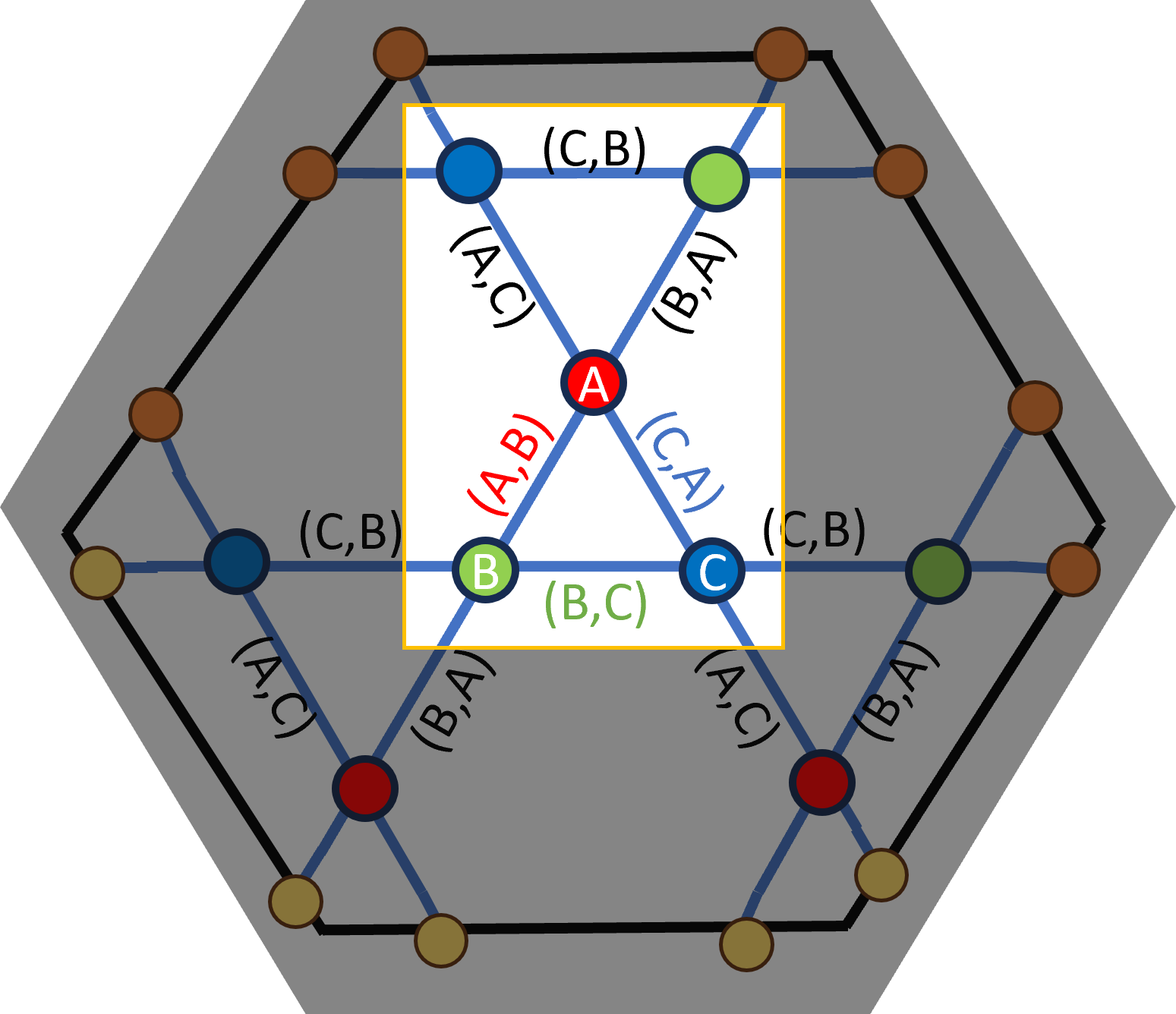}
            \caption{ModeA}
            \label{fig:blockBP:contraction:theCore:ModeA}
        \end{subfigure}
        \hfill   
        \begin{subfigure}[b]{\subwidth}
            \centering
            \includegraphics[width=\linewidth]{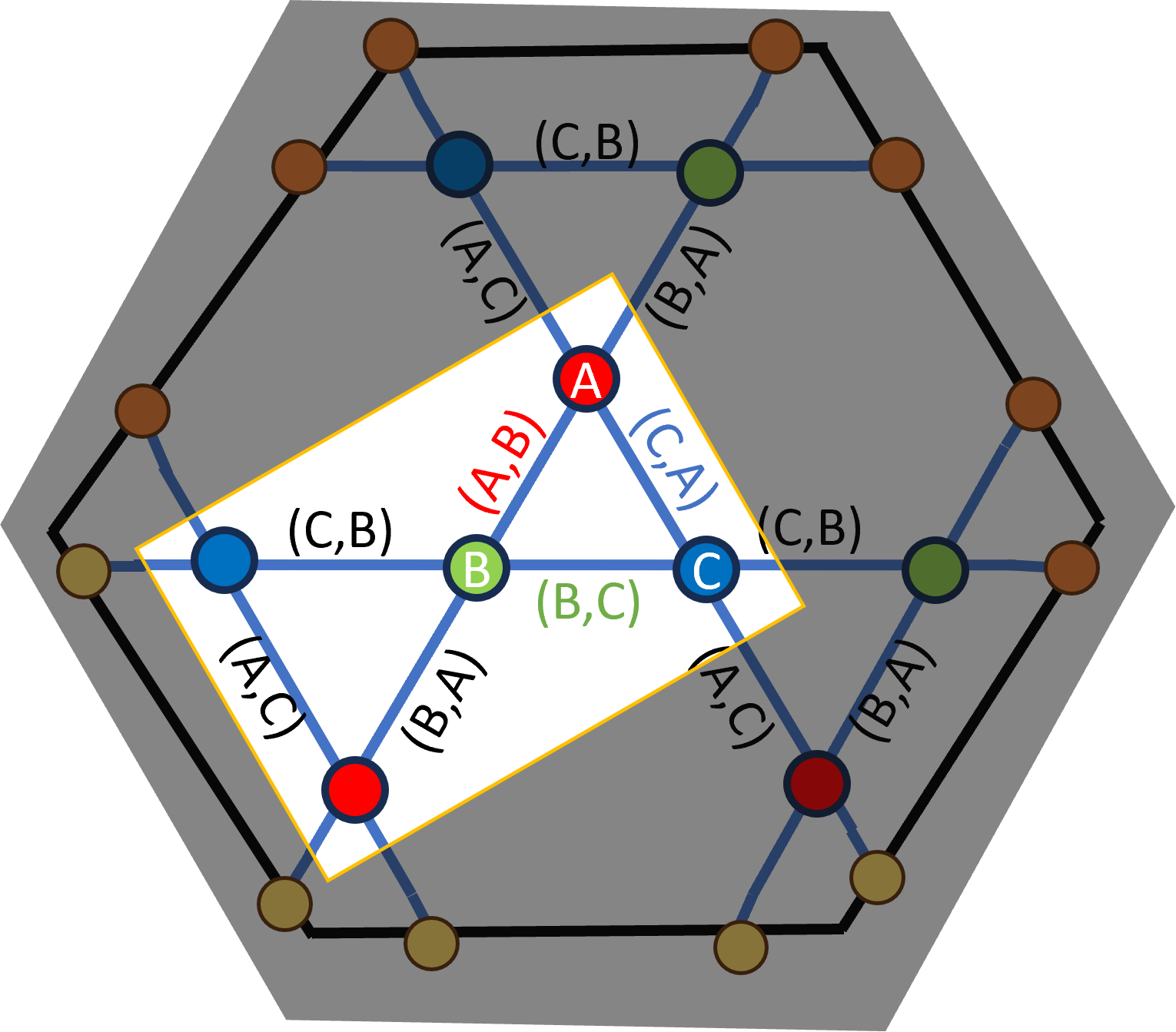}
            \caption{ModeB}
            \label{fig:blockBP:contraction:theCore:ModeB}
        \end{subfigure}
        \hfill   
        \begin{subfigure}[b]{\subwidth}
            \centering
            \includegraphics[width=\linewidth]{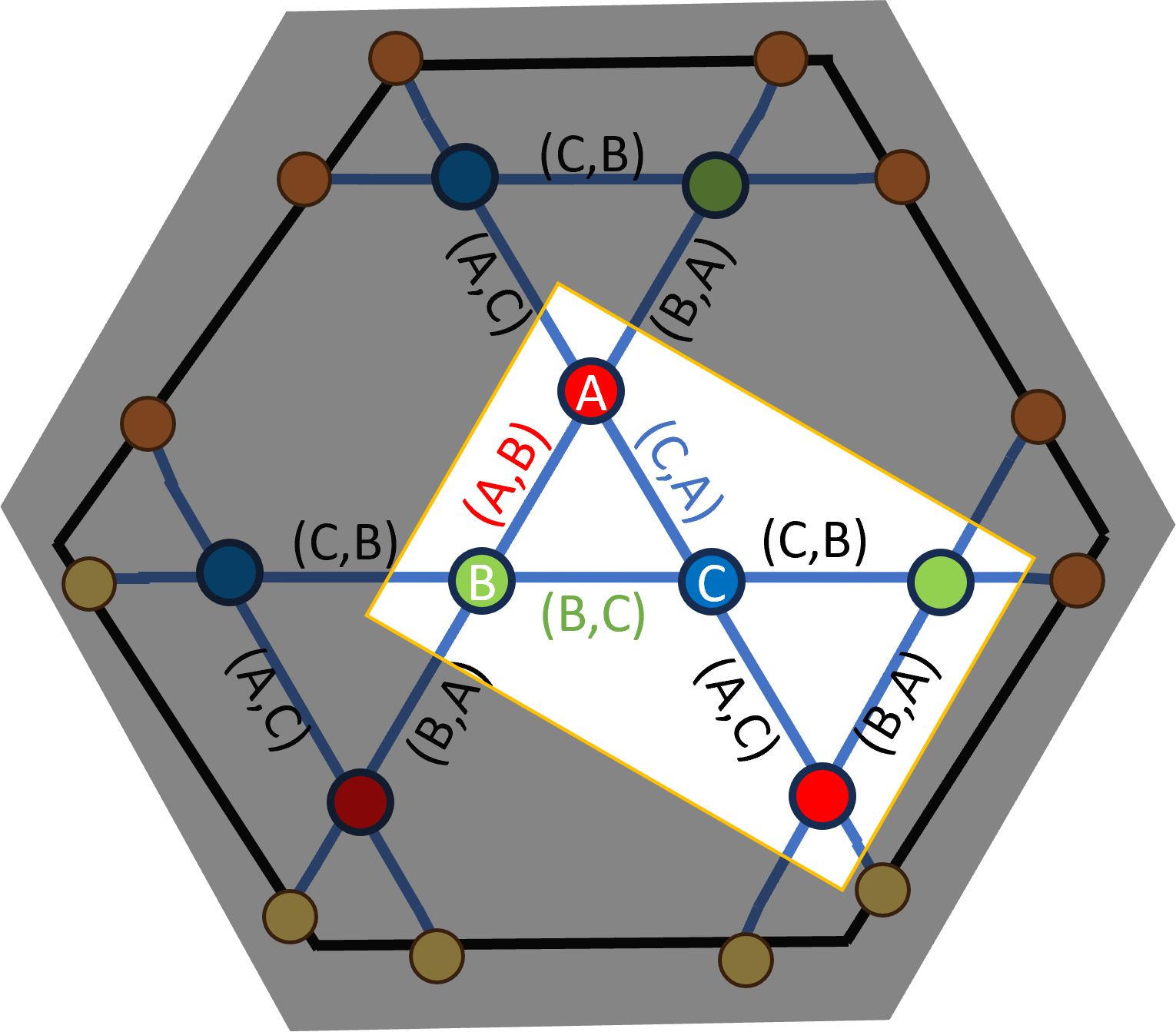}
            \caption{ModeC}
            \label{fig:blockBP:contraction:theCore:ModeC}
        \end{subfigure}
        \caption[The $3$ modes]{%
        Our core contains 3 modes. Each mode contains 1 instance of the 6 unique bonds within the 3-site unit-cell.
        }      
        \label{fig:blockBP:contraction:theCore}
    \end{minipage}
\end{figure}

One can notice that the core contains 3 repetitions of the same unit-cell. In addition, some 2-site bonds appear 3 times while some others appear only a single time.
subfig. \ref{fig:blockBP:contraction:theCore:ModeA} shows all edges and an area where they appear only once. When dealing with algorithms that require the contraction of the entire network save two neighbors, it will be beneficial to designate a middle structure where each such unique bond between pairs appear only once. This structure is called \emph{Mode}, and our core contains 3 such modes, labeled \emph{ModeA}, \emph{ModeB} and \emph{ModeC} (See Subfigures \ref{fig:blockBP:contraction:theCore:ModeA},
\ref{fig:blockBP:contraction:theCore:ModeB}
and 
\ref{fig:blockBP:contraction:theCore:ModeC} accordingly).

\subsubsection{Edge}
\label{subsec:BlockBP:kagome:edge}

\begin{figure}
    \centering
    \includegraphics[width=0.5\linewidth]{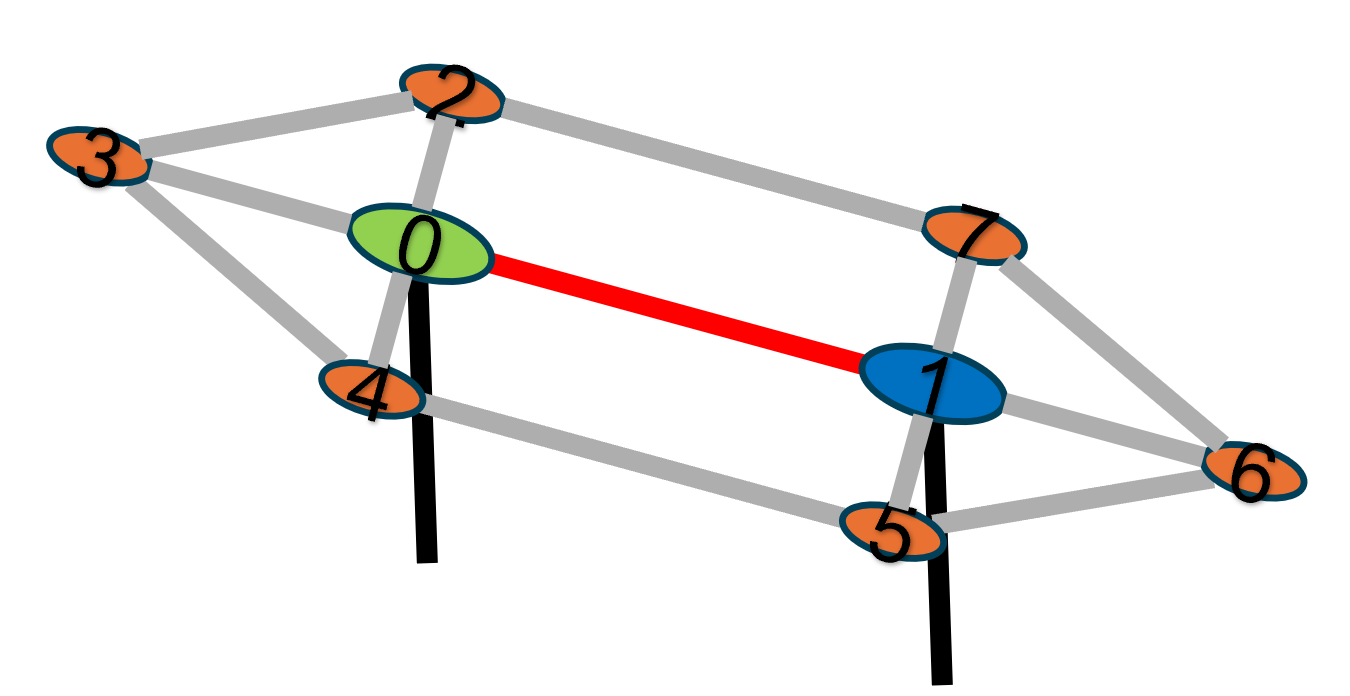}
    \caption{The Edge TN}
    \label{fig:blockBP:contraction:theEdge}
\end{figure}

The last and smallest structure that will be mentioned is the \emph{Edge}. Edge, as the name suggests, is the small TN surrounding a single edge (See Fig. \ref{fig:blockBP:contraction:theEdge}). From every mode, an edge can be achieved by exact contraction.

\subsection{Additions for numerical stability}
\label{subsec:BlockBP:kagome:numerical_stability}

Here, I present various additions to the general approach described above, that, though theoretically introduce no difference, adds numerical stability.

\subsubsection{Mantissa and Exponent:}
\label{subsec:BlockBP:kagome:numerical_stability:mantissaExponent}

In my computational code, I opted for a representation using mantissa and exponent instead of conventional floating-point numbers to enhance numerical stability. This decision was driven by the need to manage the extreme values sometimes generated during the \bubblecon{} procedure, which often results in either very large or very small global factors. Each number \( x \) is expressed in the form \( x = m \times 10^e \), where \( m \) is the mantissa and \( e \) is the exponent. The mantissa \( m \) is maintained within a normalized range, typically \( 1 \leq |m| < 10 \), ensuring that the significant digits are preserved. The exponent \( e \) scales the number appropriately, allowing for efficient representation of both large and small values. This approach not only mitigates the risk of overflow and underflow but also ensures that the arithmetic operations remain stable and accurate, thereby preserving the integrity of the computational results.

\subsubsection{BlockBP Learning-step:}
\label{subsec:BlockBP:kagome:numerical_stability:learningstep}

Another addition, is the introduction of learning-steps to the \blockBP{} algorithm \ref{alg:BlockBP_singleBlock}:
In line \ref{algo:line:replaceMessagesPeriodically},
whenever a new MPS message $msg_{new}$ is produced (See Fig. \ref{fig:blockBP:contraction:toMPS}), it replaces the old message $msg_{old}$ at the opposite block boundary. Sometimes, a more stable convergence can be reached by adding a damping parameter $\alpha\in \left[0,1\right]$ such that the MPS replacing the old MPS is not $msg_{new}$, but rather $msg_{learned}$ from Eq. \ref{eq:blockBP:kagome:numerics:learning_step1}:
\begin{equation}
\label{eq:blockBP:kagome:numerics:learning_step1}
    msg_{learned}
    = 
    \alpha \cdot msg_{new}
    +
    (1-\alpha) \cdot msg_{old}
\end{equation}
Summing MPSs is not trivial. Here we use a method from \cite{MPS:schollwock2011density} for the additions of MPSs.

\subsubsection{Canonical MPS:}
\label{subsec:BlockBP:kagome:numerical_stability:canonical}

MPSs are used throughout this work and extensively in the \blockBP{} algorithm. After producing an MPS (often with the \bubblecon{} algorithm),
I bring them into the canonical form to ensure numerical stability and computational efficiency. 
 An MPS is said to be in right-canonical form if each tensor \( A^{[i]} \) satisfies the orthogonality condition \( \sum_{j} A^{[i]}_{a,j,b} (A^{[i]}_{a',j,b})^* = \delta_{a,a'} \), where \( \delta_{a,a'} \) is the Kronecker delta, ensuring that the contraction of any two adjacent tensors results in an identity matrix. This condition simplifies computations by preserving orthogonality as operations proceed from right to left. Conversely, an MPS is in left-canonical form if each tensor \( B^{[i]} \) satisfies \( \sum_{j} (B^{[i]}_{a,j,b})^* B^{[i]}_{a,j,b'} = \delta_{b,b'} \), facilitating operations from left to right. By maintaining both these forms, we achieve the mixed canonical form where tensors to the left of a site are left-canonical and those to the right are right-canonical, therewith I can efficiently perform local optimizations and contractions. This approach reduces numerical errors while making computations quicker (sometimes absurdly so) \cite{MPS:schollwock2011density}.

\subsubsection{"Hermitize" MPSs}
\label{subsec:BlockBP:kagome:numerical_stability:hermitize}

In the simulation of quantum dynamics using double-layered PEPS (Fig. \ref{fig:bp:bpExample:sub2}; as we have in our work), it is crucial to maintain the Hermitian nature of the MPSs involved.
This requirement arises because each MPS, being double-legged, represents operators rather than pure states, necessitating Hermiticity to ensure this represent a physically valid operation. 
To achieve this, I employ a process that transforms the MPS into a Hermitian form. 

Initially, the MPS which inherently possesses a fused ket-bra structure, is treated as a Matrix Product Operator (MPO) by unfusing the physical legs, denoted as \( A_{\{I\}} \rightarrow A_{\{(i,j)\}} \). Subsequently, the Hermitian conjugate of this MPO is computed by transposing the indices and taking the complex conjugate, resulting in \( A^\dagger = (A_{\{(j,i)\}})^* \)%
. The tensors are then refolded back into the MPS format by refusing the double-legs. Finally, to ensure Hermiticity, I construct a new MPS by averaging%
\footnote{Adding MPSs using the approach from \cite{MPS:schollwock2011density}.} %
the original and its conjugate: \( B = \frac{A + A^\dagger}{2} \). This symmetrization guarantees that the resulting MPS is Hermitian, though, being a mere heuristic approach, may introduce other errors that deviate the state from what was intended.

    \chapter{Imaginary Time Evolution}
\label{chapter:ITE}

Imaginary Time Evolution (ITE) is a powerful computational technique widely used in quantum physics to find the ground states of Hamiltonians. In quantum mechanics, the Hamiltonian is an operator that represents the total energy of a system. The ground state of a Hamiltonian is the state with the lowest possible energy, and it plays a crucial role in understanding the fundamental properties of quantum systems.

ITE adapts the operations that evolve a quantum state in real time by transforming them into a process akin to cooling the system to its lowest energy configuration. This transformation involves replacing real time with imaginary time in the evolution equations. This method is particularly advantageous because it systematically reduces the energy of a given trial state, even if it starts as a random configuration, converging towards the ground state as the imaginary time progresses.
\section{Imaginary Time Evolution Fundamentals} 
\label{sec:ite:fundamentals}

\newcommand{\dt}{\delta t}

The Hamiltonian in quantum mechanics plays a key role by defining the evolution of quantum states in the \schrodinger{} picture. 
A state at time 
$t$ 
obeys \schrodinger{}'s equation: 
\begin{equation}
    H\Ket{\psi(t)} = 
    i \hbar \frac{\partial}{\partial t} \Ket{\psi(t)} 
\label{eq:ite:schrodingersEquation}
\end{equation}
where 
$H$ 
is the Hamiltonian of the system.
Therefrom we can generate an operator called the time-evolution operator, which evolves the state up to time $t$ from the state at time $0$. The time-evolution operator is $U(t)=e^{-iHt/\hbar}$. Thus we can write:

\begin{equation}
    \ket{\psi(t)}
    =
    e^{-iHt/\hbar}
    \ket{\psi(0)}
\label{eq:ite:timeEvolution}
\end{equation}
We can write the state $\ket{\psi(0)}$ in its spectral-decomposition to then replace the eigen-energies in the exponent:
Then,
\begin{equation}
    \ket{\psi(t)}
    \underset{\eqref{eq:ite:timeEvolution}}{=}
    e^{-iHt/\hbar}
    \ket{\psi(0)}
    \eqBecause{%
        \parbox{2cm}{\scriptsize \centering
            spectral\\decomposition
        }
    }
    \sum_i 
    c_i
    e^{-iHt/\hbar}
    \ket{\phi_i}
    \eqBecause{\text{eigenvalue}}
    \sum_i 
    c_i
    e^{-iE_{i}t/\hbar}
    \ket{\phi_i}
\label{eq:ite:timeEvolutionSuperPosition}
\end{equation}
where $E_i$ is the eigenenergy of state 
$
\ket{\phi_i}
$
.

Imaginary Time Evolution (ITE) involves replacing real time with imaginary time 
$
t \rightarrow -it
$ 
, thus creating the following equation%
\footnote{This equation helps in numerical works like this one, but one must be careful assuming any contested physical validity of the resulted operator}%
:
\begin{equation}
    \ket{\psi(-it)}
    =
    e^{-Ht/\hbar}
    \ket{\psi(0)}
    \underset{\eqref{eq:ite:timeEvolutionSuperPosition}}{=}
    \sum_i 
    c_i
    e^{-E_i t/\hbar}
    \ket{\phi_i}
\label{eq:ite:imaginaryTimeEvolutionSuperPosition}
\end{equation}
The higher the state's eigen-energy $E_i$ is, the smaller its coefficient 
$
    c_i
    e^{-E_i t/\hbar}
$
is. We can't expect 
$
\ket{\psi(t)}
$
to be a legitimate quantum state with norm $1$, but this can be easily solved by dividing whatever state is achieved by its norm.
If the ground state $\ket{\phi_0}$ (which has the lowest energy) has non-zero coefficient $c_0$, then it will decrease the least by applying Eq. \ref{eq:ite:imaginaryTimeEvolutionSuperPosition}.
For a large enough $t$, the initial weights $c_i$ are negligible, leaving us solely with the ground state $\ket{\phi_0}$. 
In practice the time evolution operator can only be applied for small time steps $\delta t$ at each iteration (see discussion below). 
After each such step, the state is normalized; and after sufficient iterations this will leave all states comprising $\ket{\psi(0)}$ with coefficients close to $0$, save the ground state which persists.

In many cases, the Hamiltonian $H$ is a sum of non-commuting operators %
$
    H
    =
    \sum_{k=1}^{K}
    H_k
$%
. 
This fundamentally undermines the applicability of this equation term by term, since:
$
    e^{-H t/\hbar} 
    =
    e^{-(H_1+H_2) t/\hbar} 
    \neq
    e^{-H_1 t/\hbar} 
    e^{-H_2 t/\hbar} 
$
,
if
$
    H=H_1+H_2
$
and $\left[H_1,H_2\right]\neq0$
.
Using the Suzuki-Trotter method 
\cite{QI:suzuki1991general, QI:morales2022greatly}%
,
the Hamiltonian $H$ is broken into the smallest set of pairwise non-commuting operators %
$
    H
    =
    \sum_{k=1}^{K}
    H_k
$,
such that %
$
    \left[
        H_{k}, H_{k'}
    \right] \neq 0
$
where each $H_k$ might contain a sum of several other terms $H_k = \sum_{i} h_k^{i}$, as long as they commute between themselves %
$
    \left[
        h_{k}^i, h_{k}^j
    \right] = 0
$
. 
Then,
if a $\mathcal{O}(\dt^3)$ error suffices, a symmetric product of the operators is applied. I.e.:
\begin{equation}
    e^{-H \dt/\hbar} 
    =
    e^{-
        \sum_{k}
        H_k
        \dt/\hbar
    } 
    =
    e^{-
        H_1
        \frac{\dt}{2\hbar}
    }
    e^{-
        H_2
        \frac{\dt}{2\hbar}
    }
    \cdots
    \underbrace{
        e^{-
            H_K
            \frac{\dt}{2\hbar}
        }
        e^{-
            H_K
            \frac{\dt}{2\hbar}
        }
    }_{%
        e^{-
            H_K
            \frac{\dt}{\hbar}
        }        
    }
    \cdots
    e^{-
        H_2
        \frac{\dt}{2\hbar}
    }
    e^{-
        H_1
        \frac{\dt}{2\hbar}
    }
    +
    \mathcal{O}(\dt^3)
    \label{eq:ite:2nd_order_trotterization}
\end{equation}
This product is commonly called the \nth{2} order Trotterization of the \ite{} operator.

The \nth{2} order Trotterization was used in this work, as we have found a total of six non-commuting terms in the Hamiltonian we have used. These are $2$-local operators on each unique edge in our unit-cell (see Subsec. \fullsubsubref{subsec:BlockBP:kagome:modes}).

Here I have presented the general \ite{} process on a quantum system without discussing how it is represented.
As I have established thus far, there are several ways to represent quantum systems: the most intuitive one would be to use a vector for the state, but in many-body quantum physics, the size of this vector increases exponentially making \ite{} impractical.
This is where Tensor-Networks can help, a feature that will be explored in the next section. 

\section{ITE for PEPS} 
\label{sec:ite:4PEPS}

\def\id{1\kern-0.25em\text{l}}

Given a PEPS representation of a state $\Ket{\psi}$ (see introduction in Subsec. \fullsubsubref{subsec:intro:TN:mps_and_peps:peps}), we can evolve it with imaginary time according to some Hamiltonian $H$ until it reaches the
ground state of that Hamiltonian.
$H$ is commonly expressed as a sum of nearest-neighbors interaction. 
In many cases, this causes terms acting on the same site to be non-commuting. 
We can understand why this is the case with a simple example using $2$-local pauli-gates on a system with three sites:
Say we have a Hamiltonian %
$
    H = H_1 + H_2
    = \id \otimes X \otimes X
    + Y \otimes Y \otimes \id
$%
. These two simple terms do not commute since
$
    \left[ H_1 , H_2 \right]
    = 2iY \otimes Z \otimes X \neq 0
$%
.
Thus, each application of the \ite{} process (something we frequently call an "ITE-step") may not include more than a single operator acting on a site, forcing a configuration of gates similar to a brick-wall. Fig. \ref{fig:ite:peps:product1} provides a visual example.


\tdplotsetmaincoords{60}{30} 

\setlength{\pepsLineWidth}{0.6mm}  
\def\pepsW{4} 
\def\pepsH{4}

\def\pepsSep{2.4}
\def\pepsSepZ{2.5}

\pgfmathtruncatemacro{\pepsWminusOne}{\pepsW - 1}  
\pgfmathtruncatemacro{\pepsHminusOne}{\pepsH - 1}

\newlength{\cubeWDZ}   \setlength{\cubeWDZ}{0.35mm} 
\newlength{\cubeWidth} \setlength{\cubeWidth}{0.6mm} 
\def\cubeOppacity{0.8}

\newcommand{\minTwoValues}[2]{%
    \ifnum#1<#2 #1\else #2\fi
}
\newcommand{\maxTwoValues}[2]{%
    \ifnum#1>#2 #1\else #2\fi
}

\newcommand{\clr}[1]{blue!#1}

\newcommand{\drawCubeAtPEPS}[5]{
    
    \def\xFirst{#1}
    \def\yFirst{#2}
    \def\xSecond{#3}
    \def\ySecond{#4}
    \def\z{#5}

    \def\minX{\minTwoValues{\xFirst}{\xSecond}}
    \def\maxX{\maxTwoValues{\xFirst}{\xSecond}}
    \def\minY{\minTwoValues{\yFirst}{\ySecond}}
    \def\maxY{\maxTwoValues{\yFirst}{\ySecond}}

    \pgfmathsetmacro{\xoneTemp}{\minX*\pepsSep - \cubeWidth * 0.5}
    \pgfmathsetmacro{\xtwoTemp}{\maxX*\pepsSep + \cubeWidth * 0.5}
    \pgfmathsetmacro{\yoneTemp}{\minY*\pepsSep - \cubeWidth * 0.5}
    \pgfmathsetmacro{\ytwoTemp}{\maxY*\pepsSep + \cubeWidth * 0.5}
    \pgfmathsetmacro{\zoneTemp}{\z*\pepsSepZ}
    \pgfmathsetmacro{\ztwoTemp}{\z*\pepsSepZ + \cubeWDZ }

    \edef\xone{\xoneTemp}
    \edef\xtwo{\xtwoTemp}
    \edef\yone{\yoneTemp}
    \edef\ytwo{\ytwoTemp}
    \edef\zone{\zoneTemp}
    \edef\ztwo{\ztwoTemp}

    \coordinate (A) at (\xone,\yone,\zone);
    \coordinate (B) at (\xtwo,\yone,\zone);
    \coordinate (C) at (\xtwo,\ytwo,\zone);
    \coordinate (D) at (\xone,\ytwo,\zone);
    \coordinate (E) at (\xone,\yone,\ztwo);
    \coordinate (F) at (\xtwo,\yone,\ztwo);
    \coordinate (G) at (\xtwo,\ytwo,\ztwo);
    \coordinate (H) at (\xone,\ytwo,\ztwo);

    \definecolor{bluepurple}{rgb}{0.5, 0.0, 0.5}
    
    \fill[fill=\clr{50}, opacity=\cubeOppacity] (A) -- (B) -- (C) -- (D) -- cycle; 
    \fill[fill=\clr{90}, opacity=\cubeOppacity] (A) -- (B) -- (F) -- (E) -- cycle; 
    \fill[fill=\clr{90}, opacity=\cubeOppacity] (B) -- (C) -- (G) -- (F) -- cycle; 
    \fill[fill=\clr{90}, opacity=\cubeOppacity] (C) -- (D) -- (H) -- (G) -- cycle; 
    \fill[fill=\clr{50}, opacity=\cubeOppacity] (A) -- (D) -- (H) -- (E) -- cycle; 
    \fill[fill=\clr{70}, opacity=\cubeOppacity] (E) -- (F) -- (G) -- (H) -- cycle; 
    
    \draw[thick] (A) -- (B) -- (C); 
    \draw[thick] (A) -- (E) -- (F) -- (B); 
    \draw[thick] (C) -- (G) -- (F); 
    \draw[thick] (H) -- (G); 
    \draw[thick] (E) -- (H); 

    \pgfmathsetmacro{\textCenterX}{0}
    \pgfmathsetmacro{\textCenterY}{0}
    \pgfmathsetmacro{\textCenterZ}{0}
    \pgfmathsetmacro{\textRotation}{0}

    \def\cubeText{\scriptsize $
        exp \left( -
            h_{
                (\xFirst, \yFirst  )
                }^{
                (\xSecond, \ySecond)
            }
            t
        \right) 
    $}
    
    \ifnum\numexpr\maxX-\minX>\numexpr\maxY-\minY\relax
        \pgfmathsetmacro{\tX}{(\xone + \xtwo) /2 }
        \pgfmathsetmacro{\tY}{\yone}
        \pgfmathsetmacro{\tZ}{(\zone + \ztwo) /2 }
        \pgfmathsetmacro{\tRot}{-16}
    \else
        \pgfmathsetmacro{\tX}{\xtwo}
        \pgfmathsetmacro{\tY}{(\yone + \ytwo) /2 }
        \pgfmathsetmacro{\tZ}{(\zone + \ztwo) /2 }
        \pgfmathsetmacro{\tRot}{45}
    \fi
    
    \node[anchor=center, text=white] at (\tX,\tY,\tZ) {\rotatebox{\tRot}{\cubeText}};

}

\newcommand{\drawLegs}[1]{
    \def\z{#1}
    \foreach \x in {1,...,\pepsW}
    {
        \foreach \y in {1,...,\pepsH}
        {    
            \pgfmathsetmacro{\xStart}{\pepsSep*\x}
            \pgfmathsetmacro{\yStart}{\pepsSep*\y}
            \pgfmathsetmacro{\zStart}{\pepsSepZ*\z+\cubeWDZ} 
            \pgfmathsetmacro{\xEnd}{\pepsSep*\x}
            \pgfmathsetmacro{\yEnd}{\pepsSep*\y}
            \pgfmathsetmacro{\zEnd}{\pepsSepZ*(\z+1)} 

            \draw[line width=\pepsLineWidth] (\xStart,\yStart,\zStart) -- (\xEnd,\yEnd,\zEnd);
        }
    }
}

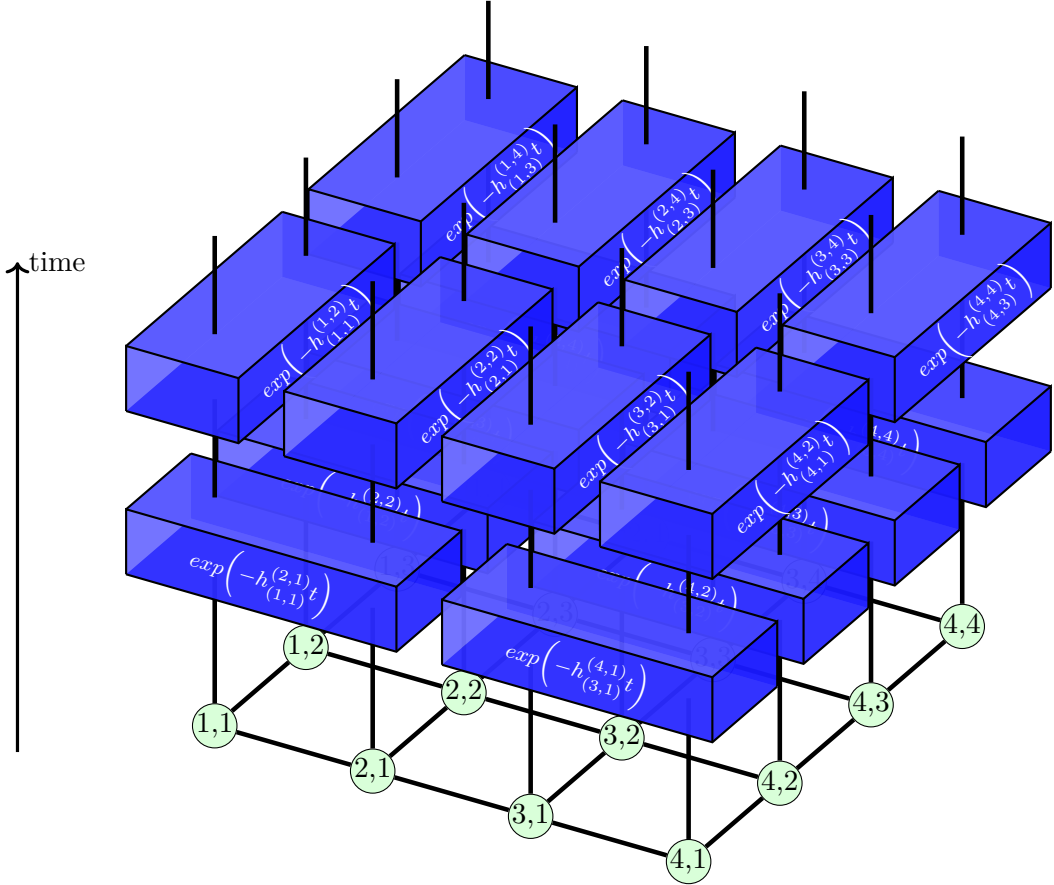
\begin{figure}[htbp]
\centering
\begin{tikzpicture}[tdplot_main_coords, scale=1.0] 
    \tikzset{tensor/.style={circle, draw=black, fill=green!15, minimum size=10pt, inner sep=0pt}}

    \foreach \x in {1,...,\pepsW}
        \foreach \y in {1,...,\pepsH}
        {
            \pgfmathsetmacro{\xpos}{\x*\pepsSep}
            \pgfmathsetmacro{\ypos}{\y*\pepsSep}
            \node[tensor, draw=black]  (\x-\y) at (\xpos,\ypos,0) {\x,\y};
        }
    
    \foreach \x in {1,...,\pepsWminusOne}
        \foreach \y in {1,...,\pepsH}
        {
            \pgfmathtruncatemacro{\nextx}{\x + 1} 
            \draw[line width=\pepsLineWidth] (\x-\y) -- (\nextx-\y);
        }
    
    \foreach \x in {1,...,\pepsW}
        \foreach \y in {1,...,\pepsHminusOne}
        {
            \pgfmathtruncatemacro{\nexty}{\y + 1} 
            \draw[line width=\pepsLineWidth] (\x-\y) -- (\x-\nexty);
        }
        
    \foreach \x in {1,...,\pepsW}
        \foreach \y in {1,...,\pepsH}
        {
            \pgfmathsetmacro{\tempLegX}{\pepsSep*\x}
            \pgfmathsetmacro{\tempLegY}{\pepsSep*\y}
            \pgfmathsetmacro{\tempLegZ}{\pepsSepZ} 

            \edef\legX{\tempLegX pt}
            \edef\legY{\tempLegY pt}
            \edef\legZ{\tempLegZ pt}

            \draw[line width=\pepsLineWidth] (\x-\y) -- (\legX,\legY,\legZ);
        }

   
    \foreach \x in {1,3,...,\pepsWminusOne}
    {
        \foreach \y in {\pepsH,...,1}
        {    
            \pgfmathtruncatemacro{\nextX}{\x + 1} 
            \drawCubeAtPEPS{\x}{\y}{\nextX}{\y}{1}
        }
    }
    \drawLegs{1}
        
    \foreach \x in {1,...,\pepsW}
        \foreach \y in {\pepsHminusOne,...,3,1}
        {    
            \pgfmathtruncatemacro{\nextY}{\y + 1} 
            \drawCubeAtPEPS{\x}{\y}{\x}{\nextY}{2}
        }
    \drawLegs{2}

    \draw[very thick, ->] (0.5,0.5,0) -- (0.5,0.5,\pepsSepZ*3) node[anchor=west] {time};
  
\end{tikzpicture}
\caption[ITE gates on PEPS]{%
    ITE steps applied on a $\pepsH{}\times{}\pepsW{}$ PEPS with Nearest-neighbors interaction. 
    For each operators $h_{i}^{j}$ describing the interaction between site $i$ and site $j$, we can construct the effective 4-legged ITE tensor 
    $
        exp \left( -
            h_{
                i
                }^{
                j
            }
            t
        \right) 
    $
    .
    An application of ITE involves contracting the tensors at two sites, via their physical legs, with their effective 4-legged ITE tensors, but we can only pack so many such tensors at a time. Specifically, looking at the contraction order as layers in time, we can never apply more than a single tensor acting on any site at a time. This results in the Jenga tower shown here.
}
\label{fig:ite:peps:product1}

\end{figure}

For each operator $h_{i}^{j}$ in the sum that constitutes the Hamiltonian, the interaction between site $i$ and site $j$ yields an effective 4-legged \ite{} tensor 
$
    exp \left( -
        h_{
            i
            }^{
            j
        }
        t
    \right) 
$
.
An application of \ite{} involves contracting the tensors at two sites, via their physical legs, with their effective 4-legged \ite{} tensors (see Fig. \ref{fig:ite:peps:product1}).
To reduce errors in the \ite{} process, we repeat the gates in a symmetric order, to adhere to the \nth{2} order Trotterization in Eq. \ref{eq:ite:2nd_order_trotterization}.

Contracting the tensors at neighboring sites $(i,j)$ with the effective \ite{} tensor via their physical legs, increases the
bond dimension $D$ of the virtual leg connecting the sites. Therefore, in order to
keep the \ite{} process tractable for many steps, we need to truncate
the bond dimension after applying the gate. In the most
general way, the \ite{} algorithm can be described by two steps that
are applied for each bond $(i,j)$ and are repeated until 
convergence to the ground state is reached \cite{TN:Foundations:orus2014practical}:
\begin{enumerate}
    \item \textbf{Evolution}: 
    $\Ket{\psi'}
    =
        exp \left( -
        h_{i}^{j}t
    \right) 
    \Ket{\psi}
    $
    
    \item \textbf{Truncation}: $\Ket{\psi} = 
      \text{truncate}(\Ket{\psi'})$
\end{enumerate}

While the evolution step is relatively straightforward, the
truncation step can be implemented in many different ways depending
on the desired accuracy and computation time.
While just performing SVD and then ignoring the least significant singular values is an easy option; %
in this work we chose to implement the Alternating Least Squares (ALS) method \cite{PEPS:ALS:reducedEnv:cirac}%
, 
where one optimizes a single tensor while keeping the rest of the tensors fixed. %
This approach allows to incorporate the surrounding environment into the optimization process, %
although the main challenge lies in computing this environment, which cannot be done exactly.

To address this challenge, we utilize \blockBP{} to approximate the larger environment and then contract the BlockTN into the smaller EdgeTN to obtain the more local environment (see Fig \ref{fig:blockBP:contraction:theEdge}). 
In the following sections, we describe the ITE update step algorithm using the environment obtained from the \blockBP{} algorithm:

\subsection{ITE update step}
\label{subsec:ite:4PEPS:iteStep}

\tdplotsetmaincoords{60}{30} 
\setlength{\pepsLineWidth}{0.6mm}  

\definecolor{mainEdgeColor}{rgb}{0.5,0.1,0.1}
\definecolor{envEdgeColor}{rgb}{0.5,0.45,0.4}

\tikzset{tensor/.style={circle, draw=black, minimum size=20pt, inner sep=0pt}}
\tikzset{envTensor/.style={tensor, fill=orange!50, minimum size=10pt}}
\tikzset{edge/.style={line width=\pepsLineWidth}}
\tikzset{envEdge/.style={edge, envEdgeColor}}
\tikzset{mainEdge/.style={edge, mainEdgeColor}}

\newcommand{\drawEdgeKet}[5]{

    \def\d{#1}    
    \def\D{#2}
    \def\tOne{#3}
    \def\tTwo{#4}
    \def\z{#5}
 
    \node[envTensor]  (e1) at ( 0  , 1, -1.2) {$e_1$};
    \node[envTensor]  (e2) at (-1  , 0, -1.2) {$e_2$};
    \node[envTensor]  (e3) at ( 0  ,-1, -1.2) {$e_3$};
    \node[envTensor]  (e4) at (\d  ,-1, -1.2) {$e_4$};
    \node[envTensor]  (e5) at (\d+1, 0, -1.2) {$e_5$};
    \node[envTensor]  (e6) at (\d  , 1, -1.2) {$e_6$};

    \foreach \i in {1,...,6}
    {
        \draw[edge] (e\i) -- ++(0,0,-0.5);
    }
    
    \foreach \i in {1,...,6}
    {
        \pgfmathtruncatemacro{\j}{mod(\i,6) + 1} 
        \draw[envEdge] (e\i) to[bend right, looseness=0.2] (e\j);
    }
    
    \node[tensor, fill=green!40]  (ti) at (0 ,0,0) {\tOne};
    \node[tensor, fill=blue!40]   (tj) at (\d,0,0) {\tTwo};

    \draw[edge] (e1) to[bend right] (ti);
    \draw[edge] (e2) to[bend left ] (ti);
    \draw[edge] (e3) to[bend left ] (ti);
    \draw[edge] (e4) to[bend left ] (tj);
    \draw[edge] (e5) to[bend right] (tj);
    \draw[edge] (e6) to[bend right] (tj);

    \draw[mainEdge] (ti) -- (tj) node[midway, above, text width=2em, align=center] {\D};
    \draw[mainEdge] (ti) -- (0 ,0,\z);
    \draw[mainEdge] (tj) -- (\d,0,\z);
       
}

A single \ite{} step using Tensor-Networks and the environment achieved by \blockBP{} is described here. A visualization is provided in Fig. \ref{fig:ite:peps:iteStep}.
\begin{figure}[htbp]
    \centering
    \setlength{\subwidth}{0.49\linewidth}  
    \begin{subfigure}[b]{1.0\subwidth}
        \centering            \definecolor{envEdgeColor}{rgb}{0.5,0.45,0.4}
\definecolor{mainEdgeColor}{rgb}{0.5,0.1,0.1}

\def\d{2}

\begin{tikzpicture}[tdplot_main_coords, scale=1.5] 

    \tikzset{tensor/.style={circle, draw=black, minimum size=10pt, inner sep=0pt}}
    \tikzset{envTensor/.style={tensor, fill=orange!50}}
    \tikzset{edge/.style={line width=\pepsLineWidth}}
    \tikzset{envEdge/.style={edge, envEdgeColor}}
    \tikzset{doubleEdge/.style={edge, double distance=1pt}}
    \tikzset{mainEdge/.style={doubleEdge, mainEdgeColor}}
    
    \node[tensor, fill=green!40]  (ti) at (0 ,0,0) {$T^{old}_i$};
    \node[tensor, fill=blue!40]   (tj) at (\d,0,0) {$T^{old}_j$};
        
    \draw[mainEdge] (ti) -- (tj) node[midway, above, text width=2em, align=center] {$D^2$};
 
    \node[envTensor]  (e1) at ( 0   , 1, 0) {$e_1$};
    \node[envTensor]  (e2) at (-1   , 0, 0) {$e_2$};
    \node[envTensor]  (e3) at ( 0   ,-1, 0) {$e_3$};
    \node[envTensor]  (e4) at (\d   ,-1, 0) {$e_4$};
    \node[envTensor]  (e5) at (\d+1 , 0, 0) {$e_5$};
    \node[envTensor]  (e6) at (\d   , 1, 0) {$e_6$};

    \foreach \i in {1,...,6}
    {
        \pgfmathtruncatemacro{\j}{mod(\i,6) + 1} 
        \draw[envEdge] (e\i) to[bend right, looseness=0.2] (e\j);
    }
    
    \foreach \i in {1,...,3}
    {
        \draw[doubleEdge] (e\i) -- (ti);
    }
    
    \foreach \i in {4,...,6}
    {
        \draw[doubleEdge] (e\i) -- (tj);
    }
  
\end{tikzpicture}
        \caption{%
            $\braket{\psi^{old}|\psi^{old}}$. The "braket" EdgeTN that we can get by contracting an entire block besides two neighboring tensors and their immediate environment. The starting bond dimension is $D$. Since its a "braket" TN, the virtual leg is actually of dimension $D^2$. The tensors $e_{1\cdots 6}$ represent the environment and form a pMPS (See Subsec. \fullsubsubref{subsec:intro:TN:mps_and_peps:pmps}).
        }
        \label{fig:ite:peps:iteStep:sub1}
    \end{subfigure}
    \hfill
    \begin{subfigure}[b]{1.0\subwidth}
        \centering
        \input{Chapters/ITE/figs/tikz_art/ite_step/ket_edge_tn}
        \caption{%
            $e^{-ht} \ket{\psi^{old}}$. The "ket"  EdgeTN acquired be replacing the "braket" $T^{old}_i$ and $T^{old}_j$ tensors with their "ket" versions $t^{old}_i$ and $t^{old}_j$, while leaving the double-layerd environment unchanged. The \ite{} operator tensor is connected to the system via the physical legs. The bond dimension is $D$.
            Since the sites are now "kets", each of the environment tensors are connected via only a single leg.
        }
        \label{fig:ite:peps:iteStep:sub2}
    \end{subfigure}
    
    \medskip 
    
    \begin{subfigure}[b]{1.0\subwidth}
        \centering  \begin{tikzpicture}[tdplot_main_coords, scale=1.5] 
    \drawEdgeKet{1}{$D'$}{$t'_i$}{$t'_j$}{-2.1}   
\end{tikzpicture}
        \caption{%
            $\ket{\psi'}$. After contracting the 4-legged ITE operator tensor with the "ket" tensors, we obtain a single tensor which we divide into two by performing SVD (see Fig. \ref{fig:intro:tn:svd}). The resulting bond dimension increases to $D'>D$.
        }
        \label{fig:ite:peps:iteStep:sub3}
    \end{subfigure}
    \hfill
    \begin{subfigure}[b]{1.0\subwidth}
        \centering            \begin{tikzpicture}[tdplot_main_coords, scale=1.5] 
    \drawEdgeKet{1}{$D$}{$t_i$}{$t_j$}{-2.1}   
\end{tikzpicture}
        \caption{%
            $\ket{\psi}$. After performing ALS the resulting tensors are very close to the exact tensors ($\ket{\psi'}$) but their bond-dimension has been lowered to $D$ while taking into account the environment tensors $e_{1\cdots 6}$.
        }
        \label{fig:ite:peps:iteStep:sub4}
    \end{subfigure}
    \caption{%
        ITE step
    }
    \label{fig:ite:peps:iteStep}
\end{figure}
Consider two neighboring sites $i$ and $j$, and the environment tensors achieved by contracting all block tensors except tensors $T_i$, $T_j$ and their immediate neighbors (Fig. \ref{fig:ite:peps:iteStep:sub1})
.
The $\ket{ket}$ tensors $t^{old}_i$ and $t^{old}_j$ are subject to a 4-legged local imaginary time step operator tensor 
$
    exp \left(
        - h^i_j t
    \right)
$
(Fig. \ref{fig:ite:peps:iteStep:sub2})
. It results
in an exact state 
$\ket{\psi'}$
with two tensors
$t'_i, t'_j$
whose mutual bond dimension $D'$ is larger than the original bond dimension
$D$
(Fig. \ref{fig:ite:peps:iteStep:sub3})%
.
The goal is to find new tensors
$t_i$, $t_j$
with bond dimension $D$ such that their resulting state 
$\ket{\psi}$
is as close as possible to the exact state
$\ket{\psi'}$%
. Therefore, we want to minimize
$
    \lVert 
        \ket{\psi} - \ket{\psi'}    
    \rVert ^2
$
. 
We define the environment of sites $i$ and $j$ by 
$N_{ij}$%
. This is a tensor that remains unchanged during the ITE on sites $i$ and $j$.

To express the target function as computable values that we can get using the Tensor-Networks framework, we first write (Eq. \ref{eq:ite:4peps:before_and_after_tn_scalars}):
\begin{equation}
\label{eq:ite:4peps:before_and_after_tn_scalars}
\begin{split}
    \lVert 
        \psi
    \rVert ^2
    & = 
    \textsc{contract} \left(
        t^*_i t^*_j N_{ij} t_i t_j
    \right)
    \\
    \lVert 
        \psi'
    \rVert ^2
    & = 
    \textsc{contract} \left(
        t'^*_i t'^*_j N_{ij} t'_i t'_j
    \right)
    \\
    \braket{\psi|\psi'}
    & = 
    \textsc{contract} \left(
        t^*_i t^*_j N_{ij} t'_i t'_j
    \right)
\end{split}
\end{equation}
Then we can write the minimization function 
\begin{equation}
\label{eq:ite:4peps:minimization_function1}
    \Phi(t_i, t_j)
    =
    \lVert 
        \ket{\psi} - \ket{\psi'}    
    \rVert ^2
    =
      \lVert \psi \rVert ^2
    + \lVert \psi' \rVert ^2
    - \braket{\psi|\psi'}    
    - \braket{\psi'|\psi}    
\end{equation}
From here, the ALS process is iterative and (as the name suggests) alternating: %
We start with 
$
    t_i=t'_i, 
    t_j=t'_j
$
.
At each iteration we fix $t_j$, and find $t_i$ that minimizes $\Phi$ (Eq. \ref{eq:ite:4peps:minimization_function1}), afterwards we fix $t_i$ and find $t_j$, and repeat until we are satisfied by the convergence of $\Phi$. 
The result is two new tensors $t_i$, $t_j$ that are close to the exact solution representing the state after local \ite{}, but with a lower bond-dimension $D$
(Fig. \ref{fig:ite:peps:iteStep:sub4}).
By simply contracting $t_{i/j}$ to its conjugate self via their physical legs, we get $T_{i/j}$ which we can replace at the original TN.

\begin{figure}
    \centering
    \includegraphics[width=0.5\linewidth]{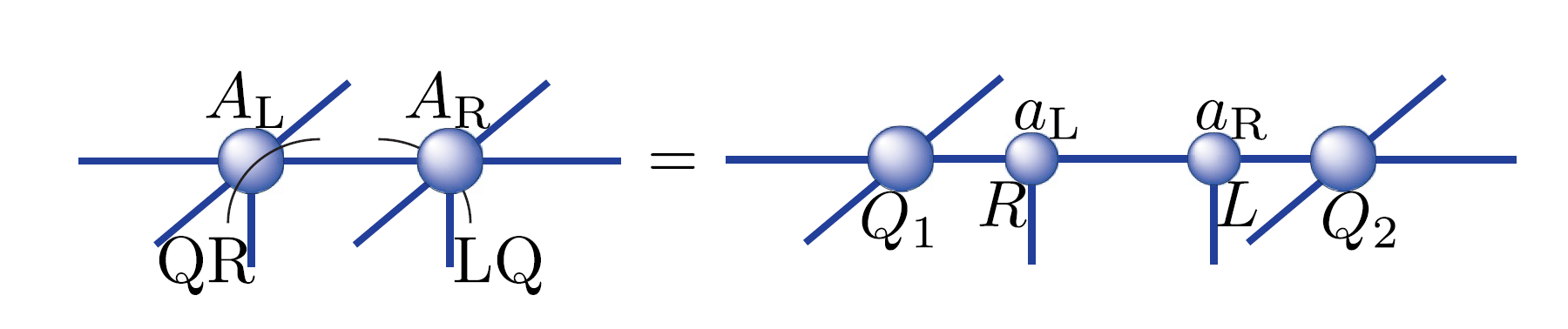}
    \caption[Reduced-Environment for ITE update]{Performing QR (and LQ) Decomposition to simplify the environment for a faster ITE update. Image from \cite{PEPS:ALS:reducedEnv:cirac}.}
    \label{fig:ite:4peps:reducedEnv}
\end{figure}

The process described here includes the entire environment of the tensors where each $5$-legged tensor has $d D'^4$ parameters to be updated. The actual process that we have implemented in this work follows the reduced-environment approach from \cite{PEPS:ALS:reducedEnv:cirac}. By performing $QR$ decomposition on each tensor and delegating the $Q$-tensor to be part of the environment while updating only the $3$-legged $R$-tensor (See Fig. \ref{fig:ite:4peps:reducedEnv}), we can reduce the amount of parameters to be updated to $d^2 D'^2$, %
simplify the environment tensor $N_{ij}$, while getting overall similar results.

\subsubsection{Time Complexity}
\label{subsec:ite:4peps:complexity}

Assuming the pMPS surrounding the lattice tensors (with physical dimension $d$) was produced via \bubblecon{} with truncation bond-dimension $\chi \propto D^2 $, then the environment tensors are connected between themselves with bond-dimension $\chi$ and to the lattice tensors with dimension $D'>D $ (each leg in the double-layered representation).
From \cite{PEPS:ALS:reducedEnv:cirac}, the update process using this reduced environment is lead by a time cost of
(Eq. \ref{eq:als:timeComplex}): %
\begin{equation}
    \label{eq:als:timeComplex}
    \texttt{time: }
    \mathcal{O}\left(
        d^2 D'^4 \chi^3
        +
        d^6 D'^6
    \right)
\end{equation}

\subsection{Full ITE}
\label{subsec:ite:4PEPS:full_ite}

Using both the \blockBP{} algorithm, helping us to acquire an approximation for the environment of two neighboring sites, and the single ITE step on PEPS described in \ref{subsec:ite:4PEPS:iteStep} --- we can now formulate a process of finding the ground state of infinite lattices in
Algorithm \ref{alg:full_ite}.
%
%
\SetKw{iIn}{ in }
\SetKw{iFor}{for}
\renewcommand{\cmnt}[1]{\text{ } \tcp{#1}}
\renewcommand{\Cmnt}[1]{\tcp{** #1 ** //}}
\def\blockParams{block\text{-}params}
\begin{algorithm}
\caption{Full ITE using Block-BP}
\label{alg:full_ite}
\DontPrintSemicolon
\LinesNumbered 
    \KwIn{$\blockParams$\cmnt{instructions for Block creation and arrangement}}
    \KwIn{$H$\cmnt{nearest-neighbors Hamiltonian terms for every edge in Block}}
    \KwIn{$\delta t$-s\cmnt{time-steps for ITE}}
    \KwOut{$Block$}
    \Cmnt{Initialize:}
    create $Block$ with random tensors according to $\blockParams$\;
    \Cmnt{Main loops:}
    \ForEach{$\delta t \iIn \delta t$-s}
    {
    \label{algo:line:full_ite:forEachDt}
        \ForEach{$edge \iIn Block$} 
        {
        \label{algo:line:full_ite:forEachEdge}
            \Cmnt{Get an updated version of the tensors connected by $edge$:}
            $Messages \gets \texttt{blockBP}(Block)$
            \cmnt{(Algorithm \ref{alg:BlockBP_singleBlock})}
            $EdgeTN \gets \texttt{contract}\left(
                Block\cup Messages \text{ except } edge \cup \text{neighbors}
            \right)$\cmnt{(Fig. \ref{fig:ite:peps:iteStep:sub1})}
            $T_i,T_j \gets \texttt{ITE-step}\left(      
                EdgeTN, \delta t, H\left[edge\right] 
            \right) \text{ with ALS}
            $
            \cmnt{(Subsec. \ref{subsec:ite:4PEPS:iteStep})}
            \Cmnt{Replace $edge$ tensors in Block:}
            Replace sites $i,j$ \iIn $Block$ according to $\blockParams$ 
            \label{algo:line:full_ite:replaceTensorsInBlock}
        }
    }
    \KwRet{$Block$}
\end{algorithm}

{
    \setlength{\parindent}{0pt} 
    
    In line \ref{algo:line:full_ite:forEachEdge},
    iterating over all edges can be done twice, the second time being a symmetric copy of the first time, to adhere to the symmetric form of the \nth{2} order Trotterization described in Eq. \ref{eq:ite:2nd_order_trotterization}.
}

In line \ref{algo:line:full_ite:replaceTensorsInBlock},
$\blockParams$ might enforce replicating the tensors into multiple sites. This happens, for example, when the block represents a unit-cell that is repeated throughout an infinite lattice.

\section{ITE for Kagome using BlockBP} 
\label{sec:ite:KagomeBlockBP}

Now that the ITE process was described for PEPS using \blockBP{}, I can describe in details the implementation thereof for the antiferromagnetic (AFM) Heisenberg model on the infinite \kagome{} lattice.

\begin{figure}
    \centering
    \includegraphics[width=0.5\linewidth]{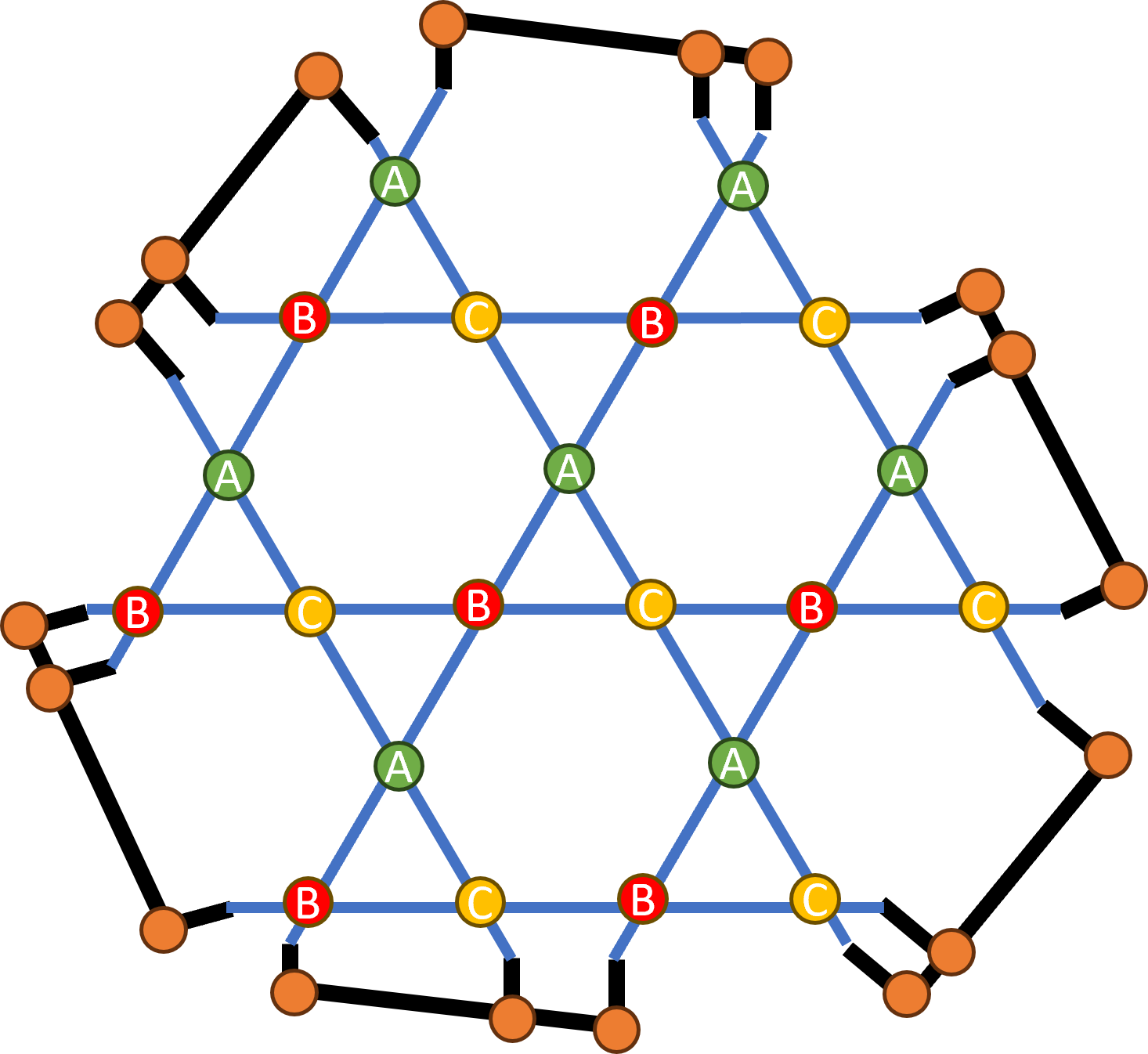}
    \caption[Hexagonal Block with repeated 3-sites unit-cell]{Hexagonal Block for the Kagome lattice with repeated 3-sites unit-cell and MPS messages}
    \label{fig:ite:4kagome:blockAndMessages}
\end{figure}

Following the General ITE algorithm for PEPS using \blockBP{} \ref{alg:full_ite}, we start by constructing a BlockTN (see Sec. %
\ref{sec:blockBP:infiniteKagomeBlock}) by repeating the 3-sites unit-cell into an hexagonal block of predefined size $N$ (see Fig. \ref{fig:ite:4kagome:blockAndMessages}).
We then surround the block with MPSs resulted from the \blockBP{} algorithm (Algorithm \ref{alg:BlockBP_singleBlock}) representing the effect of an infinite environment on the block.
We then contract the entire TN besides two neighboring sites and their immediate neighbors. If the \blockBP{} algorithm converged, the resulting periodic-MPS (see Subsec. \fullsubsubref{subsec:intro:TN:mps_and_peps:pmps}) reflects the infinite environment around those two sites.

\begin{figure}[htb]
    \centering
    \includegraphics[width=0.55\textwidth]{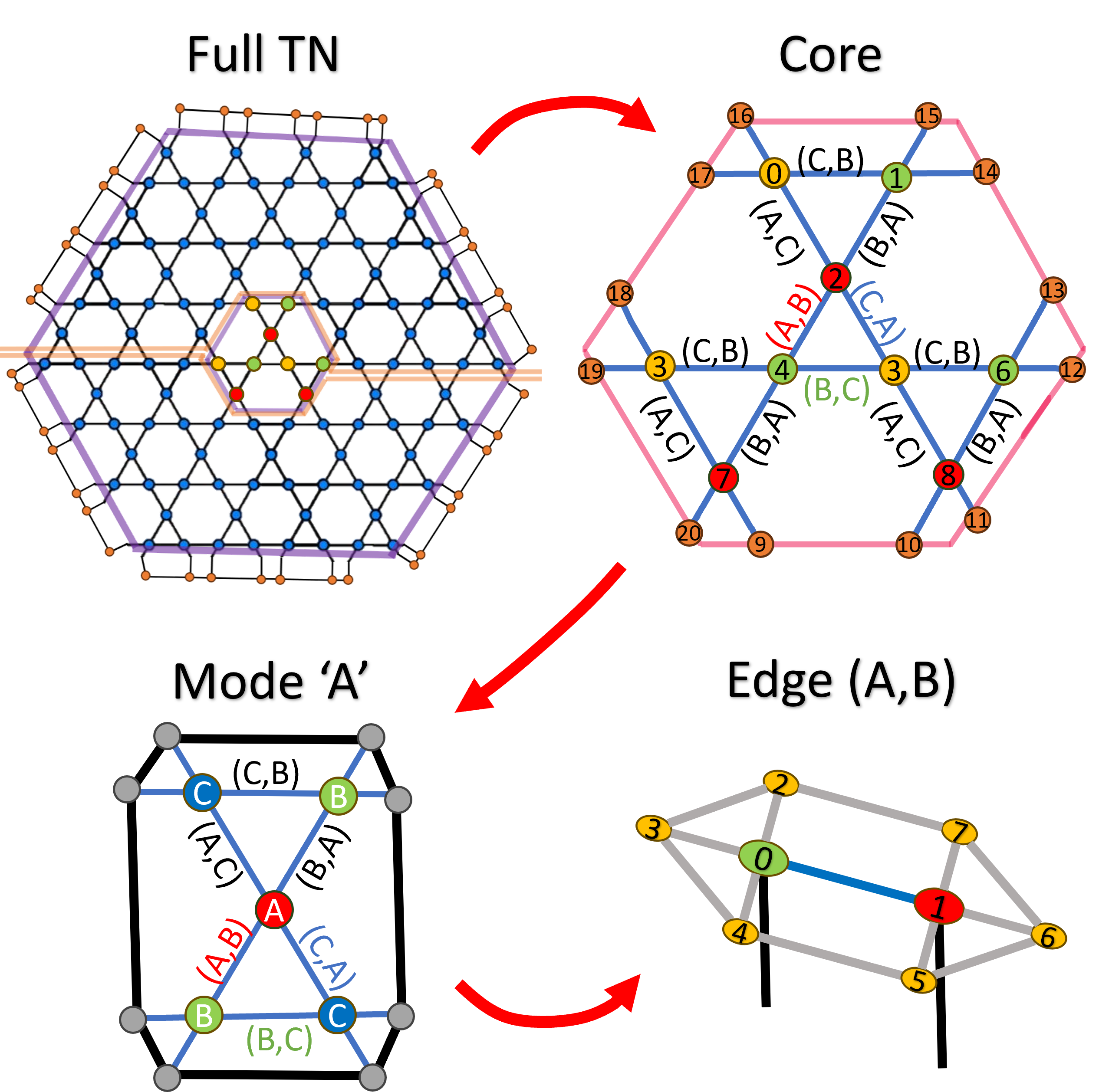}
    \caption[From BlockTN to CoreTN to ModeTN to EdgeTN]{Contracting the Tensor-Network from a large block into $2$-sites edge. The mode here is Node 'A' though it is randomly chosen between 'A' 'B' or 'C'. The Edge shown is Edge(A,B) though we can contract from any ModeTN to every EdgeTN equivalently.}
\label{fig:ite:4kagome:contractingBlockToEdge}
\end{figure}

\begin{figure}[htb]
    \centering
    \includegraphics[width=0.55\textwidth]{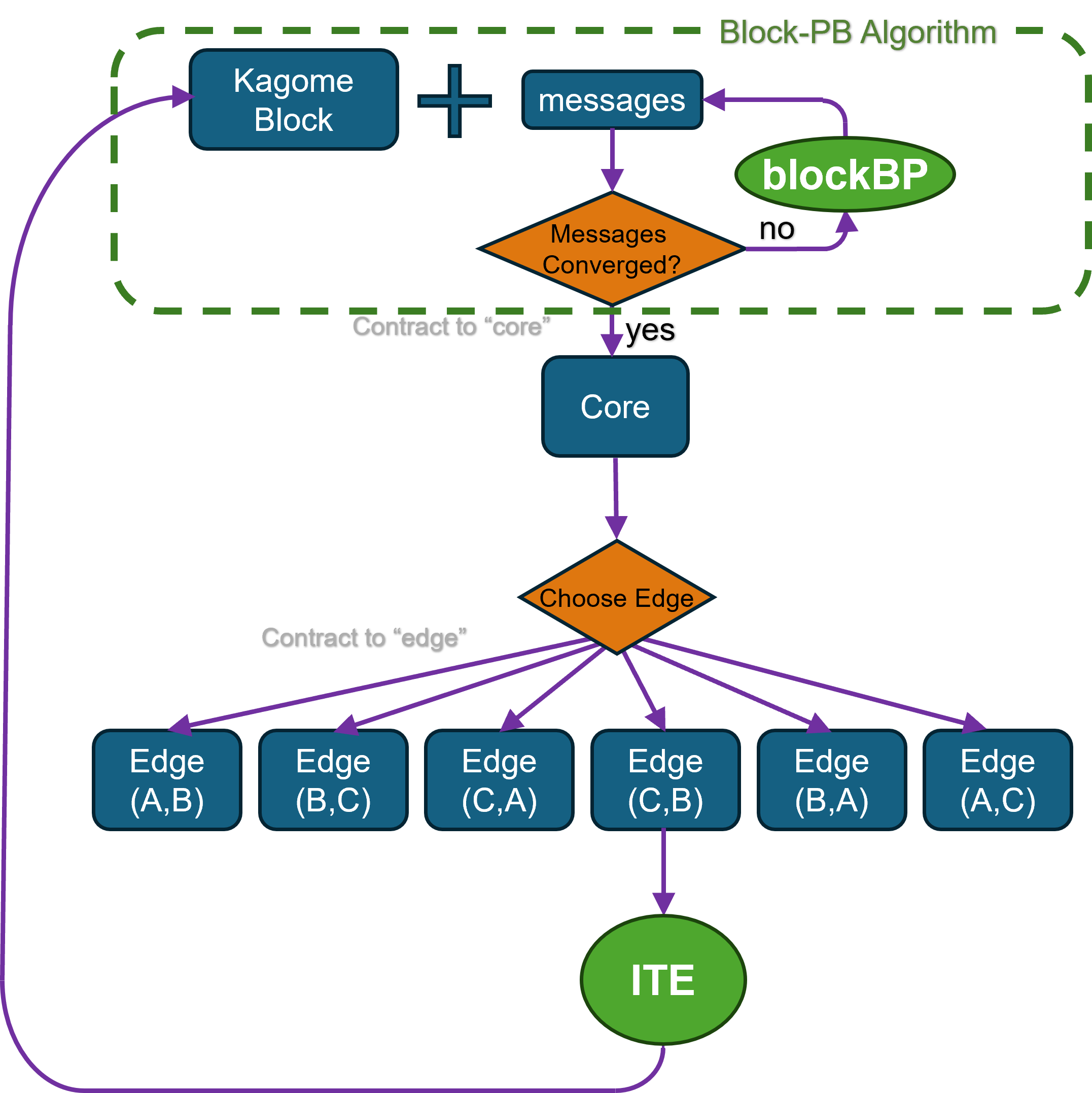}
    \caption[The ITE process]{%
        A "birds-eye view" of the ITE process on the Kagome lattice using ITE and \blockBP{}
    }
    \label{fig:ite:entireProcess}
\end{figure}

Since a single block contains multiple occurrences of the same $2$-site bonds (i.e., an edge within the repeated unit-cell), we limit the edges on which we apply ITE to only those contained within the CoreTN (See Fig. \ref{fig:blockBP:contraction:toCore:c}). Working solely on the core situated in the center of the block far from the block's boundaries, provides a bit of robustness against ill-effects caused by imperfect MPS messages or effects of broken entanglement between the messages%
\footnote{Refer to discussion at the end of section \ref{subsec:blockBP:BlockBPAlgo}.}%
. The reduction of the TN from the BlockTN to the CoreTN is described in Subsec. %
\ref{sec:blockBP:kagomeBlock:contractingKagomeBlock} and a graphical representation is shown in Fig. %
\ref{fig:ite:4kagome:contractingBlockToEdge}.
Fig. \ref{fig:ite:entireProcess} shows a summary of the entire process of finding the ground state of 2-local Hamiltonians, with ITE and \blockBP{}, tailored for the Kagome lattice using an hexagonal Block.
%

\subsubsection{Further stabilization of the algorithm}
\label{subsec:ite:4kagome:transformations}
Here, I present some alternations to the straightforward process discussed thus far, in ways that, though theoretically introduce no difference, might add numerical stability and better overall results.

As discussed in subsection \fullsubsubref{subsec:BlockBP:kagome:modes},
the CoreTN contains $3$ equivalent substructures, called Modes, each containing all unique $2$-site bonds --- once. In order to prevent bias toward one such mode, we shuffle order between those modes as an intermediate step in the contraction from the big BlockTN to the EdgeTN.

As discussed in subsection \ref{sec:blockBP:kagomeBlock:defineKagomeBlock:comparison}, one of the beneficial features of the hexagonal block is its preservation of the 6-fold rotational symmetry of the \kagome{} lattice. We leverage this property by rotating the block at each step. The rotation is chosen randomly among $120^{\circ}$, $240^{\circ}$, or \emph{no-rotation}. This approach provides additional protection against potential biases by completely transforming the entire block and all its tensors, while introducing no substantive change.


\subsubsection{Time and Space Complexity}
\label{subsec:ite:kagome:complexity}

The time and space cost of each repetition of line \ref{algo:line:full_ite:forEachDt} in Algorithm \ref{alg:full_ite} for the \kagome{} block, can be derived from those of the \blockBP{} process ( \fullsubsubref{subsec:blockBP:complexity}) and ITE update for PEPS process (\fullsubsubref{subsec:ite:4peps:complexity}), %
and adjusted for the structure of this specific block (Subsec. \ref{sec:blockBP:kagomeBlock:defineKagomeBlock}). %
The remaining hyper parameters that a user of our method is free to define are the physical dimension $d$, the bond-dimension $D$, the truncation bond-dimension for the \bubblecon{} process $\chi$ and the size of the block $N$.

Since we are using a unit-cell of 3 sites, each with 4 neighbors, and since the the number of tensors in a block grows as the square of the block-size $N$, then following Eq. \ref{eq:blockBP:spaceComplexity}, the space complexity of each \ite with \blockBP{} is (Eq. \ref{eq:ite:kagome:complexity:space}):
\begin{equation}
    \label{eq:ite:kagome:complexity:space}
    \texttt{space: }
    \mathcal{O}\left(
        d D^ 4 
        + 
        N \chi ^2 
    \right)
\end{equation}
Following Eq. \ref{eq:blockBP:timeComplexity} and Eq. \ref{eq:als:timeComplex}, the time complexity is (Eq. \ref{eq:ite:kagome:complexity:time}):
\begin{equation}
    \label{eq:ite:kagome:complexity:time}
    \texttt{time: }
    \mathcal{O}\left(
        \left(
            d^2 D^4 + N^2  
        \right)  \chi ^3 
        +
        d^6 D^6
    \right)
\end{equation}

    \chapter{Numerical Results}\label{chapter:results}

\section{Approximated contraction of the infinite Kagome lattice with BlockBP}
\label{subsec:res:representability}

In this section, we strive to show that a small block of the \kagome{} lattice (as seen in Fig. \ref{fig:ite:4kagome:blockAndMessages}) with messages achieved by applying the \blockBP{} algorithm, can represent the infinite \kagome{} lattice with good quality.

The best approximation I have for the "true" state of the infinite \kagome{} lattice, is using a very large block %
and contracting it 
using a very large truncation threshold $\chi$ %
(see Subsec. \fullsubsubref{sec:intro:bubblecon}
introducing the \bubblecon{} algorithm)%
.
Using this method I will achieve a density matrix representing the state of the system, where a specific unit-cell is repeated throughout the infinite lattice. 
The chosen unit-cell is the one obtained from the \ite{} process on the antiferromagnetic-Heisenberg Hamiltonian, with further details on that in the next section.
For this method%
\footnote{%
    This computation was performed on Google Colab%
    \cite{googlecolab}
    , which provided the necessary computational resources over several days.
}, %
I have used a block containing $3423$ tensors, pushing the random environment so far away from the central tensors, that its affect on them can be neglected. Here, the \bubblecon{} algorithm for the approximated contraction uses a truncation bond dimension $\chi=250$.
From this contraction I can get a $2$-site density matrix, which I name $\rho_{exact}$.

I will compare the resulting density matrix with the density matrices obtained by the following competing methods:
\begin{enumerate}
    \item  \textbf{Random}: With random environment, we can see how well a block represents the infinite lattice as it grows. The larger the block, the closer we are to the thermodynamic limit. Here the truncation bond dimension for the \bubblecon{} algorithm is $\chi=37$.
    
    \item  \textbf{blockBP}:
    Using \blockBP{} we can effectively simulate larger lattices using only a small block. 
     Here the truncation bond dimension for the \bubblecon{} algorithm is $\chi=37$ when used on the block with the already-converged environment messages. The messages themselves were obtained with $\chi=28$.
    
\end{enumerate}
One should note that both the "Exact" and "Random" methods are quite similar, both initiating random MPSs around the block and then contracting the entire TN using \bubblecon{}, albeit with a very different $\chi$ and TN size.

To compare between the methods, we use the fidelity measure to measure the similarity between the resultant density matrix from each method to $\rho_{exact}$. Fidelity between two density matrices is defined as (Eq. \ref{eq:metrics:fidelity}):
\begin{equation}
    \label{eq:metrics:fidelity}
    F(\rho, \sigma)
    =
    \left(
        \text{Tr} \left(
            \sqrt{
                \sqrt{\rho}
                \sigma 
                \sqrt{\rho}
            }
        \right)
    \right)^2
\end{equation}
Where the the square root of a matrix can be efficiently computed using "Blocked Schur Algorithms" \cite{metrics:MatrixSquareRoot:deadman2012blocked}.
In Fig. \ref{fig:res:representability}, the fidelity is measured between the %
density matrix obtained by the shown method (Either "BlockBP" or "Random") and that of the "Exact" method, $\rho_{exact}$. Similarity is plotted as 1-Fidelity in log-scale.
\begin{figure}[htbp]
    \centering
    \begin{subfigure}{0.49\textwidth}    
        \centering
        \includegraphics[width=1.0\linewidth]{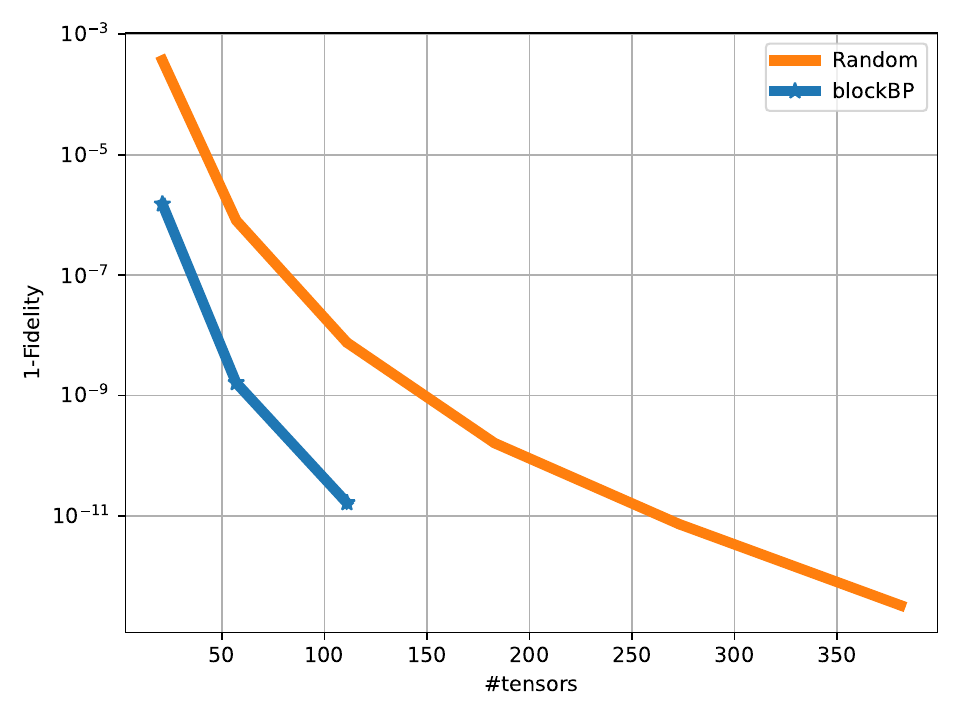}
        \caption{$D=2$.}        \label{fig:res:representability}
    \end{subfigure}
    \hfill
    \begin{subfigure}{0.49\textwidth}    
        \centering
        \includegraphics[width=1.0\linewidth]{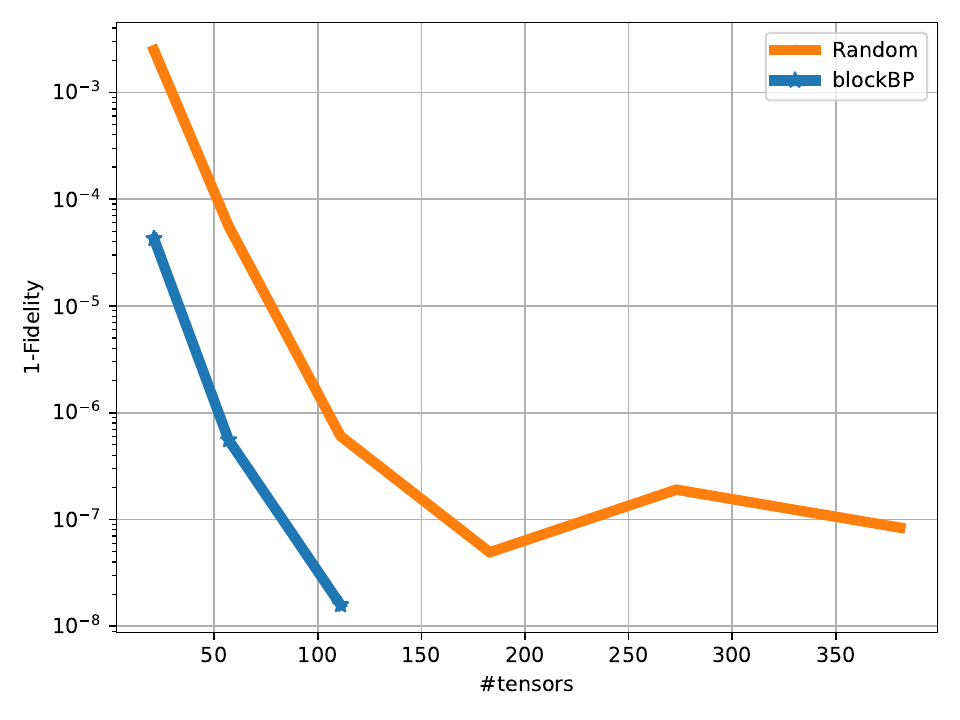}
        \caption{$D=3$.}        \label{fig:res:representability}
    \end{subfigure}
    \caption[Comparing fidelity between BlockBP and random environment]{%
        Fidelity (Eq. \ref{eq:metrics:fidelity}) between "BlockBP" or "Random" method, to the "Exact" method, as a function of the number of tensors used. Shown for two bond dimensions, $D=2,3$.
    }
    \label{fig:res:representability}
\end{figure}
We can see from this figure that \blockBP{} faithfully represents the infinite lattice with just a few PEPS tensors.

\begin{figure}[htbp]
    \centering
    \includegraphics[width=0.49\linewidth]{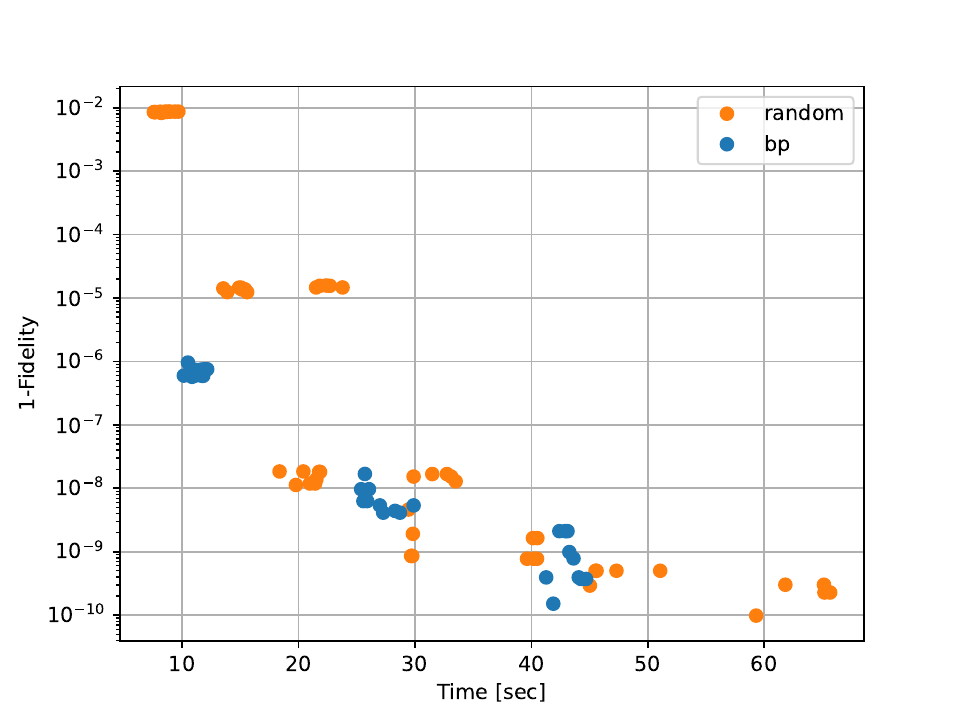}
    \caption[Fidelity vs. computation time for BlockBP and random environment]{%
        Fidelity (Eq. \ref{eq:metrics:fidelity}) between "BlockBP" or "Random" method, to the "Exact" method, as a function of the computation time. Shown for bond dimension $D=3$.
    }
    \label{fig:res:representability:timing}
\end{figure}

While this shows that a small number of tensors can sufficiently approximate the infinite lattice when using \blockBP{}, results are expected to take more time to compute because of the added complexity of generating the MPS messages (See analysis of time complexity in \fullsubsubref{subsec:blockBP:complexity}).
Fig. \ref{fig:res:representability:timing} shows the fidelity to $\rho_{exact}$ as a function of the computation time it took for each method. The code was executed for truncation bond-dimensions $\chi$ in the range $\left[D^2, 2 D^2\right]$, as higher values might produce better results at a higher computational cost (see Eq. \ref{eq:blockBP:timeComplexity}).

We can see that \blockBP{} is preferable when wanting good approximation of the infinite lattice with small computation time. As expected, when the TN grows (and computation time grows with it), the central core gets sufficiently far from the environment MPSs rendering any effect they may have, negligible. Thus, when willing to spend more computation time, using random messages with a big block can be as good and as fast as running \blockBP{} on a small block.  

\section{Approximating the ground state of antiferromagnetic Heisenberg model on the infinite Kagome lattice}
\label{subsec:res:ITEAFMH}

The main goal of this project was to study the infinite \kagome{} lattice under the Anti-FerroMagnetic Heisenberg (AFM-H) interaction (Sec. \ref{sec:intro:kagome}) using ITE and the \blockBP{} algorithm adjusted for the study of infinite lattices. 
After developing the method described in Sec. \ref{sec:ite:KagomeBlockBP}, we've implemented it on different bond-dimensions ($D$). 
Since the ground state (and hence the ground-state energy) of the AFM-H model on the infinite \kagome{} lattice is not known exactly, the goal is to see if the lowest energy we can achieve competes with published results from other state-of-the-art methods.

The AFM Heisenberg Hamiltonian can be written as a sum of nearest-neighbors terms as described in Eq. \ref{eq:res:afm-h:hamiltonian}:
\begin{equation}
    \label{eq:res:afm-h:hamiltonian}
    \mathcal{H}=
    \sum_{\langle i,j \rangle}
    \mathbf{S}_i \cdot \mathbf{S}_j
    =
    \frac{1}{4}
    \sum_{\langle i,j \rangle}
    \sigma_{i}^{x}
    \sigma_{j}^{x}
    +
    \sigma_{i}^{y}
    \sigma_{j}^{y}
    +
    \sigma_{i}^{z}
    \sigma_{j}^{z},
\end{equation}
where $\sum_{\langle i,j \rangle}$ denotes summing over all nearest-neighbor sites $i,j$.
$\mathbf{S}_i = (S_i^x, S_i^y, S_i^z)$ is the spin-1/2 operator at site $i$, where each component corresponds to the spin in the \(x\), \(y\), and \(z\) directions. The dot product \(\mathbf{S}_i \cdot \mathbf{S}_j\) expands to:
\begin{equation}
    \mathbf{S}_i \cdot \mathbf{S}_j = S_i^x S_j^x + S_i^y S_j^y + S_i^z S_j^z
\end{equation}
To convert from a sum of spin operators to a sum of Pauli matrices, we use the relation between the spin operators and the Pauli matrices. For a spin-\(\frac{1}{2}\) system, the spin operators are related to the Pauli matrices \(\sigma^x\), \(\sigma^y\), and \(\sigma^z\) by:
$
    S_i^\alpha = \frac{\hbar}{2} \sigma_i^\alpha
$
where \(\alpha = x, y, z\) and \(\hbar\) is the reduced Planck's constant we have set to $1$ throughout this study.

To quantify frustration in our lattice, we will use a measure called negativity \cite{metrics:negativity:zyczkowski1998volume, metrics:negativity:vidal2002computable}. 
Negativity is a widely used entanglement measure for quantum states represented by density matrices. It quantifies the degree of entanglement in a bipartite quantum system. Given a density matrix \(\rho\) describing a composite system \(AB\), the negativity between subsystems A and B is defined based on the partial transpose operation. The partial transpose of \(\rho\) with respect to subsystem \(B\), denoted as \(\rho^{T_B}\), is obtained by transposing only the indices associated with subsystem \(B\). The negativity \(\mathcal{N}(\rho)\) is then calculated as:
\begin{equation}
\label{eq:res:negativity}
    \mathcal{N}(\rho) 
    = \frac{\|\rho^{T_B}\|_1 - 1}{2}
    = \frac{\|\rho^{T_A}\|_1 - 1}{2}
\end{equation}
where \(\|\cdot|_1\) is the trace norm of the partially transposed density matrix, defined as the sum of the absolute values of its eigenvalues. The negativity is zero for separable states and positive for entangled states, making it a useful tool for detecting entanglement. It provides a quantitative measure of entanglement that is particularly advantageous due to its computational simplicity and its applicability to mixed states. The negativity is a non-convex function, reflecting the complex nature of quantum entanglement.
One can choose either subsystem 
$A$ or $B$ for the partial transpose operation when calculating negativity.

\subsubsection{ITE Process:}
\label{subsec:res:ITEAFMH:ITEprocess}
\begin{figure}[htbp]
    \centering
    \includegraphics[width=0.7\linewidth]{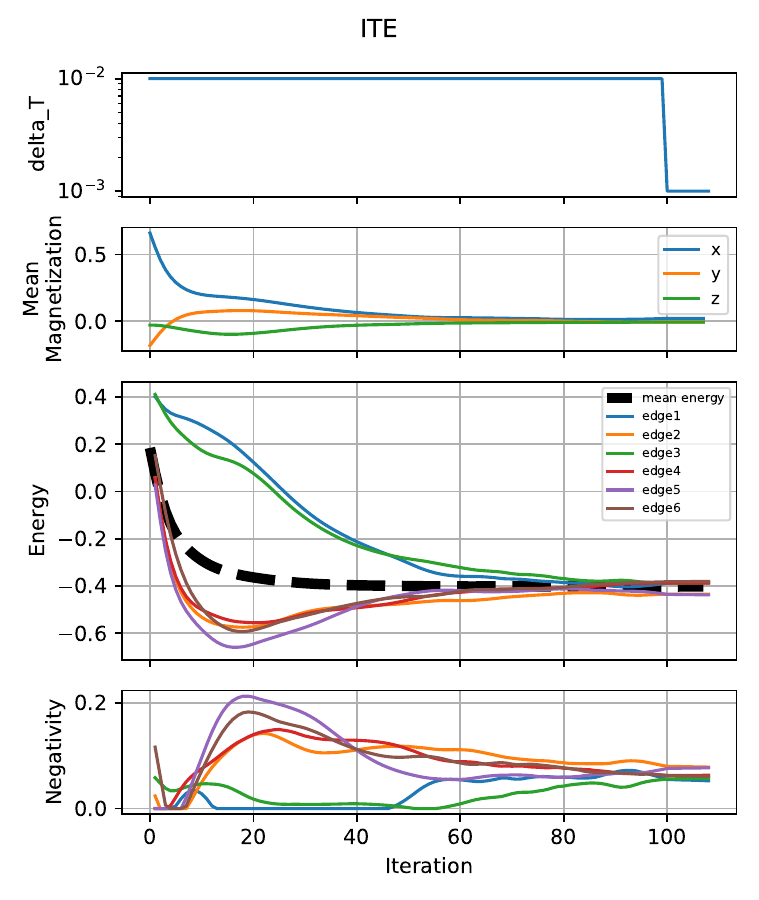}
    \caption[The ITE process with recorded results at each iteration]{%
        ITE process for $D=2$.
        \nth{1} sub-figure shows $\delta t$ per iteration.
        \nth{2} sub-figure shows the mean expectation value $\braket{\psi|\sigma^{\alpha}|\psi}$, $(\alpha\in\left[x, y, z\right])$, at each iteration.
        \nth{3} sub-figure shows the mean energy and energy per-bond measured over all edges, in relation to the AFM-H interaction (Eq. \ref{eq:res:afm-h:hamiltonian}).
        \nth{4} sub-figure shows the negativity measure (Eq. \ref{eq:res:negativity}) between any two sites for all edges in the unit-cell. The six edges are the unique bonds in the unit-cell shown in Subsec. \fullsubsubref{subsec:BlockBP:kagome:modes}.
    }
    \label{fig:res:ITEprocess}
\end{figure}
In Fig. \ref{fig:res:ITEprocess} we can see the ITE process for $D=2$: Starting from a unit-cell comprised of random tensors, we implement the ITE procedure (Sec. \ref{sec:ite:KagomeBlockBP}). 
After each ITE-step we measure the mean expectation values along the $x, y$ and $z$ directions (what can be interpreted as overall-magnetization), the mean-energy with respect to the Hamiltonian from Eq. \ref{eq:res:afm-h:hamiltonian}, and the negativity between two sites of each edge in the unit-cell according to Eq. \ref{eq:res:negativity}.

The magnetization vector (\nth{2} sub-figure) converges towards $0$, strengthening the conjecture that the ground state is not magnetically ordered.

We can see that the state starts far from the ground energy, since the mean energy (\nth{3} sub-figure) over all edges starts at a much higher value $\left(\approx 0.2 \right)$ then drops to much lower values $\left(\approx -0.4 \right)$.
After close to hundred iterations with $\delta t=0.01$, the variational energy approaches values already close to the lowest ones found using PEPS with bond-dimension $D=2$. Further iterations with lower $\delta t$s (not shown in the graph), improve this result, as we will see in the next figure (Fig. \ref{fig:res:ITE_AFM_H_compare}).
Midway between iteration $5$ and $50$ the system is in a highly frustrated state, since $4$ edges enjoy relatively low bond-energies while $2$ pay the price of high bond-energies.

The frustration can also be seen in the negativity per bond (\nth{4} sub-figure), where $4$ edges enjoy what seems like a highly entangled state, while $2$ edges seem to be in a separable state.

\subsubsection{Comparison with state-of-the-art results:}
\label{subsec:res:ITEAFMH:comparison}
\begin{figure}[htb]
    \centering
    
    \includegraphics[width=0.7\textwidth]{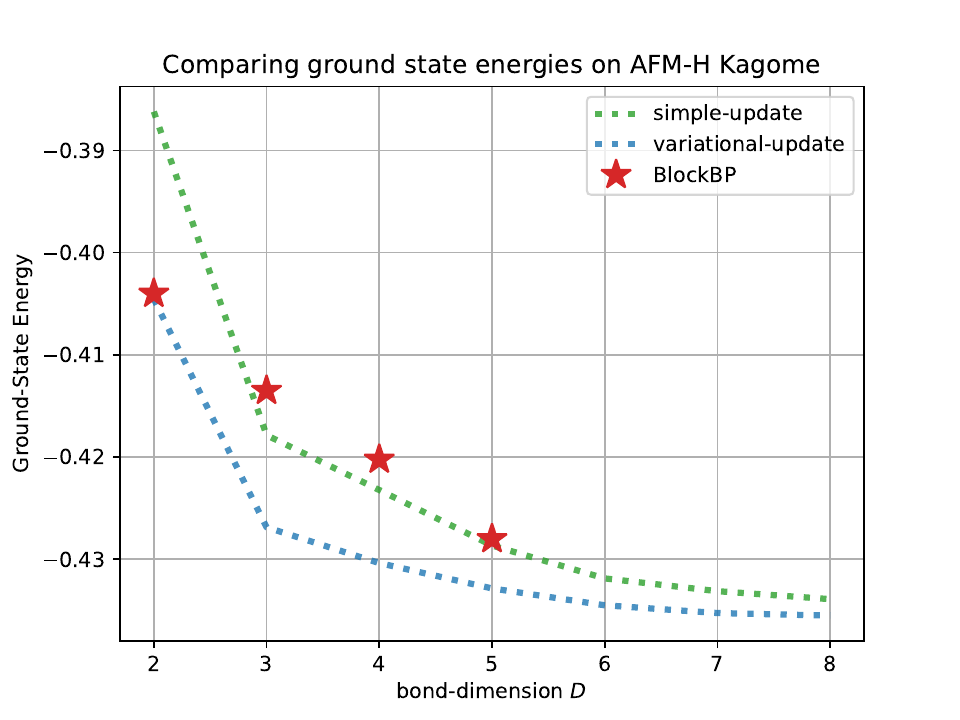}

    \vspace{10pt} 

    \begin{tabular}{c|c|c|c}
            D & SU & VU & BlockBP 
            \\  
            \hline 
            2 & -0.38620 & -0.40454 &             -0.40401 
            \\ 3 & -0.41786 & -0.42688 &             -0.41351 
            \\ 4 & -0.42323 & -0.43038 &            -0.42024 
            \\ 5 & -0.42866 & -0.43286 &            -0.42802 
            \\ 6 & -0.43188 & -0.43451 &  
            \\ 7 & -0.43313 & -0.43527 &  
            \\ 8 & -0.43391 & -0.43552 &  
            \\ 
    \end{tabular}
        
    \caption[Comparing estimations for the ground energy of AFM-H]{%
        Comparing the results of our algorithm. SU=simple-update, VU=Variational-update, both from \cite{varipepsEisert}.
    }
    \label{fig:res:ITE_AFM_H_compare}
    
\end{figure}
In Fig. \ref{fig:res:ITE_AFM_H_compare}, we compare our results to those obtained in \cite{varipepsEisert}, where a variational-update (VU) approach was employed. This approach is among the most successful methods for numerically searching for the ground state of infinite lattices using PEPS. It builds upon the Corner Transfer Matrix Renormalization Group (CTMRG) method \cite{PEPS:orus2009simulation} by integrating automatic differentiation to enhance optimization efficiency and accuracy, allowing for precise gradient calculations. Though both VU \cite{varipepsEisert} and our method use the TN framework, the former searches for the ground state by using an optimization protocol leveraging auto-differentiation tools while the latter implements ITE by successive application of two-site gates in a more traditional straight-forward fashion.

It was already shown %
\cite{ItaiArad:alkabetz2021tensor} %
that the SU method is equivalent to BP (\ref{section:blockBP:introToBeliefPropagation}) and ITE.
Here we compare our results to those from \cite{varipepsEisert} using the SU method and the novel VU method. 
We can see in Fig. \ref{fig:res:ITE_AFM_H_compare} that our method is on par with the SU and competes with VU for $D=2$, getting close to the best known ground-state energy for the AFM Heisenberg model on the infinite \kagome{} lattice using $D=2$.

The results achieved so far are better for lower $D$, suggestive of the fact that longer computation time with lower $\delta t$, can help us achieve even better results.
Since each imaginary time step operator modifies only one bond at a time, using large \(\delta t\) values can introduce significant frustration, preventing convergence to the ground state. Further simulation time with lower $\delta t$ can address this issue.

It can also be that finite blocks struggle with approximating states with critical entanglement, a subject in need for further research.

Another modification that we suggest trying to improve our results, is the addition of small Gaussian noise to the unit-cell's tensors between some iterations (As was performed by \cite{varipepsEisert} when using SU), to avoid getting stuck in local minima --- a known problem for such ITE processes.

Another factor that should be considered when comparing our method to the ones from \cite{varipepsEisert}, is that we limited ourselves to strictly working with a single repeating unit-cell of 3 sites, while \cite{varipepsEisert} worked with 9 sites. Working with more independent sites can help with convergence and avoid local minima. 
It should be noted that \cite{varipepsEisert} coarse grained their TN by fusing each triangle of 3 sites into a single tensor with physical dimension $d^3$, resulting in a square lattice. After taking this into account, they have $3 d^3 D^4$ parameters to optimize while we have $3 d D^4$ parameters.

As was discussed in Sec. \ref{subsec:blockBP:BlockBPAlgo} and shown in Fig. \ref{fig:BlockBP}(e,f), the MPS messages that surround the block cannot
encode fully the environment of an infinite TN, since the division into several messages breaks
any entanglement that could have existed between different parts of the environment. 
When our ITE process (see Subsec. \fullsubsubref{subsec:res:ITEAFMH:ITEprocess}) achieves states close to the ground energy, entanglement within the system get stronger and far-reaching.
This might explain why we witnessed MPS messages that become less hermitian as we get closer to the ground state, neccesitating the need for the "Hermitization" procedure discussed in Subsec.
\fullsubsubref{subsec:BlockBP:kagome:numerical_stability:hermitize}.
In light of these suspicions,
I acknowledge the need for further research exploring when (if at all) the "Hermitzation" procedure should be applied and what are its consequences on the approximation quality achieved by the environment messages, especially when the TN represents states with high entanglement lengths.

    \chapter{Discussion}
\label{chapter:discussion}

In this thesis, I have contributed to the development and application of the Block Belief-Propagation (\blockBP{}) algorithm for the contraction of tensor networks, with a particular focus on the antiferromagnetic Heisenberg model on the infinite Kagome lattice. This work addresses significant computational challenges in simulating large quantum systems, contributing to the broader field of quantum many-body physics.

The \blockBP{} algorithm was developed to significantly enhance the efficiency and accuracy of tensor network contractions. By coarse-graining the system into blocks and performing belief propagation between these blocks, the algorithm effectively addresses the inaccuracies often encountered with traditional belief propagation methods, especially in systems characterized by high correlations and frustration. Furthermore, since at each iteration all messages can be computed in parallel, a significant speed-up can be achieved by using multi-threading.

The Heisenberg model on the \kagome{} lattice, known for its geometric frustration and highly entangled ground state, served as an ideal testbed for the \blockBP{} algorithm. My research demonstrates that the algorithm has the potential to achieve estimates of the ground-state energies that are on par with state-of-the-art methods.

The ability to efficiently simulate large quantum systems is crucial for advancing our understanding of complex quantum phenomena. The \blockBP{} algorithm represents a significant step forward in this endeavor, offering a scalable and adaptable approach to tensor network contractions. Its application to the Kagome lattice not only provides insights into this particular system but also sets a precedent for tackling similarly challenging problems in condensed matter physics.

Though this work focuses on adjusting \blockBP{} for the study of infinite lattices, it can be easily adapted to finite systems, where it excels compared to other state-of-the-art methods. In non-uniform systems, CTMRG does not apply, while \blockBP{} can be implemented to produce fast results using its intuitive parallelizability. In periodic systems, \blockBP{} works effectively, even when boundary-MPS is not applicable.

The research opens avenues for further exploration of the \blockBP{} algorithm in other lattice geometries and Hamiltonians. Additionally, the potential integration with quantum chemistry problems and less regular structures presents exciting opportunities for future work.

To improve \blockBP{} for the study of infinite lattices, future work could focus on developing an auto-differentiation scheme for \blockBP{}. Such a scheme might perform \blockBP{} to get the environment, while the update is performed by gradient descent instead of with \ite{}. This might close the gap between the results achieved by \blockBP{} and those of the variational update.

This thesis underscores the transformative potential of innovative algorithms like \blockBP{} in the realm of quantum simulations. By bridging the gap between theoretical development and practical application, this work contributes to the ongoing quest to unravel the mysteries of quantum many-body systems, fostering a deeper understanding of the fundamental principles governing the quantum world.
    
    \begin{appendices} 

    \chapter{Appendix: Kagome Block Indexing}
\label{appendix:kagomeBlockIndexing}

This appendix complements the discussion about the structure used for the Kagome-Block in section \ref{sec:blockBP:kagomeBlock:defineKagomeBlock}, by detailing its indexing and naming schemes.

\subsubsection{Indexing Edges}

\begin{figure}[htbp]
    \centering
    \includegraphics[width=0.45\textwidth]{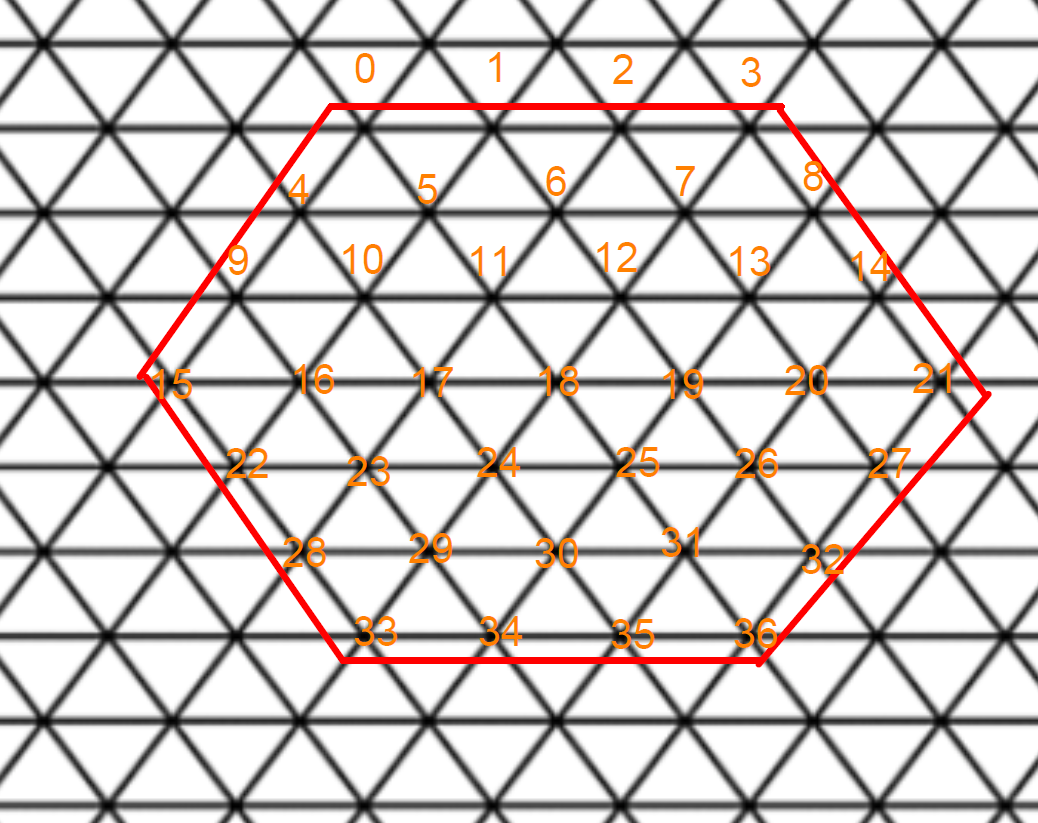}
    \caption[Indexing the block in the triangular code]{
        The indexing in the triangular code used in \cite{Itai:BlockBP}
    }
    \label{fig:indexing:kagome:triangles}
\end{figure}

\begin{figure}[htbp]
    \centering
    \includegraphics[width=0.65\textwidth]{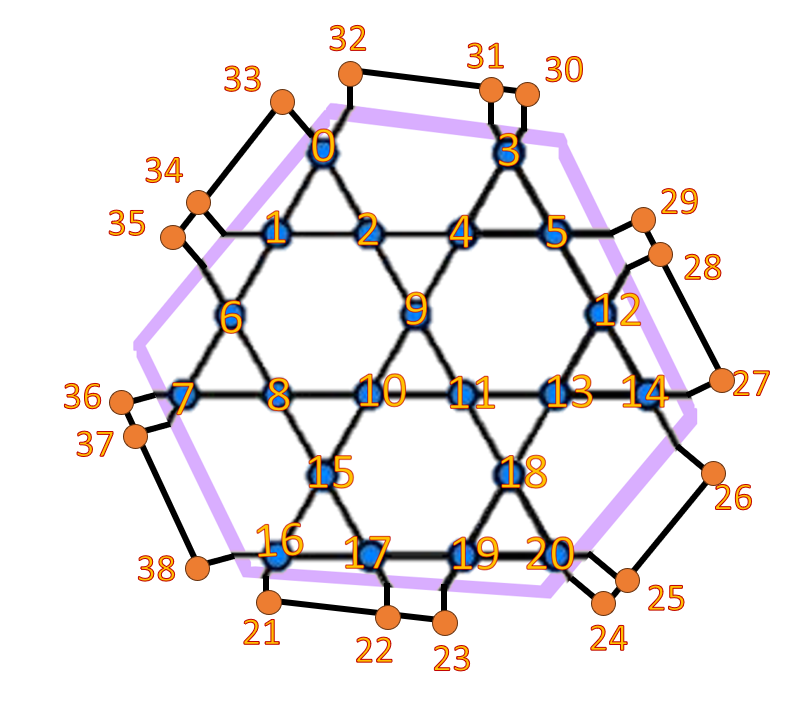}
    \caption[Indexing the block in the kagome BlockBP code]{
        The indexing in the hexagonal block for the Kagome lattice with its MPS messages
    }
    \label{fig:indexing:kagome:all-nodes}
\end{figure}

We can benefit from the work done on \blockBP{} for the triangular lattice in \cite{Itai:BlockBP}, by inheriting the structure and indexing used there (see figure \ref{fig:indexing:kagome:triangles}). Thus, we index each upper-triangle with the same indexing scheme used of the triangular \blockBP{}, (see figure \ref{fig:indexing:kagome:all-nodes}).

The tensors themselves are indexed in the order of the triangles and within the triangle, in order 
\textit{Up} $\rightarrow$
\textit{Down-Left} $\rightarrow$
\textit{Down-Right}. 
Yielding unique indexing of all the nodes within the block.

The messages are ordered \textit{counter-clockwise} starting from the bottom, therewith we get the indexing of the entire block as seen in Fig.
\ref{fig:indexing:kagome:all-nodes}.

\subsubsection{Naming Edges}

\underline{Inner edges} 
get their names from the indices of their connected sites in increasing order, so that tensors \textbf{0} and \textbf{2} are connected by edge \textbf{0-2}.

\begin{figure}[htbp]
    \centering
    \includegraphics[width=0.45\textwidth]{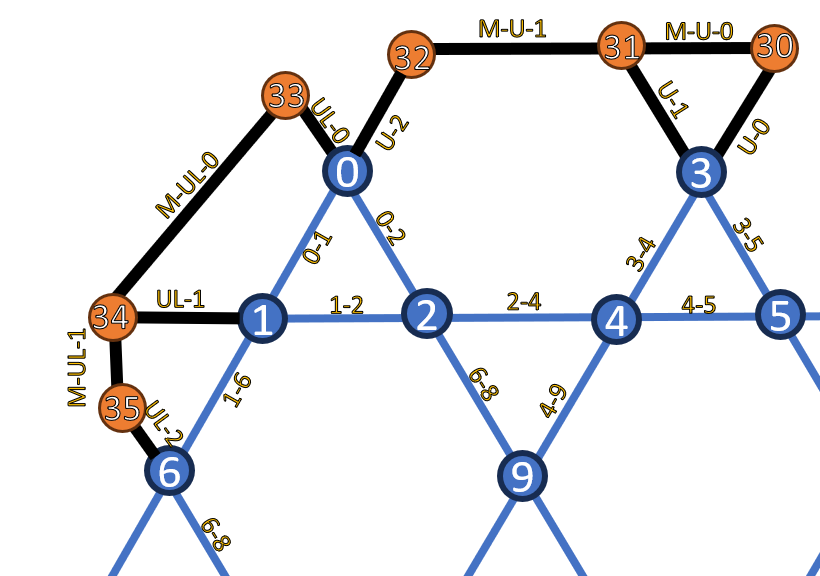}
    \caption[Indexing the edges]{
        The indexing of edges in the hexagonal block
    }
    \label{fig:indexing:kagome:edges}
\end{figure}

\noindent
\underline{Outer edges} do not inherit their name from their tensors, but rather from the ordering of their incoming MPSs (counter-clockwise order). Each name starts with the name of the block-face. 
For example, the \nth{0} tensor located at the upper-left corner has an edge going through the upper face of the block with name \textbf{U-2}, and an edge going through the upper-right face with name \textbf{UL-0} (see figure \ref{fig:indexing:kagome:edges} ).

\noindent
\underline{MPS edges} get the same names as the outgoing edges, but with a prefix \textbf{M-}.

    \chapter{Appendix: Code}
\label{appendix:code}

The code implementing the algorithms discussed in this thesis is available on
\href{https://github.com/NGBigField/KagomePeriodicBP}{github.com/NGBigField/KagomePeriodicBP}.

\end{appendices}

    \printbibliography
    
    \makeTitleHebrew

    
\end{document}